\documentclass[a4paper,11pt]{article}
\pdfoutput=1 
\usepackage{jheppub} 
\usepackage{multirow}
\usepackage{amsmath}
\usepackage{slashed}
\usepackage{amstext,amssymb}
\usepackage{graphicx}
\usepackage{graphicx,wrapfig,lipsum}
\usepackage{xmpmulti}
\usepackage{animate}
\usepackage{epstopdf}
\usepackage{xcolor,colortbl}
\usepackage{multimedia}
\usepackage{hyperref}
\usepackage{url}
\usepackage{xspace}
\usepackage{color}
\usepackage{units}

\newcommand{\mathsym}[1]{{}}

\newcommand{\baz}{\begin{array}{cc}}
\newcommand{\bad}{\begin{array}{ccc}}
\newcommand{\ba}{\begin{array}{c}}
\newcommand{\ea}{\end{array}}
\newcommand{\be}{\begin{equation}}
\newcommand{\ee}{\end{equation}}
\newcommand{\bea}{\begin{eqnarray}}
\newcommand{\eea}{\end{eqnarray}}

\newcommand{\bi}{\begin{itemize}}
\newcommand{\ei}{\end{itemize}}
\newcommand{\bmt}{\begin{pmatrix}}
\newcommand{\emt}{\end{pmatrix}}
\newcommand{\bt}{\begin{tabular}}
\newcommand{\et}{\end{tabular}}

\newcommand{\benu}{\begin{enumerate}}
\newcommand{\eenu}{\end{enumerate}}

\newcommand{\bav}{\begin{array}{cccc}}

\usepackage{wrapfig}
\definecolor{light-gray}{gray}{0.95}
\usepackage{tcolorbox}
\newcommand{\blue}[1]{{\color{blue} #1 }}
\newcommand{\red}[1]{\textcolor{red}{#1}}

\def\ltap{\ \raisebox{-.4ex}{\rlap{$\sim$}} \raisebox{.4ex}{$<$}\ }
\def\gtap{\ \raisebox{-.4ex}{\rlap{$\sim$}} \raisebox{.4ex}{$>$}\ }

\title{Neutrinoless double beta decay in Left-Right symmetric model with double seesaw}
\author[a]{Sudhanwa Patra,}
\author[b,c]{S. T. Petcov \footnote{Also at: Institute of Nuclear Research and Nuclear Energy, Bulgarian Academy of Science, 1784 Sofia, Bulgaria.},}
\author[d]{Prativa Pritimita,}
\author[a,b,e]{Purushottam Sahu}
\affiliation[a]{Department of Physics, Indian Institute of Technology Bhilai, Raipur 492015, India}
\affiliation[b]{SISSA/INFN, Via Bonomea 265, 34136 Trieste, Italy}
\affiliation[c]{Kavli IPMU (WPI), UTIAS, The University of Tokyo, Kashiwa, Chiba 277-8583, Japan}
\affiliation[d]{Department of Physics, Indian Institute of Technology Bombay, Powai, Mumbai 400076, India}
\affiliation[e]{International Centre for Theoretical Physics (ICTP),
Strada Costiera 11, Trieste 34151, Italy}
\emailAdd{sudhanwa@iitbhilai.ac.in}
\emailAdd{prativa@iitb.ac.in}
\emailAdd{psahu@ictp.it}

\abstract{
We discuss a left-right (L-R) symmetric model with 
the double seesaw mechanism at the TeV scale generating Majorana 
masses for the active left-handed (LH) flavour neutrinos 
$\nu_{\alpha L}$ and the heavy right-handed (RH) neutrinos $N_{\beta R}$, 
$\alpha,\beta = e,\mu,\tau$, which in turn  
mediate lepton number violating processes, including 
neutrinoless double beta decay. 
The Higgs sector is composed 
of two Higgs doublets $H_L$, $H_R$ and a bi-doublet $\Phi$. 
The fermion sector has 
the usual for the L-R symmetric models quarks and leptons, along with three
$SU(2)$ singlet fermion $S_{\gamma L}$. The choice of bare Majorana mass term 
for these sterile fermions induces large Majorana masses  
for the heavy RH neutrinos leading to two sets of heavy Majorana particles 
$N_j$ and $S_k$, $j,k=1,2,3$, with masses 
$m_{N_j} \ll m_{S_k}$. Working with a specific version of the model in which the 
$\nu_{\alpha L} - N_{\beta R}$ and the $N_{\beta R} - S_{\gamma L}$
Dirac mass terms are diagonal, and assuming that  
$m_{N_j} \sim (1 - 1000)$ GeV and ${\rm max}(m_{S_k}) \sim (1 - 10)$ TeV,
$m_{N_j} \ll m_{S_k}$, we study in detail 
the new ``non-standard'' contributions to the 
$0\nu\beta\beta$ decay amplitude and half-life 
arising due to the exchange of virtual $N_j$ and $S_k$. 
We find that in both cases of NO and IO light 
neutrino mass spectra, these contributions are strongly enhanced 
and are dominant at relatively small values of the lightest neutrino mass 
$m_{1(3)} \sim (10^{-4} - 10^{-2})$ eV over 
the light Majorana neutrino exchange contribution.
In large part of the parameter space, the predictions 
of the model for the 
$0\nu\beta\beta$ decay generalised effective Majorana 
mass and half-life are within the sensitivity range of the 
planned next generation of neutrinoless double beta decay 
experiments LEGEND-200 (LEGEND-1000), nEXO, KamlAND-Zen-II, CUPID, NEXT-HD.
}


\begin{document} 
\maketitle
\flushbottom
\section{Introduction}
\label{sec:intro}
Neutrino mass and mixing, which was confirmed by oscillation experiments \cite{SNO:2002tuh,Super-Kamiokande:2016yck,T2K:2019efw,DayaBay:2012fng,DoubleChooz:2011ymz} can not be understood within the Standard Model (SM) of particle physics since it predicts massless neutrinos. So, there has to be a mechanism beyond the SM which generates nonzero mass for these tiny particles. The seesaw mechanism has become quite famous for explaining the same by extending the SM in the minimal possible way. Some of the variants of this mechanism are type-I \cite{Minkowski:1977sc, Mohapatra:1979ia, Yanagida:1979as, GellMann:1980vs}, type-II\cite{Magg:1980ut, Schechter:1980gr, Cheng:1980qt, Lazarides:1980nt, Mohapatra:1980yp} and type-III  \cite{Foot:1988aq,He:2012ub} seesaw which can be achieved by adding a right-handed neutrino, a scalar triplet and a fermion triplet to the SM respectively. However, a heavy right-handed scale associated with these seesaw mechanisms renders them unverifiable at the collider experiments. Thus arises the necessity of bringing down the seesaw scale to a verifiable TeV range.
The seesaw mechanisms assume neutrinos are Majorana particles, which can be probed via the lepton number violating process of neutrinoless double beta decay \cite{PhysRevD.25.2951}. Such a rare transition occurs when two neutrons simultaneously decay into two protons and two electrons without any neutrinos. It can be induced either by light left-handed neutrinos, called the standard mechanism or by exotic particles like heavy right-handed neutrinos or sterile neutrinos, called new physics contribution. In the standard mechanism case, the experimental limits on the half-life of the decay can only be saturated by quasi-degenerate \cite{Bilenky:2001rz} light neutrinos, which are disfavored by cosmological data sets \cite{Planck:2018vyg,Abazajian:2022ofy,RoyChoudhury:2019hls,RoyChoudhury:2018gay}. On the other hand, identifying the correct neutrino mass hierarchy, considering the sum of light neutrino masses, would require a multi-ton scale detector that is beyond feasible in the near future. Any comparative future experimental observation of $0\nu\beta\beta$ decay would only be attributed to new physics contribution. 
The current lower limit on the decay half life of $\text{Ge}^{76}$ is $T^{0\nu}_{1/2} > 1.8 \times 10^{26}$ yrs at $90\%$ C.L. from GERDA~\cite{GERDA:2020xhi}. Experiments using the isotope $\text{Xe}^{136}$ like EXO-200~\cite{,EXO-200:2019rkq} and KamLAND-Zen~\cite{KamLAND-Zen:2016pfg,KamLAND-Zen:2022tow} have derived the lower bounds on half-life as $T^{0\nu}_{1/2} > 3.5 \times 10^{25}$ yrs and $T^{0\nu}_{1/2} > 1.07 \times 10^{26}$ yrs respectively. With this motivation, we consider a Left-Right symmetric model with a double seesaw mechanism \cite{PhysRevLett.56.561,PhysRevD.34.1642} as new physics and study the new contributions to $0\nu\beta\beta$ decay process.

Left-Right Symmetric Model (LRSM) \cite{Mohapatra:1974gc,Pati:1974yy,Senjanovic:1975rk,Senjanovic:1978ev} is a well-suited candidate for physics beyond SM for several reasons. To name a few, it can explain the theoretical origin of maximal parity violation in weak interaction, it can incorporate neutrino mass due to the presence of a right-handed neutrino state, it appears as a subgroup of SO(10) Grand Unified Theory, and it can be broken down to SM gauge symmetry at low energies. Moreover, it delivers rich phenomenology if the left-right symmetry breaking occurs at few TeV scale \cite{Keung:1983uu,  Ferrari:2000sp,  Schmaltz:2010xr, Nemevsek:2011hz, Chen:2011hc, Chakrabortty:2012pp, Das:2012ii, AguilarSaavedra:2012gf, Han:2012vk, Chen:2013fna, Rizzo:2014xma,  Deppisch:2014zta, Deppisch:2015qwa, Gluza:2015goa, Ng:2015hba, Patra:2015bga,Dobrescu:2015qna,PhysRevLett.115.211802, Brehmer:2015cia, Dev:2015pga,  Coloma:2015una, Deppisch:2015cua,Dev:2015kca, Mondal:2015zba, Aguilar-Saavedra:2015iew,    Lindner:2016lpp, Lindner:2016lxq, Mitra:2016kov, Anamiati:2016uxp,   Khachatryan:2014dka, Aad:2015xaa, Khachatryan:2016jqo}. The spontaneous symmetry breaking of LRSM to SM plays a vital role in generating neutrino mass through the seesaw mechanism. The seesaw scheme varies with the choice of scalars considered in the left-right model and regulates the associated phenomenology. In general, symmetry breaking can be done with the help of Higgs doublets or Higgs triplets or with the combination of both doublets and triplets. In the case of Higgs doublets, neutrinos don't get Majorana mass, and thus the model forbids any signatures of lepton number violation or lepton flavour violation. In the case of Higgs triplets, neutrino mass is generated via the type-I plus type-II seesaw mechanism. Even though Majorana mass is generated for light and heavy neutrinos in this case, the seesaw can't be probed by experiments considering the high scale associated with it.

The seesaw scale can be brought down to the TeV range in the case of a linear seesaw and inverse seesaw, some of which are discussed in ref.\cite{Hirsch:2009mx,Gu:2010xc,Dev:2009aw,Deppisch:2015cua,Humbert:2015yva,Parida:2012sq,Brdar:2018sbk,ThomasArun:2021rwf,Sahu:2022xkq,Ezzat:2021bzs}. However in case of a linear seesaw and inverse seesaw the light neutrinos are Majorana. In contrast, the heavy neutrinos are pseudo-Dirac, due to which the heavy neutrinos do not play a dominant role in lepton number violation. To study dominant new contributions to LNV and LFV decays, the Higgs and fermion sectors of LRSM have been extended in various refs \cite{Tello:2010am,Barry:2013xxa,BhupalDev:2013ntw,Hernandez:2021uxx,Brdar:2019fur}.

We explore here a double seesaw mechanism \cite{PhysRevLett.56.561,PhysRevD.34.1642} within a left-right symmetric model without Higgs triplets allowing significant lepton number violation and new physics contribution to neutrinoless double beta decay. We keep the scalar sector of the model minimal while adding only one sterile neutrino per generation in the fermion sector.
Even though the Higgs and fermion sectors are the same as in the case of a linear and inverse seesaw, the choice of bare Majorana masses for sterile neutrinos can induce large Majorana masses for heavy RH neutrino as well. The non-zero masses for RH neutrinos are generated through the double seesaw mechanism by implementing seesaw approximation twice. In the first step, the Majorana mass matrix and masses of the RH neutrinos are generated via the type-I seesaw mechanism. In this case, light neutrino mass becomes linearly dependent on a heavy sterile neutrino mass scale. This is how the double seesaw mainly differs from the canonical seesaw mechanism, where the light neutrino masses are inversely proportional to heavy RH neutrino masses. Another essential feature of our model is that we express mass relations between light and heavy Majorana neutrinos in terms of oscillation parameters and the lightest neutrino mass. Thus, it enables us to derive meaningful information about the absolute scale of the lightest neutrino mass and mass hierarchy from the new contributions to the neutrinoless double beta decay process by saturating the current experimental limits.

The plan of the paper can be summarized as follows.
In Section \ref{sec:lrsm}, we give a brief description of the left-right 
symmetric model with the double seesaw mechanism. In Section \ref{numassmixing} 
we explain the implementation of the double seesaw mechanism  
and the origin of Majorana masses for light and heavy right-handed (RH) 
and sterile neutrinos. The 
generation of the masses of gauge bosons associated with the 
$SU(2)_{\rm R}$ gauge group as well the constraints on their masses and 
on their mixing with the Standard Model gauge bosons are also considered 
in this Section. The general expression for the neutrinoless double beta 
decay half-life, including the new physics (i.e., the ``non-standard'') 
contributions is given in Section \ref{sec:0nubb}, in which we 
also, present and discuss briefly the nuclear matrix elements 
of the process and their current uncertainties.  
Detailed phenomenological analysis of the non-standard contributions 
together with numerical estimates of their magnitude are presented in 
Section \ref{sec:PA}. We also give predictions of the considered model 
for the neutrinoless double beta decay ``generalised'' effective Majorana mass 
and half-life accounting, in particular, for the uncertainties of the 
relevant nuclear matrix elements.  
Section \ref{sec:comments} contains brief comments on the 
potential lepton flavour violation and collider 
phenomenology of the considered model. 
Section \ref{sec:concl} contains a summary of the 
results obtained in the present study.
In Appendix \ref{app:lrdssm}, we give a detailed derivation of the 
the masses and mixing of the light and heavy Majorana neutrinos 
in the considered left-right symmetric model with a double seesaw 
mechanism of neutrino mass generation.

\section{Left-Right Symmetric Model with Double Seesaw}
\label{sec:lrsm}
Left-Right symmetric models were proposed with the motivation of restoring 
parity (or left-right) symmetry at a high scale \cite{Pati:1974yy,Mohapatra:1974gc,Senjanovic:1978ev,Senjanovic:1975rk}. Therefore, in the model the 
left- and right-handed fermion fields are  assigned to 
$SU(2)_L$ and $SU(2)_R$ doublets, respectively, which are related 
by a discrete symmetry. 
The complete gauge group, which is an extension of the SM 
gauge group can be written as:
\begin{align}
\mathcal{G}_{LR}\equiv SU(2)_L \times SU(2)_R \times U(1)_{B-L}\,,
\label{eq:LB-LRSM}
\end{align}
%
where $SU(3)_C$ is omitted for simplicity. The electric charge for any particle in this model is defined as 
\begin{align}
Q=T_{3L}+T_{3R}+\frac{B-L}{2}\,.
\end{align}
%
where $T_{3L}$ ($T_{3R}$) is the third component of the isospin 
associated with the $SU(2)_L$ ($SU(2)_R$) gauge group.
The model's fermion sector comprises all the Standard Model fermions plus a right-handed neutrino $N_R$. The fermion fields  with their respective quantum numbers can be written as follows:
\begin{eqnarray}
&&q_{L}=\begin{pmatrix}u_{L}\\
d_{L}\end{pmatrix}\equiv[2,1,1/3]\,, ~ q_{R}=\begin{pmatrix}u_{R}\\
d_{R}\end{pmatrix}\equiv[1,2,1/3]\,,\nonumber \\
&&\ell_{L}=\begin{pmatrix}\nu_{L}\\
e_{L}\end{pmatrix}\equiv[2,1,-1] \,, ~ \quad \ell_{R}=\begin{pmatrix} N_{R}\\
e_{R}\end{pmatrix}\equiv[1,2,-1] \,. \nonumber
\end{eqnarray}
%

The scalar sector is responsible for the spontaneous 
symmetry breaking of LRSM to SM 
and plays a crucial role in deciding the type of seesaw mechanism 
through which neutrino masses can be generated. The left-right symmetry 
breaking can be done either with the help of doublets $H_L$, $H_R$ or 
triplets $\Delta_L$, $\Delta_R$, or with the combination of both doublets and triplets. The next step of symmetry breaking, i.e., the breaking of SM symmetry 
to $U(1)_{em}$, is done with the help of the doublet  $\phi$ contained 
in the bidoublet $\Phi$. The doublets $H_L$, $H_R$ and the bidoublet $\Phi$ 
have the form,
\begin{eqnarray}
&&
 H_L=
 \begin{pmatrix} 
 h_L^+\\
 h_L^0
 \end{pmatrix}\equiv[2,1,1]  \,,\nonumber \\
&&
H_R=
\begin{pmatrix} 
h_R^+  \\
h_R^0
\end{pmatrix}\equiv[1,2,1] \,, \nonumber \\
&&
\Phi=
\begin{pmatrix} 
\phi_{1}^0     &  \phi_{2}^+ \\
\phi_{1}^-     &  \phi_{2}^0
\end{pmatrix} \equiv[2,2,0]\,,\nonumber \\
 &&
\end{eqnarray}
%
The symmetry breaking steps can be sketched as follows:
\begin{eqnarray*}
&\mbox{\bf \underline{\blue{Spontaneous symmetry breaking of LRSM:}}}& \\
&\hspace*{-2cm}  \begin{array}[t]{c} \pmb{SU(2)_L}\\ \{T_L, T_{3L} \} \\ g_L\end{array} \pmb{\times} 
\underbrace{\begin{array}[t]{c} \pmb{SU(2)_R}\\ \{T_R, T_{3R} \} \\ g_R\end{array} \pmb{\times} 
\begin{array}[t]{c} \pmb{U(1)_{B-L}}\\ {\rm B-L}    \\ g_{BL}\end{array}} \\
\\ 
%
&\hspace*{2.5cm} \downarrow  \langle H_R(1,2,1)\rangle \hspace*{0.2cm}  \hspace*{0.2cm} 
           & \\
&\hspace*{-2.0cm} \underbrace{\begin{array}[t]{c} \pmb{SU(2)_L}\\ \{T_L, T_{3L} \} \\ g\equiv g_L\end{array} \hspace*{0.2cm} \pmb{\times} \hspace*{0.2cm}
\begin{array}[t]{c} \pmb{U(1)_Y}\\ Y       \\ g^\prime \end{array}} & \\ 
&\hspace*{3.5cm} \downarrow  \langle \phi(1_{L},1/2_{Y})\rangle \subset \Phi(2_L,2_R,0_{B-L})\quad 
            & \\
&\hspace*{1.0cm}\begin{array}[t]{c} \pmb{U(1)_{\rm em}}\\ \mbox{(Q,~e) }\quad \end{array} 
\begin{array}[t]{c}  \\  \hspace*{0.0cm} \red{Q=T_{3L} +Y} \end{array} 
&
\end{eqnarray*}
%

The first step of symmetry breaking, i.e. 
$SU(2)_R \times U(1)_{B-L} \to U(1)_Y$ is achieved by assigning a non-zero 
 vacuum expectation value (VEV) $\langle H^0_R\rangle$ to the neutral 
component $h^0_R$ of $H_R$  as $v_R$.
The scale of this symmetry breaking 
determines the mass of the heavily charged and neutral gauge
bosons associated with the $SU(2)_R$ symmetry, ${W_R}$ and $Z_R$. 
The scalar doublet $H_L$ doesn't play any role but is present because of 
the left-right invariance. The electroweak symmetry breaking 
i.e., $SU(2)_L \times U(1)_Y \to U(1)_{em}$, is done by assigning 
 non-zero VEVs 
$\langle \phi^0_1 \rangle \equiv v_1$ and $\langle \phi^0_2 \rangle \equiv v_2$ 
to the neutral components of $\Phi$,
with $v_{SM} = \sqrt{v^2_1+v^2_2}\simeq$~246 GeV. 
The neutral components of the scalar bidoublet generate masses for the quarks 
and charged leptons via the following Yukawa Lagrangian:
\begin{eqnarray}
-\mathcal{L}_{Yuk} &\supset& \overline{q_{L}} \left[Y_1 \Phi + Y_2 \widetilde{\Phi} \right] q_R 
+\,\overline{\ell_{L}} \left[Y_3 \Phi + Y_4 \widetilde{\Phi} \right] \ell_R+ \mbox{h.c.}\,,
\label{eqn:LR-Yuk}
\end{eqnarray}
%
where $\widetilde{\Phi} = \sigma_2 \Phi^* \sigma_2$ and $\sigma_2$ 
is the second Pauli matrix. 
When the scalar bidoublet $\Phi$ acquires  non-zero VEVs,
\begin{equation}
\langle \Phi \rangle
=
\begin{pmatrix}
 v_1 & 0 \\
0 & v_2 
\end{pmatrix}\,,
\end{equation}
%
it gives masses to quarks and charged leptons in the following manner:
\begin{eqnarray}
&&  M_u =  Y_1 v_1 + Y_2 v_2\,, \quad \quad \nonumber \\
&&    M_d =  Y_1 v_2 + Y_2 v_1\,, \quad \quad \nonumber \\
&&  M_e =  Y_3 v_2 + Y_4 v_1 \,.\quad \quad
\label{quark_lepton}
\end{eqnarray}
%
Here $M_u$ ($M_d$) and $M_e$ are the up-type (down-type) 
quark and charged lepton mass matrices.
The Lagrangian in Eq. (\ref{eqn:LR-Yuk}) also yields Dirac mass for 
the light neutrinos as
\begin{equation}
M^\nu_D\equiv M_D = Y_3 v_1 + Y_4 v_2 \,.
\label{Dirac1}
\end{equation}
%
In contrast to Yukawa couplings, which are complex, $v_1$ and $v_2$ are here 
assumed to be real. Typically, in the context of left-right symmetric theories, 
one investigates mainly the generation of Majorana neutrino masses. 
However, we note  that when $v_2 \ll v_1$ and $|Y_3| \ll |Y_4|$ one can 
have small Dirac neutrino masses.
In this scenario, the charged lepton and neutrino masses can be written as:
\begin{eqnarray}
M_e &\approx & Y_4 v_1\,, \\
M^\nu_D &=& v_1 \left( Y_3 + M_e  \frac{v_2}{|v_1|^2} \right)\,. 
\end{eqnarray}
 The gauge couplings of ${S U}(2)_L, {S U}(2)_R$, and $U(1)_{B-L}$ are denoted as $g_L, g_R$, and $g_{B L}$ respectively. When the gauge couplings of ${S U}(2)_L$ and ${S U}(2)_R$ gauge group become equal, i.e. $g_L=g_R$, there exist two symmetry transformations between the left and right. This additional discrete left-right symmetry corresponds to either generalized parity $\mathcal{P}$ or generalized charge conjugation $\mathcal{C}$~\cite{Senjanovic:1978ev,Maiezza:2010ic}. Under the parity symmetry operation, the fields change as follows ;
\begin{eqnarray}
&&\ell_L \stackrel{\mathcal{P}}{\longleftrightarrow} \ell_R, \quad \quad q_L \stackrel{\mathcal{P}}{\longleftrightarrow} q_R,\quad ,
\\ \nonumber
&& \Phi \stackrel{\mathcal{P}}{\longleftrightarrow} \Phi^{\dagger}, \quad H_L \stackrel{\mathcal{P}}{\longleftrightarrow} H_R, \quad \widetilde{\Phi} \stackrel{\mathcal{P}}{\longleftrightarrow} \widetilde{\Phi}^{\dagger}
\label{LR_P}
\end{eqnarray}
whereas charge conjugation operation transforms the fields as  
\begin{eqnarray}
&&\ell_L \stackrel{\mathcal{C}}{\longleftrightarrow} \ell_R^c, \quad \quad q_L \stackrel{\mathcal{C}}{\longleftrightarrow} q_R^c, \quad 
\\ \nonumber
&&\Phi \stackrel{\mathcal{C}}{\longleftrightarrow} \Phi^T, \quad H_L \stackrel{\mathcal{C}}{\longleftrightarrow} H_R^* , \quad \widetilde{\Phi} \stackrel{\mathcal{C}}{\longleftrightarrow} \widetilde{\Phi}^{T}
\label{LR_C}
\end{eqnarray}
All the Left-Right symmetric models either have a $\mathcal P$ or $\mathcal C$ symmetry. It should be noted that the combination of the two symmetries, $\mathcal C \mathcal P$, does not switch the left and right-handed fields, and is not, therefore, a left-right symmetry. 
The Lagrangian in Eq. (\ref{eqn:LR-Yuk}) becomes
invariant by imposing left right symmetry with discrete $\mathcal{P}$ symmetry and it leads to hermitian Yukawa matrices as follows
\begin{equation}
Y_1=Y_1^{\dagger}, \quad Y_2=Y_2^{\dagger},\quad Y_3=Y_3^{\dagger}, \quad Y_4=Y_4^{\dagger}
\end{equation}
Therefore quark, charged lepton and Dirac mass matrices presented in Eq.  (\ref{quark_lepton}), and (\ref{Dirac1}) are hermitian matrices. 
On the other hand, if discrete $\mathcal{C}$ symmetry is imposed on the Lagrangian in Eq. (\ref{eqn:LR-Yuk}), it leads to symmetric Yukawa matrices, \begin{equation}
Y_1=Y_1^{T}, \quad Y_2=Y_2^{T},\quad Y_3=Y_3^{T}, \quad Y_4=Y_4^{T}
\end{equation}
and the corresponding mass matrices of Eq. (\ref{quark_lepton}), and (\ref{Dirac1}) become symmetric matrices. 
However, in our discussion, we consider a left-right model with discrete $\mathcal P$ symmetry.

%
\section{Neutrino Masses and Mixing}
\label{numassmixing}
%

In order to implement the double (or cascade) seesaw mechanism \cite{PhysRevLett.56.561,PhysRevD.34.1642} of neutrino mass generation within the manifest left-right symmetric model, we extend the fermion sector with the addition of one sterile neutrino $S_L \equiv[1,1,0] $ {\bf($S_L \stackrel{\mathcal{P}}{\longleftrightarrow} (S^c)_R$)} per generation . The relevant interaction Lagrangian $\mathcal{L}_{LRDSM}$ is given by:
\begin{eqnarray}
-\mathcal{L}_{LRDSM} &=& \mathcal{L}_{M_D}+\mathcal{L}_{M_{RS}}  + 
\mathcal{L}_{M_S}\,, 
\end{eqnarray}
%
where the individual terms can be expanded as follows.
\begin{itemize}
 \item $\mathcal{L}_{M_D}$ is the Dirac mass term connecting 
left-handed and right-handed neutrino fields $\nu_L-N_R$:
   \begin{eqnarray}
     \mathcal{L}_{M_D} &=& \sum_{\alpha, \beta} \overline{\nu_{\alpha L}} [M_D]_{\alpha \beta} N_{\beta R} \mbox{+ h.c.} \nonumber \\
     &&\subset \sum_{\alpha, \beta} \overline{\ell_{\alpha L}} \left( (Y_\ell)_{\alpha \beta} \Phi + (\widetilde{Y}_\ell)_{\alpha \beta} \tilde{\Phi} \right) 
     \ell_{\beta R} \mbox{+ h.c.}
     \end{eqnarray}
%
 \item  $\mathcal{L}_{M_{RS}}$  is another Dirac mass term 
connecting $N_R$ and $S_L$ and 
in the considered left-right symmetric theory
it has the form:
     \begin{eqnarray}
     \mathcal{L}_{M_{RS}} &=& \sum_{\alpha, \beta} \overline{S_{\alpha L}} [M_{RS}]_{\alpha \beta} N_{\beta R} \mbox{+ h.c.} \nonumber \\
    &&\subset \sum_{\alpha, \beta} \overline{S_{\alpha L}} (Y_{RS})_{\alpha \beta} \widetilde{H_R}^{\dagger} 
    \ell_{\beta R} \mbox{+ h.c.}
     \end{eqnarray}
%
 \item The bare Majorana mass term $\mathcal{L}_{M_S}$
for sterile neutrinos $S_L$ is given by:
      \begin{eqnarray}
      \mathcal{L}_{M_{S}} &=& \frac{1}{2} \sum_{\alpha, \beta} \overline{S^c_{\alpha R}} [M_{S}]_{\alpha \beta} S_{\beta L} \mbox{+ h.c.}
      \nonumber \\
      &&\subset \sum_{\alpha, \beta}  \frac{1}{2} (M_S)_{\alpha \beta} 
\overline{S^c_{\alpha R}} S_{\beta L} \mbox{+ h.c.}\,,
      \end{eqnarray}
%
\end{itemize}
%
where $S^c_{\alpha R} \equiv C(\overline{S_{\alpha L}})^T$, 
$C$ being the charge conjugation matrix  
($C^{-1} \gamma_\mu C = -\, \gamma^T_\mu$).
 We have taken into account the scalar fields' VEVs as $\langle H^0_{R} \rangle = v_{R}$ and $\langle H^0_{L} \rangle = 0$, which prevents the mass term from linking $\nu_L-S^c_{R}$ through the interaction $\sum_{\alpha, \beta} \overline{\ell_{\alpha L}} (Y_{LS})_{\alpha \beta} \widetilde{H_L}
    S^c_{\beta R} \mbox{+ h.c.}$ despite being permitted by gauge symmetry.

%
\subsection{The Double Seesaw Approximation}
%

After the spontaneous symmetry breaking, the complete $9 \times 9$ 
neutral fermion mass matrix in the flavor basis of $\left(\nu_L, N^c_R, S_L \right)$ can be written as:
\begin{equation}
 \mathcal{M}_{LRDSM}=
 \left[ 
\begin{array}{c  c} 
  \begin{array}{c c} 
     {\bf 0} & M_D \\ 
     M^T_D & {\bf 0}
  \end{array} & 
  \begin{array}{c} 
     {\bf 0} \\ M_{RS}
  \end{array} \\ 
  \begin{array}{c c} 
     \,~{\bf 0}  &\quad  M^T_{RS}
  \end{array} & 
  \begin{array}{c} 
     {M_S}
  \end{array} \\
 \end{array} 
\right]
\label{eqn:dss1}       
\end{equation}
%
 We assume in what follows that  $|M_D| \ll |M_{RS}| \ll |M_s|$. 
This allows us to apply to the mass matrix  $\mathcal{M}_{LRDSM}$  
twice the seesaw approximate block diagonalization 
procedure for getting the mass matrices of light and heavy neutrinos, 
as discussed below.

\begin{itemize}
\item {\bf First Seesaw Approximation:}
 We implement the first seesaw block diagonalization procedure 
on the lower right $6\times 6$ sub-matrix of  $\mathcal{M}_{LRDSM}$ 
as indicated below.
\begin{eqnarray}
&& \mathcal{M}^{}_{LRDSM} 
=
 \left[ 
\begin{array}{c | c} 
  \begin{array}{c} 
     {\bf 0}
  \end{array} & 
  \begin{array}{c c} 
     M_D & {\bf 0} 
  \end{array} \\ 
  \hline
  \begin{array}{c} 
     M^T_D  \\  {\bf 0}
  \end{array} & 
  \begin{array}{cc} 
     {\blue {0}} & \blue{M_{RS}} \\
     \blue{M^T_{RS}}  &  \blue{M_S}
  \end{array} \\
 \end{array} 
\right]  \nonumber \\
&&\hspace*{-0.2cm}
 \mathop{\xrightarrow{\hspace*{1.5cm}}}^{M_S > M_{RS} \gg M_D}_{\mbox{1st seesaw}} 
  \left[ 
\begin{array}{c | c} 
  \begin{array}{c c} 
     {\bf 0}
  \end{array} & 
  \begin{array}{c c} 
     M_D & {\bf 0} 
  \end{array} \\ 
  \hline
  \begin{array}{c} 
     M^T_D  \\  {\bf 0}
  \end{array} & 
  \begin{array}{cc} 
     \blue{- M_{RS} M^{-1}_S M^T_{RS}} & {\bf 0} \\
     {\bf 0}  &  \blue{M_S}
  \end{array} \\
 \end{array} 
\right]
\label{eqn:dss-b}  
\end{eqnarray}
%
\item {\bf Second Seesaw Approximation:-}
%
Denoting $- M_{RS} M^{-1}_S M^T_{RS}= M_R$, 
 which is the expression for the 
mass matrix for right-handed neutrinos, we repeat the diagonalization 
procedure with seesaw condition, $|{M_R}| \gg |M_D|$. 
We get the resultant matrix structure as
\begin{eqnarray}
 \left[ 
\begin{array}{c | c}  
  \begin{array}{c c} 
     {\bf 0} & \blue{M_D} \\ 
     \blue{M^T_D} & \blue{M_R}
  \end{array} &          \begin{array}{c} 
                       {\bf 0}  \\  {\bf 0}
                         \end{array}                \\
  \hline
  \begin{array}{c} 
   \hspace*{-0.0cm} {\bf 0}  \hspace*{0.6cm} {\bf 0}
  \end{array}          & {M_S} 
 \end{array} 
\right]
 \mathop{\xrightarrow{\hspace*{1.5cm}}}^{M_R \gg  M_D}_{\mbox{2nd seesaw}} 
 \left[ 
\begin{array}{c | c | c}  
     \blue{\tiny -M_D M^{-1}_R M^T_D} & 0  & 0\\
  \hline
  0  & \blue{M_{R}} & 0 \\
  \hline 
  0 & 0 & \blue{M_S}
 \end{array} 
\right] 
\label{eqn:dss-c}       
\end{eqnarray}
\end{itemize}
%
 Using the above results, the light neutrino, the heavy neutrino 
and sterile fermions mass matrices  
$m_\nu$,  $m_N$ and $m_S$  can be expressed as: 
\begin{eqnarray}
&&m_\nu \cong 
- M_D \left( -M_{RS} M^{-1}_S M^T_{RS} \right)^{-1} M^T_D\, \nonumber \\
  &&\hspace*{0.5cm}
=\frac{M_D}{M^T_{RS}}  M_S \frac{M^T_D}{M_{RS}}  ,\nonumber \\
&& m_N \equiv M_R \cong -M_{RS} M^{-1}_S M^T_{RS} ,\nonumber \\
&& m_S \cong M_S\, .
\label{massformulae}
\end{eqnarray}
\begin{table}[h!]
 \centering
	\begin{tabular}{|c|c|c||c|c|c||c|c|}
	\hline \hline
    $M_D$ & $M_{RS}$ & $M_S$ & $m_\nu$(eV) & $m_{N}$ & $m_{S}$ & $V^{\nu N}$ & $V^{N S}$ \\
	\hline
	$10^{-4}$  & $10^{3}$  & $10^{4}$ & $0.1$  & $10^{2}$ & $10^{4}$ & $10^{-6}$ & $0.1$ \\
    $10^{-5}$  & $10^{2}$  & $10^{3}$ & $0.01$  & $10$ & $10^{3}$ & $10^{-6}$ & $0.1$ \\    $10^{-5}$  & $10^{1}$  & $10^{2}$ & $0.1$  & $1$ & $10^{2}$ & $10^{-5}$ & $0.1$ \\
    \hline \hline
	\end{tabular}
	\label{tab:tab1}
\caption{A representative set of model parameters in left-right symmetric models and the order of magnitude estimation of various neutrino masses within the double seesaw mechanism. All the masses are expressed in units of GeV except the light neutrino masses, which are in the eV scale.}
\label{tab:dss}
\end{table}

In the double seesaw 
expression  for the light neutrino Majorana mass matrix 
as given in Eq. (\ref{massformulae}), different choices of $M_D$ and $M_{RS}$ 
 are possible. Following \cite{Smirnov:1993af,Altarelli:2004za,Lindner:2005pk,Ludl:2015tha,Bajc:2016eiw,Smirnov:2018luj}, 
we have considered in the present article the case of 
$M_D$ and $M_{RS}$ being proportional to identity such that  $M_D = k_d I$ 
and $M_{RS}= k_{rs}I$ , where $k_d$ and $k_{rs}$ are real 
constants with $|k_d| < |k_{rs}|$. 
This means, $ M_D M^{-1}_{RS} = \frac{k_d}{k_{rs}}I$.
As discussed in \cite{Lindner:2005pk,Brdar:2018sbk}, the equality and 
simultaneous diagonal structures of $M_D$ and $M_{RS}$  may arise as a 
consequence of $Z_2 \times Z_2$ symmetry \cite{Ludl:2015tha}. With the 
introduction of additional permutation symmetry in the diagonal elements of 
$M_D$ and $M_{RS}$, one can get equal diagonal elements. 
As we have indicated, these kinds of considerations have been reasoned 
for the double seesaw mechanism, e.g.,  
in the references \cite{Smirnov:1993af,Altarelli:2004za,Lindner:2005pk,Ludl:2015tha,Bajc:2016eiw,Smirnov:2018luj}.

 With the choices for the forms of $M_D$ 
and $M_{RS}$ made above, the relation between 
light neutrino  and sterile neutrino mass matrices $m_\nu$ and $m_S$
can be written as $ m_{\nu}= \frac{k^2_d}{k^2_{rs}} m_S $. 
The mass matrix $m_N $ can also be determined from Eq. (\ref{massformulae})
and the relationship between 
light neutrino  and heavy right-handed neutrino mass matrices 
$m_{\nu}$ and $m_N$ has the form $ m_{N}= - k^2_d\frac{1}{m_{\nu}}$ .

In the basis in which the charged lepton mass matrix is diagonal 
we will work with in what follows,
the light neutrino Majorana mass matrix is diagonalized with the help of a 
unitary mixing matrix -- the Pontecorvo, Maki, Nakagawa, Sakata (PMNS) 
mixing matrix 
$U_{\rm PMNS} \equiv U_{\nu}$ 
\cite{Pontecorvo:1957qd,Maki:1962mu,Pontecorvo:1967fh}:
$$m^{\rm diag}_\nu= U^\dagger_{\rm PMNS} m_\nu U^*_{\rm PMNS} 
= \mbox{diag}\left(m_1,m_2,m_3 \right)\,,~~m_i > 0\,,$$
%
so the physical masses $m_i$ are related to the mass matrix $m_\nu$ in 
the flavour basis as 
$$m_\nu = U_{\rm PMNS} m^{\rm diag}_\nu U^T_{\rm PMNS}\,.$$
%
 The right-handed neutrino Majorana mass matrix $m_N$ is diagonalized as 
$\widehat{m_N} = {U_N}^\dagger m_N {U_N}^*$, 
$\widehat{m_N} =  \mbox{diag}(m_{N_1},m_{N_2},m_{N_3})$, 
$m_{N_j}$ being the mass of the heavy RH Majorana neutrino $N_j$, $j=1,2,3$.
It proves convenient to work with positive masses 
of $N_j$, $m_{N_j} > 0$. 
Given the relation $ m_{N}= - k^2_d\frac{1}{m_{\nu}}$ and the 
positivity of the eigenvalues of $m_\nu$, 
the requirement that the eigenvalues of $m_N$ are also positive
implies that the unitary transformation matrices 
diagonalizing the light neutrino and the heavy right-handed neutrino 
mass matrices $m_\nu$ and $m_N$ are related in the following way:
\begin{eqnarray}
 U_N = i\,U^*_{\nu} \equiv i\,U^*_{PMNS}\,.
\label{eq:UN}
 \end{eqnarray} 
%
 Since $m_S = (k^2_{rs}/k^2_d)m_\nu$,              
the diagonalization of sterile neutrino Majorana mass matrix $m_S$, 
$\widehat{m_S} = {U_S}^\dagger m_S {U_S}^*$,
where $\widehat{m_S} =  \mbox{diag}(m_{S_1},m_{S_2},m_{S_3})$, $m_{S_k} > 0$, 
$k=1,2,3$, can be performed with the help of the same mixing matrix $U_{PMNS}$:
\begin{eqnarray}
 U_S = U_{\nu}\equiv U_{PMNS}\,. 
\label{eq:US}
 \end{eqnarray}
%
 Thus, in the considered 
scenario, 
the light neutrino masses $m_i$, the heavy RH neutrino masses 
$m_{N_j}$ and the sterile neutrino masses ($m_{S_k}$) are related  as follows:
\begin{equation}
m_{i} = \frac{k^2_d}{m_{N_i}} = \frac{k^2_d}{k^2_{rs}}\, m_{S_i}\,,~i=1,2,3\,.
\label{eq:massrel}
\end{equation}
%

 In what follows, we will use the standard parametrization of 
the PMNS matrix (see, e.g., \cite{ParticleDataGroup:2018ovx}):
\bea
&&U_{\rm {PMNS}}= \nonumber \\
&& \hspace*{-0.3cm} 
\begin{pmatrix} c_{13}c_{12}&c_{13}s_{12}&s_{13}e^{-i\delta}\\
-c_{23}s_{12}-c_{12}s_{13}s_{23}e^{i\delta}&c_{12}c_{23}-s_{12}s_{13}s_{23}e^{i\delta}&s_{23}c_{13}\\
s_{12}s_{23}-c_{12}c_{23}s_{13}e^{i\delta}&-c_{12}s_{23}-s_{12}s_{13}c_{23}e^{i\delta}&c_{13}c_{23}
\end{pmatrix} \mbox{P}
\label{PMNS} 
\eea
%
where the mixing angles are denoted by $s_{ij}=\sin \theta_{ij}$, 
$c_{ij}=\cos \theta_{ij}$,  
 $0 \leq \theta_{ij}\leq \pi/2$,
$\delta$ is the Dirac CP violation phase, $0 \leq \delta \leq 2\pi$,
$\mbox{P} $ is the diagonal phase matrix 
containing the two Majorana CP violation phases
$\alpha$ and $\beta$ \cite{Bilenky:1980cx},
$\mbox{P}=\mbox{diag}\left(1, e^{i\alpha/2}, e^{i \beta/2} \right)$. 
The Majorana phases take values in the interval 
$[0, \pi]$. The experimental values of 
different oscillation parameters for 
the light neutrino mass spectrum 
with normal ordering (NO) and inverted ordering (IO) 
(see, e.g., \cite{ParticleDataGroup:2018ovx}) 
are taken from Ref.~\cite{PhysRevD.101.116013} and are 
presented in Table \ref{tab2_oscipara}. 
\begin{table}[]
\centering
\begin{tabular}{|l|c|c|}
\hline
Parameter & Best fit values & 3$\sigma$ range
\\
\hline\hline
$\Delta m^2_{21}\: [10^{-5}\,\mbox{eV}]$ & 7.34 & 6.92--7.90 \\
\hline
$|\Delta m^2_{31}|\: [10^{-3}\,\mbox{eV}]$ (NO) &  2.485 &  2.389--2.578\\
$|\Delta m^2_{32}|\: [10^{-3}\,\mbox{eV}]$ (IO)&  2.465  &  2.374--2.556 \\
\hline
$\sin^2\theta_{12} / 10^{-1}$ (NO) & 3.05 & 2.65--3.47\\
$\sin^2\theta_{12} / 10^{-1}$(IO) & 3.03 & 2.64--3.45\\
\hline
  $\sin^2\theta_{23} / 10^{-1}$ (NO)
	  &	5.45 & 4.36--5.95 \\
  $\sin^2\theta_{23} / 10^{-1}$ (I))
	  & 5.51 & 4.39--5.96 \\
\hline 
$\sin^2\theta_{13} / 10^{-2}$ (NO) & 2.22 & 2.01--2.41 \\
$\sin^2\theta_{13} / 10^{-2}$ (IO) & 2.23 & 2.03--2.43 \\
    \hline
  \end{tabular}
  \caption{The current updated estimates of experimental values of 
Neutrino oscillation parameters for global best fits and $3\sigma$ range taken from~\cite{PhysRevD.101.116013}.}
  \label{tab2_oscipara}  
\end{table}

\begin{itemize}
 \item \underline{Masses of light neutrinos}

 It proves convenient to express the masses of the two heavier 
neutrinos in terms of the lightest neutrino mass and the neutrino 
mass squared differences measured in neutrino oscillation experiments.
In the case of NO light neutrino mass spectrum, 
$m_1< m_2 < m_3$, we have: 
 \begin{eqnarray}
 &&m_1 = \mbox{lightest neutrino mass} \,,\qquad  \nonumber \\
 &&m_2 = \sqrt{m_1^2 +\Delta m_{\rm sol}^2}\,,\qquad \nonumber \\
 &&m_3 = \sqrt{m_1^2 +\Delta m_{\rm atm}^2}\,,
\label{eq1:NH_m2_m3}
\end{eqnarray}
%
where $\Delta m_{\rm sol}^2=\Delta m_{21}^2$ and 
$\Delta m_{\rm atm}^2=\Delta m_{31}^2$.
Similarly, for inverted mass ordering, $m_3 < m_1 < m_2$, we get: 
\begin{eqnarray}
 &&m_3 = \mbox{lightest neutrino mass} \,,\qquad  \nonumber \\
 &&m_1 = \sqrt{m_3^2 -\Delta m_{\rm sol}^2-\Delta m_{\rm atm}^2}\,,
\qquad \nonumber \\
 &&m_2 = \sqrt{m_3^2  -\Delta m_{\rm atm}^2 }\;, 
\label{eq1:IH_m1_m2}
\end{eqnarray}
%
with $\Delta m_{\rm sol}^2=\Delta m_{21}^2$ and 
$\Delta m_{\rm atm}^2=\Delta m_{32}^2$.

 Depending on the value of the lightest neutrino mass 
the neutrino mass spectrum can also be normal hierarchical (NH) 
when $m_1 \ll m_2 < m_3$, inverted hierarchical (IH) 
if $m_3 \ll m_1 < m_2$, or else quasi-degenerate (QD)
when $m_1 \cong m_2 \cong m_3$, $m^2_{1,2,3} >> \Delta m^2_{31(23)}$, 
i.e., $m_{1,2,3}\gtap 0.1$ eV.

 All considered types of neutrino mass spectrum are compatible with the existing data. The best upper limit on the lightest 
neutrino mass $m_{1(3)}$ has been obtained in the KATRIN experiment. It is in the range of the QD spectrum and effectively 
reads: $m_{1,2,3} < 0.80$ eV (90\% C.L.).

\item \underline {Masses of heavy RH neutrinos}
 
 Since $m_i$ and $m_{N_j}$ are inversely proportional to each other, 
for NO light neutrino mass spectrum, $m_1 < m_2 < m_3$, 
$m_{N_1}$ has to be the largest RH neutrino mass. 
We can express $m_{N_2}$ and $m_{N_3}$ in terms of $m_{N_1}$ 
and the light neutrino masses:
 \begin{eqnarray}
 m_{N_2} = \frac{m_{1}}{m_{2}} m_{N_1}\,, \quad
m_{N_3} = \frac{m_{1}}{m_{3}} m_{N_1}\,, \quad m_{N_3} <  m_{N_2} <  m_{N_1}\,.
\label{eqn:Mne1} 
\end{eqnarray}
%
In the case of IO spectrum, $m_3 < m_1 < m_2$, 
$m_{N_3}$ is the largest mass. The mass relations in this case become:
  \begin{eqnarray}
m_{N_1} = \frac{m_{3}}{m_{1}} m_{N_3}\,, \quad 
 m_{N_2} = \frac{m_{3}}{m_{2}} m_{N_3}\,, \quad  m_{N_1} <  m_{N_2} <  m_{N_3}\,. 
\label{eqn:Mne2} 
\end{eqnarray}

 \item \underline {Masses of sterile neutrinos}
 
 Since $m_i$ and $m_{S_k}$ are directly proportional to each other, 
in the NO case  $m_{S_3}$ is the heaviest sterile neutrino mass 
and the analogous mass relations read: 
\begin{eqnarray}
 m_{S_1} = \frac{m_{1}}{m_{3}} m_{S_3}\,, \quad 
 m_{S_2} = \frac{m_{2}}{m_{3}} m_{S_3}\,, \quad  m_{S_1} <  m_{S_2} < m_{S_3}\,.
\label{eqn:MSe1} 
\end{eqnarray}
%
For the IO spectrum we have $m_{S_3} <  m_{S_1} <  m_{S_2}$ and
\begin{eqnarray}
m_{S_1} = \frac{m_{1}}{m_{2}} m_{S_2}\,, \quad
m_{S_3} = \frac{m_{3}}{m_{2}} m_{S_2}\,, \quad  m_{S_3} <  m_{S_1} < m_{S_2}\,. 
\label{eqn:MSe2} 
\end{eqnarray}

\item \underline {Neutrino Mixing}
\end{itemize}

The diagonalisation of $\mathcal{M_{\rm LRDSM}}$ 
 leads to the following relation between 
the fields of the neutral fermions written in the flavour 
(weak interaction eigenstate) basis and in the mass eigenstate basis:
\begin{eqnarray}
&& \begin{pmatrix}
\nu_{\alpha L}\\ N^c_{\beta L}\\ S_{\gamma L}
\end{pmatrix}
= 
\begin{pmatrix}
V_{\alpha i}^{\nu \nu} & V_{\alpha j}^{\nu N} & V_{\alpha k}^{\nu S}\\V_{\beta i}^{ N \nu} & V_{\beta j}^{NN} & V_{\beta k}^{N S}\\V_{\gamma i}^{S \nu} & V_{\gamma j}^{S N} & V_{\gamma k}^{SS}
\end{pmatrix}               
\begin{pmatrix}
\nu_{iL}\\ N^c_{jL}\\ S_{kL}
\end{pmatrix} \,.
\label{nuNS}
\end{eqnarray}
%
 Here  $N^c_{\beta L} \equiv C(\overline{N_{\beta R}})^T$, 
 $N^c_{j L} \equiv C(\overline{N_{j R}})^T = N_{jL}$ 
($N_j$ are Majorana fields),
$C$ being the charge conjugation matrix, 
the indices $\alpha, \beta, \gamma$ run over three generations of 
light left-handed neutrinos, heavy right-handed neutrinos and 
sterile neutrinos in flavor basis respectively, 
whereas the indices $i,j,k$ run over corresponding mass eigenstates.
 The mixing matrix elements in Eq. (\ref{nuNS}) 
are given in Eq. (\ref{mixingmatrix1}) in 
 Appendix \ref{app:lrdssm}.
The mixing between the right-handed neutrinos and sterile neutrinos 
$(N^c_L-S_L)$ is given by the term,
\begin{equation}
V^{N S} \propto M_{RS} M^{-1}_S 
\end{equation}
%
while the mixing between the fields of the 
left-handed flavour neutrinos and the  
heavy right-handed neutrinos $(\nu_L-N^c_L)$ is determined by
\begin{equation}
V^{\nu N} \propto M_D M^{-1}_R = -M_D {M^T_{RS}}^{-1}  M_{S} M^{-1}_{RS} 
\end{equation}
%
The mixing between sterile and light neutrinos ($\nu_L - S_L$) is vanishing, 
$V_{\alpha k}^{\nu S}\cong 0$ and $V_{\gamma i}^{S \nu} \cong 0$.

The possible sets of numerical values for different mixing matrices, masses 
and mixing that can give rise to dominant contributions to LNV decays are 
listed in Table \ref{tab:dss}. By choosing one representative 
set of model parameters from 
the table, we get the mixing as given below. 
\begin{eqnarray}
\begin{pmatrix}
V_{\alpha i}^{\nu \nu} & V_{\alpha j}^{\nu N} & V_{\alpha k}^{\nu S}\\V_{\beta i}^{ N \nu} & V_{\beta j}^{NN} & V_{\beta k}^{N S}\\V_{\gamma i}^{S \nu} & V_{\gamma j}^{S N} & V_{\gamma k}^{SS}
\end{pmatrix}  \simeq \begin{pmatrix}
{\cal O}(1.0)       & {\cal O}(10^{-6})     & 0  \\
{\cal O}(10^{-6})      & {\cal O}(1.0)    & {\cal O}(0.1)    \\
0       & {\cal O}(0.1)    & {\cal O}(1.0)  
\end{pmatrix}
\label{mixing}
\end{eqnarray} 
%
In the above matrix, the non-zero elements come from 
$V_{\alpha i}^{\nu \nu}$, $V_{\beta j}^{N N}$,  $V_{\beta k}^{N S}$, 
$V_{\gamma j}^{S N}$ and $V_{\gamma k}^{S S}$ while all other terms are 
negligibly small. These non-zero mixings would contribute sizeably to 
the predicted neutrinoless double beta decay rate.
Thus, Eq. (\ref{nuNS}) can be rewritten for the 
fields of flavour neutrinos
$\nu_{\alpha L}$ and the heavy RH neutrinos $N^c_{\beta L}$  as:
\begin{eqnarray}
\nu_{\alpha L} &\cong &V_{\alpha i}^{\nu \nu} \nu_{iL} + V_{\alpha j}^{\nu N} N_{jL}\,,
\nonumber \\
 N_{\beta L}^c &=& V_{\beta i}^{ N \nu} \nu_{iL}+ V_{\beta j}^{N N} N_{jL} 
+ V_{\beta k}^{N S} S_{kL}\,. 
 \label{contibution}
\end{eqnarray}
%

 As we have indicated, in the considered model we have
$|V_{\alpha j}^{\nu N}|\sim 10^{-6}$. Correspondingly,
the contribution to the $0\nu\beta\beta$ decay amplitude 
arising from the coupling of $N_{jL}$ 
to the electron in the LH (i.e., V-A) charged lepton current 
involves the factor $(V_{e j}^{\nu N})^2$ and  
is negligible.

%
\subsection{Gauge Boson Masses}
\label{sec:gbmass}
%

We briefly summarize here the gauge bosons masses and mixing in our model 
which will be used in estimating half-life of neutrinoless double beta decay 
process. Besides the SM gauge bosons $W_L^{\pm}$ and $Z$, there are 
right-handed gauge bosons $W_R^{\pm}$ and $Z^{\prime}$ which get their masses 
from left-right symmetry breaking. 
Following ref~\cite{Senjanovic:1978ev} and choosing VEVs of the Higgs fields as,
\begin{equation}
  \langle H^0_{R} \rangle = v_{R} \,,
\quad \langle \phi^0_{1,2} \rangle = v_{1,2} \,,
\end{equation}
%
the mass matrix for charged gauge bosons, 
in the basis $(W_L^+  \ W_R^+)$ can be written as,
\begin{equation*}
  \mathbb M_{CGB}
=
\left\lgroup
\begin{matrix} \frac{g_L^2 v^2}{2} & & - g_L g_R v_1 v_2 \\ - g_L g_R v_1 v_2 & & \frac{g_R^2}{2}(\frac{1}{2}v_R^2 + v^2) \end{matrix}
\right\rgroup \,,
\end{equation*}
%
where $v^2 = v_1^2 + v_2^2$ 
 and $g_R = g_L$. 
The physical mass for extra charged 
gauge boson is given by 
\begin{eqnarray}
M_{W_R} &\simeq & \frac{1}{2} g_R v_R \,. 
\end{eqnarray}
%
The mixing angle between $W_R$ and $W_L$ is defined as 
$$\tan 2 \theta_{LR} \approx 8 \frac{g_L}{g_R} \frac{v_1 v_2}{v_R^2}$$.
 
The neutral gauge boson mass matrix is given by 
\begin{equation}
\mathbb M_{NGB} =
\left\lgroup
 \begin{matrix} 
\frac{g_{L}^2 v^2}{2} && - \frac{g_L  g_{R}}{2} v^2 && 0 \\ 
- \frac{g_L g_{R}}{2} v^2 && \frac{g_R^2}{2} (\frac{1}{2}v_R^2 + v^2) && - \frac{g_R g_{BL}}{4} v_R^2  \\ 
0 && - \frac{g_R g_{BL}}{4} v_R^2 && \frac{g_{BL}^2 v^2_R}{4} \end{matrix}
\right\rgroup \,. \nonumber
\end{equation}
%
 As can be easily checked, this mass matrix has one zero eigenvalue 
corresponding to the photon $A_\mu$.
After few simplification, the mass eigenstates  $Z_\mu$, $Z^{'}_\mu$ 
and $A_\mu$ are related to the weak eigenstates 
$(W^0_{L\mu}, W^0_{R\mu}, Z_{BL\mu})$ in the following way,
\begin{eqnarray}
W_{L\mu}^0 &=& \cos \theta_W Z_{L\mu} +\sin \theta_W A_{\mu}\,, \nonumber \\
W_{R\mu}^0 &=& \cos \theta_R Z_{R\mu} -\sin \theta_W \sin \theta_R Z_{L\mu} 
+ \cos \theta_W \sin \theta_R A_{\mu}\,, \nonumber \\
Z_{BL\mu}^0 &=& -\sin \theta_R Z_{R\mu}- \sin \theta_W \cos \theta_R Z_{L\mu} + 
\cos \theta_W \cos \theta_R A_\mu\,, \nonumber 
\end{eqnarray}
%
where 
\begin{eqnarray}
Z_{L\mu} &\equiv& Z_\mu \cos\xi + Z^{'}_\mu \sin\xi\,,
\nonumber \\
 Z_{R\mu} &\equiv& -\, Z_\mu \sin\xi + Z^{'}_\mu \cos\xi\,.
\end{eqnarray}
%
Here, the mixing angles are defined as $\tan \theta_R = g_{BL}/g_R$, 
$\tan \theta_W=g_Y/g_L$ with $g_Y = g_{BL}g_R/\sqrt{g^2_{BL}+g_R^2}$,  
while the mixing angle between the $Z$ and the heavy $Z^{'}$ reads: 
\begin{equation}
\tan 2 \xi \approx \frac{v_1 v_2}{v^2_R} \ \frac{-4 g_R^2 \sqrt{g_L^2g_R^2 + g^2_{BL}(g_L^2+g_R^2)}}{(g_{BL}^2+g_R^2)^2}\,.
\end{equation}
%
The physical mass for extra neutral gauge boson $Z^\prime$ is given by:
\begin{eqnarray}
M^2_{Z^{'}} &\simeq & \frac{1}{2} \left(g^2_{BL} + g^2_R \right) 
   \bigg[v^2_R +\frac{g^2_R v^2}{g^2_R+g^2_{BL}} \bigg]
\end{eqnarray}
%
The value of $\tan 2 \xi$ has to be smaller than $10^{-3}$ in order to 
satisfy the electroweak precision constraints in the limit 
$v_R^2 \gg v_1^2 +v_2^2$. With $v^2=v^2_1+v^2_2\simeq (246$~GeV)$^2$
and $g_R\simeq g_L = 0.653$, we have 
$\tan\theta_W=\frac{g_{BL}}{\sqrt{g^2_{BL}+g^2_{R}}}$, 
which implies 
$\frac{g^2_{BL}}{g^2_{L/R}}=\frac{\sin^2\theta_W}{1-2\sin^2\theta_W}\approx0.43$, 
where $sin^2\theta_W=0.231$. 
 Using this result and $g_L = g_R$, we get for 
the angle describing the  $Z-Z^{'}$ mixing: 
$|\tan 2\xi| \cong 2.67 v_1v_2/v^2_R$. The upper limit 
$|\tan 2\xi| < 10^{-3}$ implies:
\begin{equation}
\dfrac{v_1\,v_2}{v^2_R} < 3.75\times 10^{-4}\,.
\label{eq:v1v2ov2R}
\end{equation}
%
This in turn leads to the following upper limit on the 
$W_L - W_R$ mixing angle $\theta_{LR}$:
\begin{equation}
\theta_{LR} \cong 4\,\dfrac{v_1\,v_2}{v^2_R} < 1.50\times 10^{-3}\,.
\label{eq:thLR}
\end{equation}
%

  The left-handed gauge boson masses are similar 
to those of the SM gauge bosons with 
$g_Y = \left(g_R g_{BL} \right)/\left( g^2_R+g^2_{BL}\right)^{1/2} $, 
while the masses of the extra heavy gauge bosons are related as follows,
\begin{eqnarray}
M_{W_R} &\simeq & \frac{1}{2} g_R v_R \,, \\
M_{Z^{'}} &\simeq & \frac{\sqrt{g^2_{BL} + g^2_R}}{g_R} M_{W_R} 
{\simeq} 1.2 \, M_{W_R}\,.
\end{eqnarray}
%
The current experimental bound on $M_{W_R} > 5$ TeV is 
obtained in high energy 
collider experiments at LHC~\cite{ATLAS:2018dcj,ATLAS:2019isd,CMS:2018agk},
while the  low energy precision measurements \cite{Li:2020wxi,Dekens:2021bro} 
imply a lower bound on the $Z^{'}$ mass, i.e. $M_{Z^{'}}>6$ TeV.

%
\section{Neutrinoless double beta decay}
\label{sec:0nubb}
%

 Neutrinoless double beta decay process can be induced by the exchange of 
light active Majorana neutrinos, which is usually referred to as  
``the standard mechanism'', or by some other lepton number violating 
``non-standard mechanism'' associated with BSM physics.
In this section, we discuss the
standard and the new physics contributions to $0\nu\beta\beta$ decay 
amplitude and rate that arise in our model 
due to the exchange of the light Majorana neutrinos $\nu_i$, 
heavy Majorana neutrinos $N_{1,2,3}$ 
and sterile Majorana neutrinos $S_{1,2,3}$.

The charged current (CC) interaction Lagrangian for leptons and quarks, 
relevant for our further discussion,  
are given by:
\begin{eqnarray}
\mathcal{L}^{\rm \ell}_{\rm CC}
 &=& \sum_{\alpha=e, \mu, \tau}
\bigg[\frac{g_L}{\sqrt{2}} \overline{\ell}_{\alpha L} \gamma_\mu {\nu}_{\alpha L} W^{\mu}_L 
      +\frac{g_R}{\sqrt{2}} \overline{\ell}_{\alpha R} \gamma_\mu {N}_{\alpha R} W^{\mu}_R \bigg] + \text{h.c.} 
      \nonumber \\
&=& \frac{g_L}{\sqrt{2}}
\overline{e}_{L} \gamma_\mu {\nu}_{e L} W^{\mu}_L  
   + \frac{g_R}{\sqrt{2}}\overline{e}_{R} \gamma_\mu {N}_{e R} W^{\mu}_R + \text{h.c.} + \cdots 
\label{eqn:ccint-flavor-lepton}
\\
\mathcal{L}^{\rm q}_{\rm CC} &=& \bigg[\frac{g_L}{\sqrt{2}}\, \overline{u}_{\,L}\, \gamma_\mu {d}_{\,L}\, W^{\mu}_L 
      +\frac{g_R}{\sqrt{2}}\, \overline{u}_{R}\, \gamma_\mu d_{R}\, W^{\mu}_R \bigg] + \text{h.c.} 
\label{eqn:ccint-flavor-quark}
\end{eqnarray}
%
Using Eq. (\ref{contibution}), $\mathcal{L}^{\rm \ell}_{\rm CC}$
be rewritten as:
\begin{eqnarray}
\mathcal{L}^{\rm \ell}_{\rm CC}&=& \frac{g_L}{\sqrt{2}}
\bigg[ \overline{e}_{L} \gamma_\mu 
        \{\mbox{V}^{\nu\nu}_{e i} \nu_i + 
                      \mbox{V}^{\nu N}_{e i} N_i \} 
              W^{\mu}_L \bigg] +\mbox{h.c.} \nonumber \\
    & &  + \frac{g_R}{\sqrt{2}}
\bigg[ \overline{e}_{R} \gamma_\mu 
\{\mbox{V}^{N \nu}_{e i}\nu_i +
      \mbox{V}^{N N}_{e i}N_i +
                      \mbox{V}^{N S}_{e i} S_i \} 
              W^{\mu}_R \bigg] + \mbox{h.c.}
\label{eqn:ccint-mass}
\end{eqnarray}
%

In the present model, the heavy neutrino masses are around $1-1000$~GeV, 
$\nu_L-N^c_L$ mixing  $|V_{\nu N}| \leq 10^{-6}$ and $\nu_L-S_L$ mixing 
is vanishing. Other contributions involving the light-heavy neutrino mixings 
and the $W_L-W_R$ mixing are negligible. 
The  lepton Lagrangian that is relevant for the dominant contributions 
to $0\nu\beta\beta$ decay rate is:
\begin{eqnarray}
\mathcal{L}^{\rm \ell}_{\rm CC}&=& \frac{g_L}{\sqrt{2}}\,
\bigg[ \overline{e}_{\,L}\, \gamma_\mu 
        \{\mbox{V}^{\nu\nu}_{e\, i}\, \nu_i  
                      \}\, 
              W^{\mu}_L \bigg] +\mbox{h.c.} \nonumber \\
    & &  + \frac{g_R}{\sqrt{2}}
\bigg[ \overline{e}_{R} \gamma_\mu 
\{ \mbox{V}^{N\, N}_{e j}N_j +
                      \mbox{V}^{N S}_{e k} S_k \} 
              W^{\mu}_R \bigg] + \mbox{h.c.}
\label{eqn:ccint-mass}
\end{eqnarray}
%
Thus, in the considered model, the dominant contributions to 
the $0\nu\beta\beta$ decay amplitude are given by:
\begin{itemize}
\item the standard mechanism due to the exchange of light neutrino $\nu_i$, 
mediated by left-handed gauge boson $W_L$, i.e. due to purely left-handed 
(LH) CC interaction;
\item new contributions due to the exchange of heavy neutrinos $N_{1,2,3}$ 
and sterile neutrinos $S_{1,2,3}$, mediated by right-handed gauge boson $W_R$, 
i.e., due to purely right-handed (RH) CC interaction. 
 The contribution due to exchange of virtual $S_{1,2,3}$ 
is possible due to the mixing between $N^c_L$ and $S_L$.
\end{itemize}

 The so-called $<\lambda>$- and $<\eta>$- mechanism contributions 
$0\nu\beta\beta$ decay amplitude arising  
from the product of LH and RH lepton currents 
\cite{Doi:1985dx}
are sub-dominant being strongly suppressed.
The $<\lambda>$-mechanism contribution involves the factor 
$|V^{N \nu}_{ei}|(M_{W_L}/M_{W_R})^2 < 2.6\times 10^{-10}$, 
where we have used $|V^{N \nu}_{ei}| = 10^{-6}$, 
$M_{W_L} = 80.38$ GeV and $M_{W_R} > 5$ TeV, 
while the $<\eta>$-mechanism contribution is suppressed by the factor 
$|V^{N \nu}_{ei} \sin\theta_{LR}| < 10^{-9}$. As a consequence, 
we neglect these contributions in the analysis which follows.

\begin{figure*}[!ht]
       \includegraphics[width=1.0\textwidth]{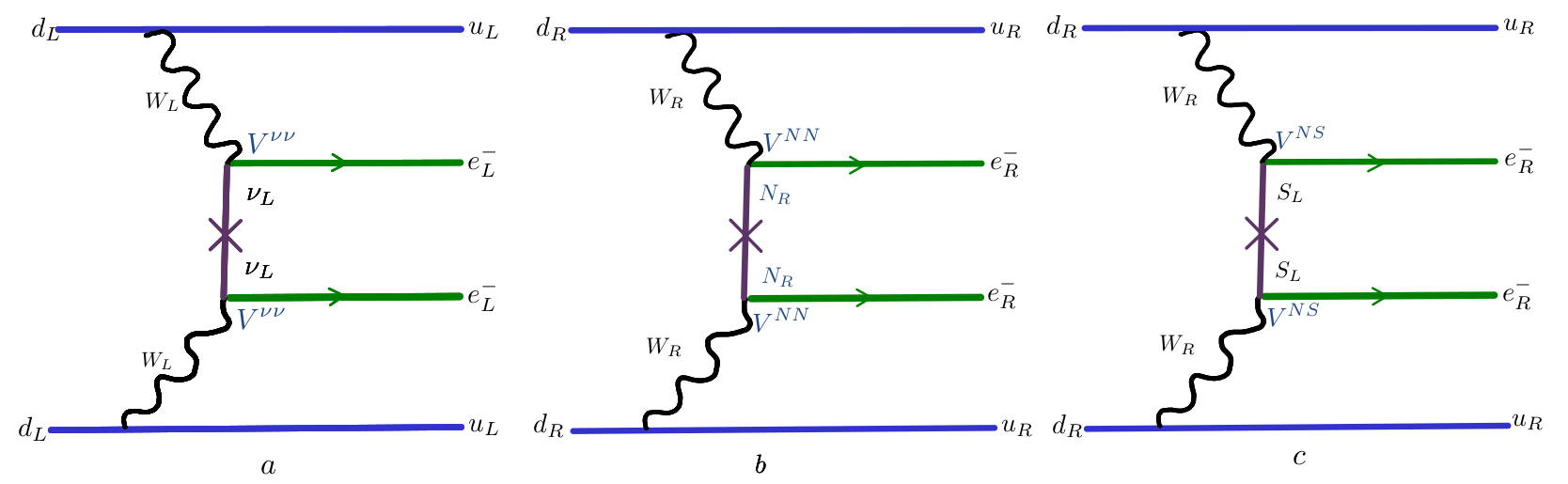}
\caption{Feynman diagrams for the process of neutrinoless double beta decay 
mediated by the exchane of virtual (a) light Majorana neutrinos $\nu_i$
(the standard mechanism), (b) heavy  neutrinos $N_R$ 
(heavy Majorana neutrinos $N_{1,2,3}$) and 
(c) heavy sterile neutrinos $S_L$ (heavy Majorana neutrinos $S_{1,2,3}$). 
}
\label{FD}
\end{figure*}

The Feynman diagrams for the dominant contributions of interest  
to the $0\nu\beta\beta$ decay amplitude are shown in Fig. \ref{FD}, 
where the first diagram from the left corresponds to the standard mechanism, while 
the second and third diagrams correspond to the new contributions mediated 
by $N_{1,2,3}$ and $S_{1,2,3}$, respectively.
%

When $0\nu\beta \beta$ decay is mediated by only light Majorana neutrinos 
$\nu_i$, the inverse half-life for this process can be expressed as,
\begin{eqnarray}
 \left[T_{1/2}^{0\nu}\right]^{-1} &=& g^4_{\rm A}\,G^{0\nu}_{01}~|{\cal M}^{0\nu}_\nu\big|^2 ~ |\eta_\nu|^2   
\label{std_halflife}
\end{eqnarray}
%
where $g_{\rm A}$ is the axial coupling constant, 
 $G^{0\nu}_{01}$ is the phase space factor, 
${\cal M}^{0\nu}_\nu$ is the Nuclear Matrix Elements (NME) for 
light neutrino exchange and $\eta_\nu$ is a dimensionless particle physics 
parameter that is a measure of lepton number violation. 
Considering both the standard mechanism and the new contributions 
to this decay process in our model, the inverse half life can be written as:
\begin{eqnarray}
 \left[T_{1/2}^{0\nu}\right]^{-1} &=& g^4_{\rm A}\,G^{0\nu}_{01}\bigg[ |{\cal M}^{0\nu}_\nu \cdot \eta_\nu|^2 + 
 |{\cal M}^{0\nu}_N \cdot \left(\eta_{N}+ \eta_{S} \right) \big|^2 \bigg]
\label{eq:Hlife-a} ,
\end{eqnarray}
%
where ${\cal M}^{0\nu}_N$ is the Nuclear Matrix Elements (NME) for the 
heavy neutrino exchange and 
 $\eta_{N}$ and $\eta_{S}$ are lepton number violating parameters associated 
with the exchange of the heavy neutrinos $N_{1,2,3}$ and $S_{1,2,3}$.

In the analysis and the numerical estimates which follow, we will use 
a mildly quenched value of the axial coupling constant $g_{\rm A} = 1.00$, 
the unquenched value being, as is well known, $g_{\rm A} = 1.27$.
If it turns out that  $g_{\rm A}$ is actually not quenched, that will reduce 
the estimates of the $0\nu\beta\beta$ decay half-lives made in the present study by a factor of 2.60. 

The interference term of the light neutrino $\nu_{1,2,3}$ and 
the heavy neutrinos $N_{1,2,3}$ and $S_{1,2,3}$ exchange contributions
to the $0\nu\beta\beta$ decay amplitude,
which are generated respectively by LH ($V-A$) CC and RH ($V+A$) 
CC interactions, is strongly suppressed, being proportional to the 
electron mass~\cite{Halprin:1983ez} (see also~\cite{Faessler:2011qw}) 
and we have neglected it in Eq. (\ref{eq:Hlife-a}). 

 The values of $G^{0\nu}_{01}$ and the NMEs for both light and heavy 
neutrino exchange mechanism are distinct for different isotopes and 
can be found, e.g., in \cite{Ejiri:2019ezh}. 
 We present in Table \ref{tab:nucl-matrix} the values obtained 
by six different groups of authors using different methods of NME 
calculation. Of particular importance 
for the estimates of the relative magnitude of the new non-standard 
contributions in the $0\nu\beta\beta$ decay amplitude
with respect to the contribution of the standard mechanism 
is the ratio ${\cal M}^{0\nu}_N/{\cal M}^{0\nu}_\nu$.
As it follows from Table \ref{tab:nucl-matrix}, 
the ratio  ${\cal M}^{0\nu}_N/{\cal M}^{0\nu}_\nu$ 
predicted by each of the six cited groups using different 
methods of NME calculation   
is essentially the same for the four isotopes 
$^{76}$Ge,  $^{82}$Se,  $^{130}$Te, $^{136}$Xe -- 
it varies with the isotope by not more than 
$\sim 15\%$. At the same time, for a given isotope 
the ratio of interest obtained by the six different 
methods of NME calculation quoted in 
Table \ref{tab:nucl-matrix} varies by a factor 
of up to $\sim 3.5$. In view of this we will use 
the values of the NMEs for $^{76}$Ge as reference values 
in our numerical analysis. For the minimal and maximal values 
of the ratio  ${\cal M}^{0\nu}_N/{\cal M}^{0\nu}_\nu$ 
for $^{76}$Ge we get from Table \ref{tab:nucl-matrix}:
\begin{equation}
22.2 \lesssim \dfrac{{\cal M}^{0\nu}_N}{{\cal M}^{0\nu}_\nu} 
\lesssim 76.3\,,~~~{\rm  ^{76}Ge}\,. 
\label{eq:MnuMN}
\end{equation}
%
They correspond respectively to ${\cal M}^{0\nu}_\nu = 4.68$ and 5.26.

\begin{table}[h]
 \centering
\vspace{10pt}
 \begin{tabular}{|l|cc|cc|cc|cc|c|}
 \hline \hline
  &  $^{76}$Ge 
& & $^{82}$Se  & & $^{130}$Te & &$^{136}$Xe &  \\[2mm] 
{methods}  & {${\cal M}^{0\nu}_\nu$}  
 & {${\cal M}^{0\nu}_N$} & {${\cal M}^{0\nu}_\nu$} & {${\cal M}^{0\nu}_N$}&{${\cal M}^{0\nu}_\nu$} &{${\cal M}^{0\nu}_N$} &{${\cal M}^{0\nu}_\nu$}  & {${\cal M}^{0\nu}_N$} \\ 
 \hline 
 dQRPA \cite{Fang:2018tui} & $3.12$  & $187.3$ & $2.86$  & $175.9$& $2.90$& $191.4$ &$1.11$ & $66.9$\\ 
  \hline
  QRPA-Tu \cite{PhysRevC.87.045501,PhysRevD.90.096010}  & $5.16$  & $287.0$ &$4.64$  & $262.0$&$ 3.89$ & $264.0$ & $2.18 $ & $152.0$ \\ 
  \hline
  QRPA-Jy \cite{PhysRevC.91.024613}  & $5.26$  & $401.3$ & $3.73$ &$287.1$ &$4.00$ &$338.3$ & $2.91 $ & $186.3$ \\  
  \hline
   IBM-2 \cite{PhysRevC.87.014315} & $4.68$  & $104$ & $3.73$ & $82.9$ & $3.70 $& $91.8 $ & $3.05 $ & $72.6 $ \\ 
  \hline
  CDFT \cite{PhysRevC.90.054309,PhysRevC.91.024316,PhysRevC.95.024305} & $ 6.04$ & $209.1 $  & $5.30 $ & $189.3 $ & $4.89 $ & $193.8 $ & $4.24 $ & $166.3 $ \\ 
\hline 
 ISM \cite{Menendez:2017fdf} & $2.89 $ & $130 $  & $2.73 $ & $121 $ & $2.76 $ & $146 $ & $ 2.28$ & $116 $ \\ 
\hline
\hline
$G^{0\nu}_{01}$ $[{10^{-14} \rm yrs}^{-1}] $ \cite{Horoi:2017gmj} & $0.22$ & & 1 & & $1.4$ & &$1.5$ &\\
\hline \hline
 \end{tabular}
 \caption{Values of Nuclear Matrix Elements for various isotopes calculated 
by different methods for light and heavy neutrino exchange. 
Here QRPA-Jy uses CD-Bonn short range correlations (SRC) and the rest 
use Argonne SRC, with minimally quenched $g_A = 1$. The last row shows the 
phase space factor $G^{0\nu}_{01}$ for various 
isotopes \cite{Horoi:2017gmj,Ejiri:2019ezh}.}
 \label{tab:nucl-matrix}
\end{table} 
%

The dimensionless particle physics parameters $\eta_\nu$, 
$\eta_N$ and $\eta_s$ in Eq. (\ref{eq:Hlife-a}) are functions of 
neutrino masses, 
mixing parameters and CPV phases and can be expressed as:
\begin{eqnarray}
\label{eq:eta_LL} 
|\mathcal{\eta}_{\nu}| = \sum_{i=1,2,3} \frac{{\mbox{V}^{\nu \nu}_{ei}}^2\, m_{\nu_i}}{m_e} \,, \\
 \label{eq:etaN}
 |\mathcal{\eta}_{N}| = m_p \left(\frac{M_{W_L}}{M_{W_R}}\right )^4 \sum_{j=1,2,3} \frac{{\mbox{V}^{N N}_{ej}}^2}{m_{N_j}} \,, \\                   
\label{eq:etaS}  
|\mathcal{\eta}_{S}| = m_p \left(\frac{M_{W_L}}{M_{W_R}}\right )^4 \sum_{k=1,2,3} \frac{{\mbox{V}^{N S}_{ek}}^2}{m_{S_k}} \,.
\end{eqnarray}
%
where $m_e$ and $m_p$ are the electron and proton masses. 
The quantity  
$m_e|\mathcal{\eta}_{\nu}| \equiv | m^\nu_{\beta\beta, L}|$ is the effective 
Majorana mass (EMM) associated with the standard mechanism of  
$0\nu\beta\beta$ decay (see, e.g., \cite{Bilenky:1987ty,Bilenky:2001rz}). 

 In \cite{Cirigliano:2018hja}
it was noticed that there exists a short distance (contact) 
contribution to the $0\nu\beta\beta$ decay amplitude 
even in the case of light neutrino exchange. 
The magnitude of this contribution was investigated in a number of studies.
Using the results of the estimates of the $nn\rightarrow ppee$ 
amplitude derived in \cite{Cirigliano:2020dmx}
the magnitude of this contribution relative to the standard light 
neutrino exchange one was calculated for the neutrinoless double beta decay 
of $^{48}$Ca in \cite{Wirth:2021pij}
and for  $^{76}$Ge,  $^{130}$Te and $^{136}$Xe in \cite{Weiss:2021rig}.
Both groups of authors find a positive contribution enhancing the standard one 
by about 43\% and 30\% respectively for  $^{48}$Ca 
and $^{76}$Ge,  $^{130}$Te, $^{136}$Xe. 
These effects are accounted for in our analysis by the much larger 
uncertainties in the NMEs included in the analysis.

The mixing parameters in Eqs. (\ref{eq:eta_LL}) - (\ref{eq:etaS}) are given in 
Appendix \ref{app:lrdssm}. In the framework of our model we have: 
${V}^{\nu \nu} \approx U_\nu$, 
${V}^{N N}\approx U_N$ and 
${V}^{N S}\equiv M_{RS} M^{-1}_S U_S$.
We recall that 
 $U_N = i\,U^*_\nu$, 
$U_S = U_\nu$ and 
$U_\nu \equiv U_{PMNS}$ (Eqs. (\ref{eq:UN}) and (\ref{eq:US})).

 The expressions for  $|\mathcal{\eta}_{N}|$ and  $|\mathcal{\eta}_{S}|$
in Eqs. (\ref{eq:etaN}) and Eqs. (\ref{eq:etaS}) are obtained under the condition
$\langle p^2 \rangle \ll  M^2_i$, where $\sqrt{\langle p^2 \rangle}$ 
is the average momentum  exchanged in the process of $0\nu\beta\beta$ decay 
and $M_i$ here is a  generic notation for the masses of 
$N_{1,2,3}$ and $S_{1,2,3}$. The chiral structure of the matrix elements 
involving virtual  $N_{1,2,3}$ and $S_{1,2,3}$ propagators in the case of the 
heavy neutrino exchange contribution is given by:
\begin{eqnarray}
& &P_{R}\frac{\slashed{p}+M_i}{p^2-M_i^2}P_{R} = \frac{M_i}{p^2-M_i^2}P_{R}\,,  
\end{eqnarray}
%
where $P_{R} = (1 + \gamma_5)/2$ is the RH projection operator.
A typical value of the neutrino virtuality is 
$\langle p^2 \rangle \cong (190 \,\mbox{MeV})^2$ 
(see, e.g., \cite{Babic:2018ikc}). Thus, in the case of interest, we have 
$ p^2 \ll  M^2_i$ and the heavy state propagators reduce 
to a good approximation to $1/M_i$.

 It proves convenient for our further analysis 
to rewrite the inverse half-life in terms of one 
particle physics parameter -- generalised effective Majorana mass (GEMM) -- 
that contains the lepton number violating information in it:
\begin{eqnarray}
\left[T_{1/2}^{0\nu}\right]^{-1} &=& G^{0\nu}_{01}\bigg[ |{\cal M}^{0\nu}_\nu  \eta_\nu|^2 + 
{\cal M}^{0\nu}_N \big| \eta_{N} + \eta_{S}  \big|^2  \bigg] \nonumber \\
&=&  G^{0\nu}_{01} \bigg| \frac{{\cal M}^{0\nu}_\nu}{m_e} \bigg|^2   \bigg[\big|m^{\nu}_{\beta \beta, L} \big|^2 
+  \big|m^{N}_{\beta \beta, R} +  m^{S}_{\beta \beta, R} \big|^2 \bigg] \nonumber \\
&=&G^{0\nu}_{01} \bigg |\frac{{\cal M}^{0\nu}_\nu}{m_e}\bigg |^2  
|m^{\rm eff}_{\beta \beta, L,R}|^2\,,
\label{halflife:no-int}
\end{eqnarray}
%
where \cite{Babic:2018ikc}
\begin{eqnarray}
m^{N}_{\beta \beta ,R} &=& \sum_{j}  
m_p m_e \frac{\mathcal{M}_N^{0\nu}}{\mathcal{M}_\nu^{0\nu}}
\frac{M^4_{W_L}}{M^4_{W_R}}  \frac{{\mbox{V}^{N N}_{ej}}^2}{m_{N_{j}}}\,,
\label{eqn:mee_NR}   
\nonumber \\
 m^{S}_{\beta \beta ,R} &=& \sum_{k} 
m_p m_e \frac{\mathcal{M}_N^{0\nu}}{\mathcal{M}_\nu^{0\nu}}
\frac{M^4_{W_L}}{M^4_{W_R}}  \frac{{\mbox{V}^{N S}_{ek}}^2}{m_{S_{k}}}\,.
\label{eqn:mee_SR} 
\end{eqnarray}
%

 It follows from Eqs. (\ref{eq:etaN}), (\ref{eq:etaS}) 
and (\ref{eqn:mee_NR}) that 
the new physics contributions to the $0\nu\beta\beta$ decay amplitude
are suppressed, in particular, by the factor  
$M^4_{W_L}/M^4_{W_R} < (1.6\times 10^{-2})^4$,
where  $M_{W_L} = 80.38$~GeV is the SM $W$-bosons mass and  
we have used the lower bound $M_{W_R} > 5$ TeV  
\cite{ATLAS:2018dcj,ATLAS:2019isd,CMS:2018agk,Li:2020wxi,Dekens:2021bro}.
Fixing  $M_{W_R}$ at, e.g., = 5.5 TeV, we have for the ratio 
$\left( M^4_{W_L}/M^4_{W_R} \right)  \sim {\cal O}(10^{-8})$. 
Taking further the masses of $S_k$ and $N_j$ in the ranges 
respectively of $(10^2 - 10^4)$ GeV and $(1 - 10^2)$ GeV,
the mixing ${V}^{N N}_{ej} \approx U_N$ and 
${V}^{N S}_{ek}\equiv M_{RS} M^{-1}_S U_S$ from Appendix \ref{app:lrdssm}, 
one finds that the new physics contributions can be  
in the $0.01-0.1$~eV range (see Table \ref{tab:half-life}), 
i.e., within the experimental search sensitivity.

We note that, since the dominant contributions to 
$0 \nu\beta\beta$ decay arises from more than one contribution, 
it is also possible that there might be interference between them 
in the decay rate of the process. The interference of light 
neutrino ($\nu_i$) contribution due to purely $V-A$ interaction 
involving LH  currents 
with either of the heavy neutrino $N_j$ and $S_k$ contributions, 
which are generated by  $(V+A)$ interaction with RH currents,  
is suppressed, as we have indicated earlier. However, the interference 
between the contributions of the heavy neutrinos $N_j$ and $S_k$   
both involving RH currents, in general, can't be neglected.  
 In the case when this interference is not taken into consideration, 
the generalised effective Majorana mass 
is determined by the sum of individual 
contributions of the three types of neutrinos $\nu_i$, $N_j$ and $S_k$:
 \begin{equation}
|m^{\rm eff}_{\beta \beta, L,R}| \equiv m^{\nu+N+S}_{ee} = 
\left (\big|m^{\nu}_{\beta \beta, L} \big|^2 
 +  \big|m^{N}_{\beta \beta, R} \big|^2 + \big| m^{S}_{\beta \beta, R} \big|^2 
\right)^{\frac{1}{2}}\,.
\label{eq:no-int}
 \end{equation}
%
 Accounting for the interference, the 
generalised effective Majorana mass can be written as:
 \begin{eqnarray}
%
|m^{\rm eff}_{\beta \beta, L,R}| \equiv m^{\nu+|N+S|}_{ee} &=& 
\left( \big|m^{\nu}_{\beta \beta, L} \big|^2 
 +  \big|m^{N}_{\beta \beta, R} + m^{S}_{\beta \beta, R} \big|^2 \right )^{\frac{1}{2}}
  \nonumber \\
 &=& \left( (m^{\nu+N+S}_{ee})^2 +  
2 \text{Re} (m^{N}_{\beta \beta, R} \cdot m^{S^{*}}_{\beta \beta, R})
\right)^{\frac{1}{2}}\,.
 \label{eq:int}
 \end{eqnarray}   
%
 In order to assess the relevance of the interference term 
$2 \text{Re} (m^{N}_{\beta \beta, R} \cdot m^{S^{*}}_{\beta \beta, R})$
in our study, we consider both the cases  of neglecting it and of taking 
it into account.

 We express next the three terms 
in the generalised effective Majorana mass in terms of 
the PMNS mixing angles, Dirac and Majorana CPV phases present 
in the PMNS matrix, the three light neutrino masses and, 
in the case of the non-standard contributions, the masses of 
$N_{1,2,3}$ and of $S_{1,2,3}$.

The  effective Majorana mass term for standard mechanism can be written as 
(see, e.g., \cite{Bilenky:1987ty,Bilenky:2001rz}):
\begin{subequations}
\begin{eqnarray}
\left| m^\nu_{\beta\beta, L} \right| &=& \left|\sum_{i=1}^{3} U_{ei}^2m_i \right|
=  \left|m_1c_{12}^2c_{13}^2 + m_2s_{12}^2c_{13}^2e^{i\alpha} + m_3s_{13}^2e^{i(\beta-2\delta)} \right|
 \end{eqnarray}
  \label{effmassnu}
\end{subequations}
%
 where $m_1,m_2,m_3$ are masses of the light Majorana neutrinos 
$\nu_{1,2,3}$ and we have used the standard parametrization of the PMNS matrix.
 Defining 
\begin{equation}
C_N =  m_e\,m_p\,\frac{M^{0\nu}_N}{M^{0\nu}_\nu} \frac{ M^4_{W_L}}{M^4_{W_R}}\,, 
\label{eq:CN}
\end{equation}
%
the expression for $m_{\beta \beta, R}^N$ can be cast in the form:
\begin{eqnarray}
\left| m^N_{\beta\beta,R} \right| &=&\frac{C_N}{{m_{N_1}}} \left| \bigg[U_{e1}^2 +
\frac{U_{e2}^2 e^{i\, \alpha}{m_{N_1}}}{m_{N_2}} 
+\frac{U_{e3}^2 e^{i\, \beta}{m_{N_1}}}{m_{N_3}} 
\bigg] \right| \nonumber \\
&=& \frac{C_N}{{m_{N_1}}} \left| \bigg[U_{e1}^2 +
\frac{U_{e2}^2 e^{i\, \alpha}{m_2}}{m_1} 
+\frac{U_{e3}^2 e^{i\, \beta}{m_3}}{m_1} 
\bigg] \right|  \nonumber \\
&=& \frac{C_N}{{m_{N_1}m_1}} \left| m^\nu_{\beta\beta,L} \right|\,, 
\hspace{1cm}\text{NO case}\,,
\label{eq:M1} 
\end{eqnarray}
\begin{eqnarray}
\left| m^N_{\beta\beta,R} \right|&=&\frac{C_N}{{m_{N_3}}}\left| \bigg[\frac{U_{e1}^2 {m_{N_3}}}{m_{N_1}} +
\frac{U_{e2}^2 e^{i\, \alpha}{m_{N_3}}}{m_{N_2}} 
+U_{e3}^2 e^{i\, \beta} \bigg]\right|  \nonumber \\
&=& \frac{C_N}{{m_{N_3}}}\left| \bigg[\frac{U_{e1}^2 m_{1}}{m_{3}} +
  \frac{U_{e2}^2 e^{i\, \alpha} m_{2}}{m_{3}}
 +U_{e3}^2 e^{i\, \beta} \bigg]\right|  \nonumber \\
&=& \frac{C_N}{{m_{N_3}m_3}} \left| m^\nu_{\beta\beta,L} \right|\,, 
\hspace{1cm}\text{IO case}\,, 
\label{eq:M2} 
\end{eqnarray}
%
 where we have used Eqs. (\ref{eqn:Mne1}) and (\ref{eqn:Mne2}).
We see that in the considered setting 
the contribution due to exchange of the heavy Majorana neutrinos 
$N_{1,2,3}$ is proportional to the standard contribution due to the 
light Majorana neutrino exchange, 
$|m^N_{\beta\beta,R}| \propto |m^\nu_{\beta\beta,L}|$.

 We consider next $m^S_{\beta\beta, R}$. 
It follows from 
Eq. (\ref{eq:massrel}) that $m_{N_i}= \frac{k^2_{rs}}{m_{S_i}}$.
As it is described in Appendix (\ref{app:lrdssm}), 
the mixing ${V}^{N S}_{ek}\equiv M_{RS} M^{-1}_S U_S$. 
We can diagonalize $M_S$ as 
$M_S = {U_S}M^{D}_S {U_S}^T$. 
Since $M_{RS}=k_{rs}I$, the mixing 
\begin{eqnarray}
&&{V}^{N S}_{ek} = k_{rs} ({U_S}M^{D}_S {U_S}^T)^{-1} U_S\, \nonumber \\
&&\hspace*{0.95cm}=k_{rs}{U^{*}_S} \,\text {diag}(1/m_{S_1},1/m_{S_2},1/m_{S_3})
\label{massmixing}
\end{eqnarray}
%
 Using the relation 
$m_{N_i}= \frac{k^2_{rs}}{m_{S_i}}$
and Eqs. (\ref{eqn:Mne1}) - (\ref{eqn:MSe2}), 
the expression for $m^S_{\beta \beta, R}$ can be written as:
\begin{eqnarray}
\left| m^S_{\beta\beta, R} \right|&=& C_N\, k^2_{rs}\left| \bigg[\frac{U_{e1}^2}{m^3_{S_1}} +
\frac{U_{e2}^2 e^{i\, \alpha}}{m^3_{S_2}} 
+\frac{U_{e3}^2 e^{i\, \beta}}{m^3_{S_3}} 
\bigg]\right|   \nonumber \\
&=& \left| C_N \bigg[\frac{U_{e1}^2 \,  m_{N_1}}{m^2_{S_1}} +
\frac{U_{e2}^2 e^{i\, \alpha} \,  m_{N_2}}{m^2_{S_2}} 
+\frac{U_{e3}^2 e^{i\, \beta} \,  m_{N_3}}{m^2_{S_3}} 
\bigg]\right|   \nonumber \\
%
%
&=& \left| \frac{C_N \,  m_{N_1} \,m_1 \,m^2_3}{m^2_{S_3}\,m^3_1} 
\bigg[U_{e1}^2 +
U_{e2}^2 e^{i\, \alpha}\,\frac{m^3_1}{m^3_{2}}
+ U_{e3}^2 e^{i\, \beta}\, \frac{m^3_1}{m^3_3}
\bigg]\right|\,~~~\text{ NO case}\,, 
\label{eq:M3} 
\end{eqnarray}
\begin{eqnarray}
\left| m^S_{\beta\beta,R} \right| &=& \left|C_N \, m_{N_3} \bigg[\frac{U_{e1}^2}{m^2_{S_1}} \frac{m_{3}}{m_{1}} +
\frac{U_{e2}^2 e^{i\, \alpha}}{m^2_{S_2}}\frac{m_{3}}{m_{2}} 
+\frac{U_{e3}^2 e^{i\, \beta}}{m^2_{S_3}} 
\bigg]\right|   
\nonumber \\
%
%
&=& \left|\frac{C_N \, m_{N_3}\,m_3\,m^2_2}{m^2_{S_2}\,m^3_3} 
 \bigg[U_{e1}^2\,\frac{m^3_3}{m^3_1} +
 U_{e2}^2 e^{i\, \alpha}\, \frac{m^3_3}{m^3_{2}}
 + U_{e3}^2 e^{i\, \beta}
 \bigg]\right|\,,~~~\text{ IO case}\,. 
\label{eq:M4} 
\end{eqnarray}
%

 It follows from Eqs. (\ref{eq:M1}) - (\ref{eq:M4})
that $|m^N_{\beta\beta,R}|$ and $|m^S_{\beta\beta,R}|$
exhibit very unusual dependence on the lightest neutrino mass 
$m_{1(3)}$:  $|m^N_{\beta\beta,R}|\propto 1/m_{1(3)}$ 
and  $|m^S_{\beta\beta,R}|\propto 1/m^2_{1(3)}$.
Correspondingly, the new physics contributions to the 
$0 \nu\beta\beta$ decay amplitude are strongly enhanced at relatively 
small values of $m_{1(3)}$. We will discuss this 
dependence in greater detail in the next Section. 
We will show, in particular, that in the considered scenario with  
$m_{S_k} \sim (10^2 - 10^4)$ GeV and $m_{N_j}\sim (1-100)$ GeV of interest,  
the lightest neutrino mass $m_{1(3)}$  cannot be smaller than 
$\sim 10^{-4}$ eV. We will also show that due to 
the indicated enhancement the new contributions   
dominate over the standard mechanism contribution for 
$m_{1(3)}\sim (10^{-4} - 10^{-2})$ eV 
\footnote{
{ The effects of the heavy Majorana neutrino exchange in 
$0 \nu\beta\beta$ decay amplitude in a left-right symmetric model 
setting were studied recently J. de Vries et al., JHEP 11 (2022) 056 \cite{deVries:2022nyh}. 
However, the version of the left-right symmetric model considered by us and 
in J. de Vries et al., JHEP 11 (2022) 056 \cite{deVries:2022nyh}, differ significantly and 
practically there is no overlap in what concerns the results on the contributions of interest 
of the heavy Majorana 
exchange to the $0 \nu\beta\beta$ decay amplitude.}
}.

\begin{table}[h!]
\begin{center}
	\begin{tabular}{cc|c|c}
	\hline
	Isotope & $T_{1/2}^{0 \nu}~\text{ yrs}$ & $m_{\beta \beta}^{0 \nu}~[\text{eV}]$ & Collaboration \\
	\hline
	$^{76}$Ge	& $> 1.8\times10^{26}$	& $< (0.08 - 0.18)$ 	& GERDA~\cite{GERDA:2020xhi} 			\\
	$^{76}$Ge	& $> 2.7\times10^{25}$	& $< (0.2 - 0.433)$ 	& MAJORANA DEMONSTRATOR~\cite{Majorana:2019nbd} 			\\
		& $> 8.3\times10^{25}$	& $< (0.113 - 0.269)$ 	& ~\cite{Majorana:2022udl} 			\\
	$^{82}$Se	& $> 3.5\times10^{24}$	& $< (0.311 - 0.638)$ 	& CUPID-0~\cite{PhysRevLett.123.032501} 			\\
	$^{130}$Te	& $> 2.2\times10^{25}$	& $< (0.09 - 0.305)$ 	& CUORE~\cite{CUORE:2021mvw} 			\\
	$^{136}$Xe	& $> 3.5 \times10^{25}$	& $< (0.093 - 0.286)$ 	& EXO~\cite{EXO-200:2019rkq} 				\\
	$^{136}$Xe	& $> 1.07\times10^{26}$	& $< (0.061 - 0.165)$ 				& KamLAND-Zen~\cite{KamLAND-Zen:2016pfg} 		\\
		& $> 2.3\times10^{26}$	& $< (0.036 - 0.156)$ 				& ~\cite{KamLAND-Zen:2022tow} 		\\
	\\
	\hline
	\label{tab:half-life}
	\end{tabular}	
\caption{The current lower limits on the half life $T_{1/2}^{0 \nu}$ and upper 
limits on the effective mass parameter $m_{\beta\beta}^{0 \nu}$ of neutrinoless 
double beta decay for different isotopes. The range for the 
effective Majorana mass parameter comes from uncertainties in the 
nuclear matrix element.}
\end{center}
\end{table}

%
\section{Phenomenological analysis}
\label{sec:PA}
%

 In the present Section, we will discuss the effects of the new physics 
contributions to the $0 \nu\beta\beta$ decay amplitude 
on the predictions for the effective Majorana mass 
and the $0 \nu\beta\beta$ decay half-life. 
We recall that if $0 \nu\beta\beta$ decay will be observed, 
the data on the  half-life of $0 \nu\beta\beta$ 
decay generated by the standard mechanism can provide
important  information 
on the absolute scale of light neutrino masses and 
on the neutrino mass ordering \cite{Pascoli:2002xq,Pascoli:2001by}.
With additional input data about the values of the lightest neutrino mass 
$m_{1(3)}$ (or the sum of the neutrino masses), it might be possible to get 
information about the values of the Majorana phases in the PMNS matrix 
as well \cite{Bilenky:2001rz,Pascoli:2005zb}.
In what follows, we will investigate, in particular, how the quoted  
results are possibly modified by the new contributions to the 
 $0 \nu\beta\beta$ decay amplitude.

%
\subsection{Mass parameter ranges}
\noindent
%

 We note first that there exist rather stringent constrains on coupling 
and masses of the heavy Majorana neutrinos associated with the low-scale 
type I seesaw mechanism of neutrino mass generation which have been 
comprehensively discussed in \cite{Bolton:2019pcu}.
In the model studied by us the 
heavy Majorana neutrinos have masses greater than 1 GeV. 
The couplings of the heavy Majorana neutrino states in the 
left-handed (V-A) charged lepton current are suppressed, 
being smaller than $\sim 10^{-6}$. Their couplings in the right-handed 
(V+A) charged current are not suppressed being $\sim U_{PMNS}$, 
but the contribution of the 
(V+A) charged current interaction to the rates of experimentally 
measured observables is suppressed by the factor 
$(M_{W_L}/M_{W_R})^4 < 10^{-8}$, where $M_{W_L} = 80.38$ GeV in the mass of the 
Standard Model $W^\pm$ boson, while $M_{W_R}$ is the mass of its 
$SU(2)_R$ counterpart, and we have used the constraint  $M_{W_R} > 5$ TeV
following from the LHC data. As a consequence, 
the low energy experimental constrains on the heavy Majorana neutrinos 
summarised in   \cite{Bolton:2019pcu}
are satisfied in the model considered by us and do not lead to 
additional restrictions on the couplings and/or masses of these states.

The new non-standard contributions to the 
 $0 \nu\beta\beta$ decay amplitude,
$|m^N_{\beta\beta,R}|$ and $|m^S_{\beta\beta,R}|$, 
as it follows from Eqs. (\ref{eq:M1}) - (\ref{eq:M4}), 
have very peculiar dependence on the lightest neutrino 
mass $m_{1(3)}$. They are strongly enhanced and, 
as we are going to show below, are considerably larger 
that the standard mechanism contribution $|m^\nu_{\beta\beta,L}|$
at $m_{1(3)} \ltap 10^{-3}$ eV, where 
$|m^N_{\beta\beta,R}| > |m^\nu_{\beta\beta,L}|$,
$|m^S_{\beta\beta,R}| >> |m^\nu_{\beta\beta,L}|$ 
and $|m^S_{\beta\beta,R}| >> |m^N_{\beta\beta,R}|$.
At $m_{1(3)} \gtap 5\times 10^{-2}$ eV, however, we have 
$|m^N_{\beta\beta,R}|,|m^S_{\beta\beta,R}| \ll |m^\nu_{\beta\beta,L}|$.
This implies that the most stringent conservative 
experimental upper limit on  $|m^{0\nu}_{\beta\beta}| < 0.156$ eV 
reported by the KamLAND-Zen collaboration [26] 
(see Table 4) applies to $|m^\nu_{\beta\beta,L}|$
since it corresponds to light neutrino masses 
$m_{1,2,3}\gtap 0.1$ eV. Actually, it follows from 
the quoted upper limit that 
\cite{ParticleDataGroup:2018ovx,Penedo:2018kpc}
$m_{1,2,3} \ltap 0.156/(\cos 2\theta_{12} - \sin^2\theta_{13}) \cong 0.55$ eV,
where we have used the $3\sigma$ allowed ranges of 
$\sin^2\theta_{12}$ and $\sin^2\theta_{13}$
given in Table \ref{tab2_oscipara} neglecting the minor 
differences in the ranges corresponding to NO and IO neutrino 
mass spectra. Thus, the largest light neutrino mass 
$m_{3(2)}$ is allowed to vary approximately between 
$\sqrt{\Delta m^2_{31(23)}}\cong 5\times 10^{-2}$ eV and 
0.55 eV.

The Cosmic Microwave Background (CMB)
data of WMAP and PLANCK experiments, combined with
supernovae and other cosmological and astrophysical data 
can be used to obtain information in the form of an upper 
limit on the sum of neutrinos masses and thus on  $m_{3(2)}$ 
(see e.g., ref.~\cite{Vagnozzi:2017ovm}). 
Depending on the model complexity and the input data used 
one typically finds
\cite{ParticleDataGroup:2020ssz} (see also \cite{Capozzi:2017ipn}): 
$\sum_j m_j <  (0.12 - 0.54)$ eV (95\% CL). 
The quoted conservative upper limit on $\sum_j m_j$ 
implies  $m_{3(2)}\lesssim 0.18$ eV. In our phenomenological and numerical 
analysis, we will use somewhat larger values of  $m_{3(2)}$, 
keeping in mind the existence of more stringent limits.
We recall further that in the model considered by us 
$m_{N_i} = k^2_d/m_i$, $m_{S_i} = (k^2_{rs}/k^2_d)m_i$, 
where   $k_d$ and $k_{rs}$ are real constant parameters.
Correspondingly, in the case of NO light neutrino mass spectrum, 
$m_1 < m_2 < m_3$, we have $m_{N_3}< m_{N_2} < m_{N_1}$ and 
$m_{S_1} < m_{S_2} < m_{S_3}$. For IO spectrum, $m_3 < m_1 < m_2$,  
we have instead: $m_{N_2}< m_{N_1} < m_{N_3}$ and 
$m_{S_3} < m_{S_1} < m_{S_2}$. In the double seesaw model 
under discussion, we should always have in the NO (IO) case   
$m_{N_{1(2)}} \ll m_{S_{3(2)}}$, i.e., $m_{S_{3(2)}} \gtap 10\,m_{N_{1(2)}}$.

In what follows, we will consider the values of 
$m_{S_{3(2)}}$ and $m_{N_{1(3)}}$
in the intervals $(1 - 10)$ TeV  and $(10^2 - 10^3)$ GeV, respectively, 
while the mass of the lightest RH Majorana neutrino $N_{3(2)}$ 
will be assumed to satisfy $m_{N_{3(2)}} \geq 1 $ GeV.
The minimal value of $m_{N_{3(2)}}$ of 1 GeV should correspond to the 
maximal allowed value of $m_{3(2)} \cong 0.55$ eV considered by us.
As a consequence, we have: 
$k^2_d = {\rm min}(m_{N_{3(2)}})\,{\rm max}(m_{3(2)}) = 0.55$ eV GeV.  
We get similar value of $k^2_d$ if we use $m_{N_{3(2)}} = 10$ GeV 
and $m_{3(2)} \cong 0.05$ eV.
Given the value of $k^2_d$, the requirement that the mass 
of the heaviest RH Majorana neutrino 
$m_{N_{1(3)}}$ should not exceed $10^3$ GeV implies a lower limit on the mass 
of the lightest Majorana neutrino $m_{1(3)}$: 
$m_{1(3)} = k^2_d/m_{N_{1(3)}} \gtap 0.55\times 10^{-3}$ eV.
Thus, for consistency with the chosen ranges of value of 
the heavy Majorana fermions in the model, the value of the lightest 
neutrino mass should not be smaller than about $5.5\times 10^{-4}$ eV.
In the numerical analysis, we will perform 
we will exploit the range $m_{1(3)} = (10^{-4} - 1.0)$ eV.
 
In the analysis which follows, we will use the values of the 
neutrino oscillation parameters given in Table \ref{tab2_oscipara}.
We set the Dirac phase $\delta = 0$. 
The Majorana phases $\alpha$ and $\beta$ are varied in the interval 
$[0,\pi]$. For the parameters  $M_{W_R}$, $m_{N_{1(3)}}$ $m_{S_{3(2)}}$ 
and the ratio $M^{0\nu}_N/M^{0\nu}_\nu$ the following reference values 
will be utilised: $M_{W_R} = 5.5$ TeV, $m_{N_{1(3)}} = 300$ GeV, 
$m_{S_{3(2)}} = 3$ TeV and $M^{0\nu}_N/M^{0\nu}_\nu \cong 22.2 - 76.3$ 
(concerning $M^{0\nu}_N/M^{0\nu}_\nu$, see Eq. (\ref{eq:MnuMN}) and 
the discussion related to it).

%
\subsection{Light neutrino contribution}
\label{sec:nuexch}
%

The phenomenology of the light neutrino contribution to 
the $0 \nu\beta\beta$ decay half-life, including the properties 
of the corresponding effective Majorana mass $| m^\nu_{\beta\beta,L}|$
have been extensively studied and are well known 
(see, e.g., \cite{ParticleDataGroup:2018ovx}). 
In this subsection, we summarise the main features of $| m^\nu_{\beta\beta,L}|$.

{\bf Normal Ordering}

In this case $| m^\nu_{\beta\beta,L}|$ (see 
Eq. \ref{effmassnu}) can be rewritten in terms of neutrino 
mass square differences as,
\begin{equation}
\left| m^\nu_{\beta\beta,L} \right|=
 \left| m_1 c_{12}^2c_{13}^2 +
 \sqrt{m^2_1 + \Delta m^2_{21}} s_{12}^2c_{13}^2e^{i\alpha} +
\sqrt{m^2_1 + \Delta m^2_{31}}s_{13}^2e^{i\beta} \right|\,.
\label{eq:mnuNO}
\end{equation}
%
The best fit values and the 3$\sigma$ allowed ranges of  
$s_{12}^2 \equiv \sin^2\theta_{12}$, $s_{13}^2 \equiv \sin^2\theta_{13}$ 
and of  $\Delta m^2_{21}$ and $\Delta m^2_{31}$ are given 
in Table \ref{tab2_oscipara}.

The case of hierarchical 
light neutrino mass spectrum corresponds to $m_1\ll m_2 < m_3$. 
In this case 
$m_2 \approx \sqrt{\Delta m^2_{21}} \approx 8.57\times 10^{-3}$ eV and
$m_3 \approx \sqrt{\Delta m^2_{31}} \approx 4.98\times 10^{-2}$ eV, 
and thus  $m_1 \lesssim 8\times 10^{-4}$ eV. 
Depending on the values of the Majorana phases, 
$| m^\nu_{\beta\beta,L}|$ can take values in the interval
$(0.4 - 4.8)\times 10^{-3}$ eV, where we have used the 
the $3\sigma$ allowed ranges of the relevant oscillation parameters.
At $m_1 = 10^{-4}$ eV we have:
$0.91\times 10^{-3}~{\rm eV} \lesssim | m^\nu_{\beta\beta,L}| 
\lesssim 4.37\times 10^{-3}$ eV. 

The effective Majorana mass $| m^\nu_{\beta\beta,L}|$ exhibits strong dependence 
on the values of the Majorana phases $\alpha$ and $\beta$
in the case of NO neutrino mass spectrum with partial hierarchy 
corresponding to $m_1 = (10^{-3} - 10^{-2})$ eV.
Indeed, for $\alpha = \pi$ and $\beta = 0$,
$| m^\nu_{\beta\beta,L}|$
is strongly suppressed for values of $m_1$ lying 
in the interval $(1.3 - 9.0)\times 10^{-3}$ eV, 
where  $| m^\nu_{\beta\beta,L}| \lesssim 2\times 10^{-4} $ eV 
due to cancellations (partial or complete) 
between the three terms in the expression of 
$| m^\nu_{\beta\beta,L}|$. 
Using the best fit values of the neutrino oscillation parameters, 
we find that a complete cancellation takes place 
and  $| m^\nu_{\beta\beta,L}| = 0$ at $m_1 \cong 2.26\times 10^{-3}$ eV. 
At the same time, at $m_1 = 2.26~(9.0)\times 10^{-3}$ eV, 
for example,  $| m^\nu_{\beta\beta,L}|\approx 5~(10)\times 10^{-3}$ eV
if  $\alpha = 0$ and $\beta = 0$.

As $m_1$ increases beyond $10^{-2}$ eV, 
$| m^\nu_{\beta\beta,L}|$ increases almost linearly with $m_1$ 
and at $m_1\cong 0.1$ eV enters the quasi-degenerate (QD)
neutrino mass spectrum region where 
$| m^\nu_{\beta\beta,L}| \gtap 0.05$ eV.

{\bf Inverted Ordering}

In this case we have:
\begin{equation}
\left| m^\nu_{\beta\beta,L} \right|=
 \left|\sqrt{m^2_3 + \Delta m^2_{23} - \Delta m^2_{21}}\, c_{12}^2\,c_{13}^2 +
 \sqrt{m^2_3 + \Delta m^2_{23}}\, s_{12}^2\,c_{13}^2\,e^{i\alpha} +
m_3\,s_{13}^2\,e^{i\beta}) \right|\,.
\label{eq:mnuIO}
\end{equation}
%
The behavior of $| m^\nu_{\beta\beta,L}|$ as a function of 
the lightest neutrino mass $m_3$ is very different 
from the behavior in the NO case.
Given the fact that 
$m_2 = \sqrt{m^2_3 + \Delta m^2_{23}}\gtap 5\times 10^{-2}$ eV, 
$m_1 = \sqrt{m^2_3 + \Delta m^2_{23} - \Delta m^2_{21}}\gtap 5\times 10^{-2}$ eV, 
$s^2_{13} \cong  0.022$ and at $3\sigma$ we have 
$(c^2_{12} - s^2_{12}) \gtap 0.31$, complete 
cancellation between the three terms in Eq. (\ref{eq:mnuIO})
is not possible. Actually, at $m^2_3 \ll \Delta m^2_{23}$, or 
equivalently,  
at $m_3 \lesssim 1.6 \times 10^{-2}$ eV, 
$| m^\nu_{\beta\beta,L}|$ practically does not depend on $m_3$. 
At these values of $m_3$ we have 
$\sqrt{\Delta m^2_{23}} \cos2\theta_{12} \lesssim | m^\nu_{\beta\beta,L}| 
\lesssim \sqrt{\Delta m^2_{23}}$. 
Using the $3\sigma$ allowed ranges of $\sqrt{\Delta m^2_{23}}$
and $\cos2\theta_{12}$ from Table \ref{tab2_oscipara} we get:
$1.51\times 10^{-2}~{\rm eV} \lesssim | m^\nu_{\beta\beta,L}| 
\lesssim 5.06 \times 10^{-2}~{\rm eV}$. 

If the  $0 \nu\beta\beta$ decay were generated by the standard mechanism only, 
the fact that in the case of hierarchical light neutrino mass spectrum 
the minimal value of $| m^\nu_{\beta\beta,L}|$ for the IH spectrum 
is approximately by a factor of 3.4 larger than the maximal 
value of  $| m^\nu_{\beta\beta,L}|$ for the NH spectrum 
opens up the possibility of obtaining information about the type of 
neutrino mass spectrum from a measurement of $| m^\nu_{\beta\beta,L}|$ 
\cite{Pascoli:2002xq}.

As $m_3$ increases beyond $1.6 \times 10^{-2}$ eV, 
$|m^\nu_{\beta\beta,L}|$ also increases and at 
$m_3 \cong 0.1$ eV enters the QD region where
$|m^\nu_{\beta\beta,L}|\gtap 0.03$ eV
growing linearly with $m_3$.

%
\subsection{The contribution due to the exchange of $N_{1.2.3}$}
\label{sec:Nexch}
%
%

\begin{figure*}[!ht]
\centering
       \includegraphics[width=0.5\textwidth]{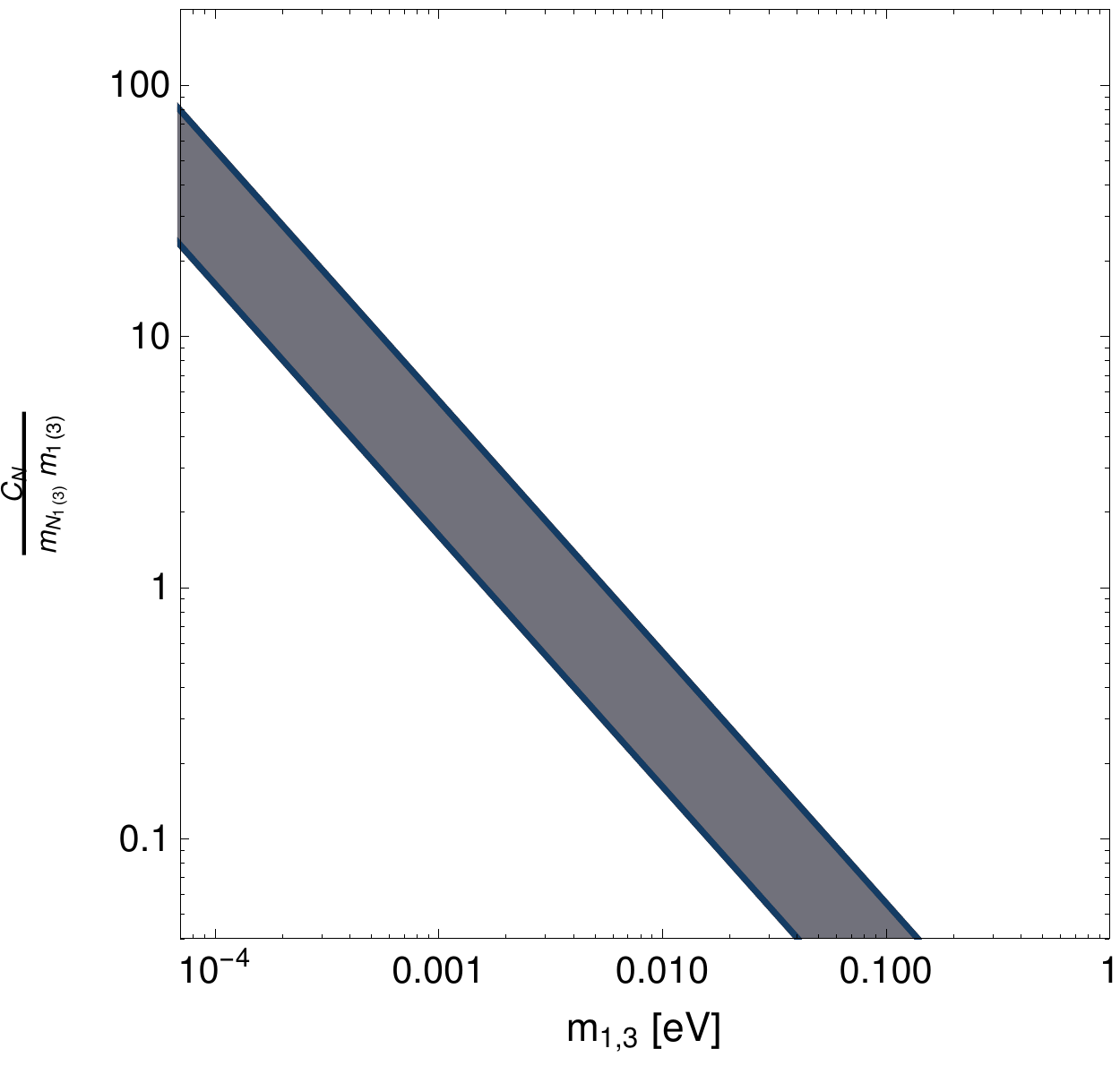}
\caption{ The plot for $C_N/(m_{N_{1(3)}}m_{1(3)}$ with the change of lightest neutrino mass ($m_{1(3)}$) for the
reference value of $m_{N_{1(3)}} = 300 $  GeV and that band corresponds to varying $M^{0\nu}_N /M^{0\nu}_\nu$ in the interval $22.2 \leq {\rm M^{0\nu}_N/M_{\nu}^{0\nu}} \leq 76.3$  as given
in Eq. (\ref{eq:MnuMN}).}
\label{CN}
\end{figure*}

The contribution due to the exchange of virtual $N_{1,2,3}$ 
in the NO and IO cases 
are given respectively in Eqs. (\ref{eq:M1}) and (\ref{eq:M2}) 
can be cast in the form:
\begin{eqnarray}
\left| m^N_{\beta\beta,R} \right| 
&=& \frac{C_N}{{m_{N_{1(3)}}m_{1(3)}}} \left| m^\nu_{\beta\beta,L} \right|\,, 
\hspace{1cm}\text{NO (IO)~case}\,,
\label{eqn:mnNOIO} 
\end{eqnarray}
%
where $C_N$ is defined in Eq. (\ref{eq:CN}) 
and $|m^\nu_{\beta\beta,L}|$ is the effective Majorana mass 
associated with the standard mechanism discussed in the 
preceding sub-section. Using the $M_{W_L} = 80.38$ GeV, 
the reference values of 
$M_{W_R} = 5.5$ TeV 
we get:
\begin{equation} 
\frac{C_N}{m_{N_{1(3)}}\,m_{1(3)}} \cong 
0.729\, \Big(\frac{m_{N_{1(3)}}}{300~{\rm GeV}}\Big)^{-1}\,
 \Big(\frac{m_{1(3)}}{10^{-4}~{\rm eV}}\Big)^{-1}\, 
\frac{M^{0\nu}_N}{M^{0\nu}_\nu}\,.
\label{eq:CNmnmi}
\end{equation}
%
Taking into account the minimal and maximal reference values of 
$M^{0\nu}_N/M^{0\nu}$ in Eq. (\ref{eq:MnuMN}) and 
fixing $m_{N_{1(3)}}$ to the reference value of 300 GeV, the variation of the factor   
 $C_N/(m_{N_{1(3)}}m_{1(3)})$ with the change of lightest neutrino mass is displayed in Fig.\ref{CN}. We also estimate 
the possible range of values of the factor  
$C_N/(m_{N_{1(3)}}m_{1(3)})$ for four values of the lightest 
neutrino mass $m_{1(3)}$ as follows :\\
$C_N/(m_{N_{1(3)}}m_{1(3)}) \cong (16.2 - 55.6)$ for $m_{1(3)} = 10^{-4}$ eV,\\ 
$C_N/(m_{N_{1(3)}}m_{1(3)}) \cong (1.62 - 5.56)$ for $m_{1(3)} = 10^{-3}$ eV,\\ 
$C_N/(m_{N_{1(3)}}m_{1(3)}) \cong (0.32 - 1.11)$ for 
$m_{1(3)} = 5\times 10^{-3}$ eV,\\ 
$C_N/(m_{N_{1(3)}}m_{1(3)}) \cong (0.16 - 0.56)$ for $m_{1(3)} = 10^{-2}$ eV. \\

It is clear from these estimates that
for  $ 10^{-4}{\rm eV} \leq m_{1(3)} \leq 10^{-3}$ eV, 
the  contribution due to exchange of virtual $N_{1,2,3}$ is larger than 
the standard mechanism contribution: 
$| m^N_{\beta\beta,R}| > | m^\nu_{\beta\beta,L}|$. 
For $m_{1(3)}\sim (10^{-4} - 5\times 10^{-4})$ eV we have actually:
$| m^N_{\beta\beta,R}| \gg | m^\nu_{\beta\beta,L}|$. 
In this interval of values of $m_{1}$ 
in the NO case, $| m^N_{\beta\beta,R}|$ lies in the region 
corresponding to the IO neutrino mass spectrum 
if only the standard mechanism 
(i.e., only light Majorana neutrino exchange)
were operative in $0 \nu\beta\beta$ decay.   
The predicted values of 
$| m^N_{\beta\beta,R}|$ in the IO case 
are larger than the experimental limits on 
effective Majorana mass reported by the GERDA and KamLAND-Zen 
experiments (see Table \ref{tab:half-life}).

In the region  $m_{1(3)}\sim (10^{-3} - 10^{-2})$ eV 
we have roughly $| m^N_{\beta\beta,R}| \sim | m^\nu_{\beta\beta,L}|$ 
(see below), with $| m^N_{\beta\beta,R}|$ decreasing as $1/m_{1(3)}$. 
In the NO case, $| m^\nu_{\beta\beta,L}|$ can be strongly suppressed,
i.e., depending on the values of the Majorana phases 
it can have value $| m^\nu_{\beta\beta,L}| \leq 10^{-4}$ eV,
and in this case $| m^N_{\beta\beta,R}|$ will also be 
suppressed. At $m_{1} = 10^{-3}$ eV though at which 
$| m^\nu_{\beta\beta,L}| \cong 3\times 10^{-4}$ eV, $| m^N_{\beta\beta,R}|$ 
can be somewhat larger than $| m^\nu_{\beta\beta,L}|$  
owing to the relevant NME element ratio 
and can have a value $| m^N_{\beta\beta,R}| \cong 1.5\times 10^{-3}$ eV.
In the IO case, $| m^N_{\beta\beta,R}| \gtap | m^\nu_{\beta\beta,L}|$ 
in the discussed region. It can be larger than $| m^\nu_{\beta\beta,L}|$ 
by a factor of 2.

At  $m_{1(3)} > 10^{-2}$ eV, $| m^\nu_{\beta\beta,L}| \gtap | m^N_{\beta\beta,R}|$
and at  $m_{1(3)} \geq 5\times 10^{-2}$ eV, we have 
 $| m^\nu_{\beta\beta,L}| >> | m^N_{\beta\beta,R}|$ and 
the contribution due to the exchange of $N_{1.2.3}$ is 
subleading and practically negligible in both NO and IO cases.

For values of $m_{N_{1(3)}}$ smaller (larger)
than the considered 300 GeV, 
$|m^N_{\beta\beta,R}|$ will have  values which are larger (smaller) 
than those discussed above by the factor
$300~{\rm GeV}/m_{N_{1(3)}}$. Since in the considered scenario 
the mass of the lightest $N_{j}$ 
is assumed to satisfy $m_{N_{3(2)}} \geq 1$ GeV and 
is given by $m_{N_{3(2)}} = (m_{1(3)}/m_{3(2)}) m_{N_{1(3)}}$, 
$m_{1(3)}\gtap 5.5\times 10^{-4}$ eV, 
and from the data it follows that $m_{3(2)} \gtap 5\times 10^{-2}$ eV,
for consistency one should have also $m_{N_{1(3)}}\gtap 100$ GeV. 

%
\subsection{The contribution due to the exchange of $S_{1.2.3}$}
\label{sec:Sexch}
%

The  important parameter 
for the contribution due to the exchange of $S_{1.2.3}$ 
in the NO (IO) case is the dimensionful factor
\begin{equation} 
C^{\rm NO(IO)}_{S} \equiv \frac{C_N\,m_{N_{1(3)}}\,m_{1(3)}\,m^2_{3(2)}} 
{m^2_{S_{3(2)}}\, m^3_{1(3)}}\,,~\hspace{1cm}\text{NO (IO)}\,.
\label{eq:CNS}
\end{equation}
%
Taking into account Eq. (\ref{eq:CNmnmi}), 
$C^{\rm NO(IO)}_{S}$ can be cast in the form:
\begin{equation} 
C^{\rm NO(IO)}_{S} = 0.729\times 10^{-6}~{\rm eV}\, 
\frac{m_{N_{1(3)}}}{300~{\rm GeV}}\, 
\left (\frac{m_{S_{3(2)}}}{3~{\rm TeV}}\right)^{-2}\,
\left(1 + \frac{\Delta m^2_{31(23)}}{m^2_{1(3)}}\right)\, 
\frac{{\rm M^{0\nu}_N}}{{\rm M^{0\nu}_\nu}}\,,~~\text{NO (IO)}\,.
\label{eq:CNS2}
\end{equation}
%

\begin{figure*}[!ht]
\centering
       \includegraphics[width=0.5\textwidth]{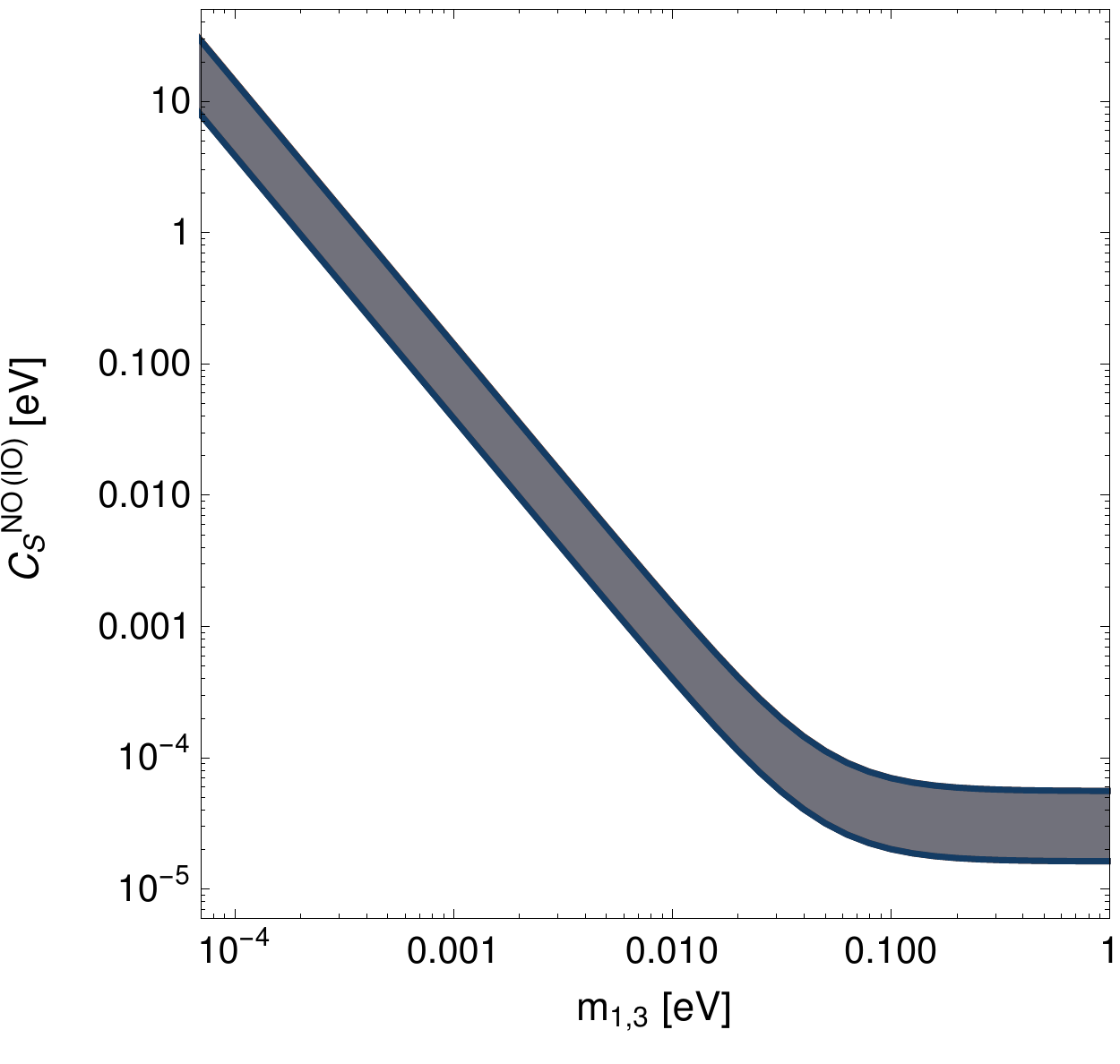}
\caption{ Variation of $C_S$ with the lightest neutrino mass ($m_1$ for NO and $m_3$ for IO) obtained by 
setting $m_{N_{1(3)}}$, $m_{S_{3(2)}}$, $M_{W_R}$
to the reference values
of 300 GeV, 3 TeV, 5.5 TeV respectively and
$\Delta m^2_{31(23)} \cong 2.5\times 10^{-3}~{\rm eV^2}$
(see Table \ref{tab2_oscipara}). The solid band is obtained by varying the ratio $M^{0\nu}_N/M_{\nu}^{0\nu}$ in the range, 
$22.2 \leq {\rm M^{0\nu}_N/M_{\nu}^{0\nu}} \leq 76.3$  as given in Eq. (\ref{eq:MnuMN}).}
\label{CS}
\end{figure*}

Setting $m_{N_{1(3)}}$, $m_{S_{3(2)}}$, $M_{W_R}$ 
to the reference values 
of 300 GeV, 3 TeV, 5.5 TeV respectively and using the values of 
$\Delta m^2_{31(23)} \cong 2.5\times 10^{-3}~{\rm eV^2}$ 
(see Table \ref{tab2_oscipara}) and  
${\rm M^{0\nu}_N/M^{0\nu}}$ as given in Eq. (\ref{eq:MnuMN}),  the variation of $C^{\rm NO(IO)}_{S}$ with the change of lightest neutrino mass is shown in Fig.\ref{CS}.  Using these reference model parameters we calculate the factor $C^{\rm NO(IO)}_{S}$ 
for different values of lightest neutrino mass as given below :\\ 
$C^{\rm NO(IO)}_{S} \cong (4.05 - 13.90)$ eV for $m_{1(3)} = 10^{-4}$ eV,\\ 
$C^{\rm NO(IO)}_{S} \cong (0.040 - 0.139)$ eV for $m_{1(3)} = 10^{-3}$ eV,\\ 
$C^{\rm NO(IO)}_{S} \cong (4.21\times 10^{-4} - 1.45\times 10^{-3})$ eV for 
$m_{1(3)} = 10^{-2}$ eV,\\ 
$C^{\rm NO(IO)}_{S} \cong (3.24\times 10^{-5} - 1.11\times 10^{-4})$ eV for 
$m_{1(3)} = 5\times 10^{-2}$ eV,\\ 
$C^{\rm NO(IO)}_{S} \cong (2.0 - 6.9)\times 10^{-5})$ eV for 
$m_{1(3)} = 10^{-1}$ eV.

It follows from these numerical estimates that $C^{\rm NO(IO)}_{S}$, and thus
$|m^S_{\beta\beta,R}|$, decreases  rapidly with the increase of 
$m_{1(3)}$ in the interval  $(10^{-4} - 5\times 10^{-2})$ eV.

We recall that the contributions due to the exchange of $S_{1.2.3}$ 
in the NO and IO cases are given by
\begin{eqnarray} 
\left| m^S_{\beta\beta,R} \right|
&=& C^{\rm NO}_{\rm S} \left|U_{e1}^2 +
U_{e2}^2 e^{i\, \alpha}\,\frac{m^3_1}{m^3_{2}}
+ U_{e3}^2 e^{i\, \beta}\, \frac{m^3_1}{m^3_3}
\right|\,~~~\text{ NO case}\,, 
\label{eq:mSNO} 
\end{eqnarray}
\begin{eqnarray}
\left| m^S_{\beta\beta,R} \right|
&=& C^{\rm IO}_{\rm S} 
\left|U_{e1}^2 \frac{m^3_3}{m^3_{1}}+
U_{e2}^2 e^{i\, \alpha}\,\frac{m^3_3}{m^3_{2}}
+ U_{e3}^2 e^{i\, \beta}\,
\right|\,~~~\text{ IO case}\,, 
\label{eq:mSIO} 
\end{eqnarray}
%
and that $|U_{e1}|^2 \cong 0.7$ and 
$|U_{e3}|^2 \cong 0.022$.

Consider the NO case. 
We note first that with the increasing of $m_1$ beyond $10^{-2}$ eV, 
the contribution $| m^S_{\beta\beta,R}|$ to the 
$0 \nu\beta\beta$ decay amplitude becomes sub-dominant and 
negligible. 
For $m_{1} = (10^{-4} - 10^{-2})$ eV, 
the ratio  $m^3_1/m^3_3 \ll 1$, 
while $m^3_1/m^3_2 \ll 1$ in the interval 
$m_{1} = (10^{-4} - 4.5\times 10^{-3})$ eV. 
This implies that for $m_1 \ltap 4.5\times 10^{-3}$ eV,
the second and third terms in the expression
(\ref{eq:mSNO}) for $| m^S_{\beta\beta,R}|$
are practically negligible  
and $| m^S_{\beta\beta,R}| \cong  C^{\rm NO}_{\rm S} |U_{e1}|^2$ 
with essentially no dependence on the Majorana phases.
In the interval $m_1 = (4.5\times 10^{-3} - 10^{-2})$ eV 
the ratio  $m^3_1/m^3_2$ increases with $m_1$ and 
at $m_1 =  10^{-2}$ eV we have  $m^3_1/m^3_2 \cong 0.44$. 
Thus,  even at this value the maximal effect of the Majorana phases 
is to change the value of $|U_{e1}|^2\cong 0.7$ to  
$|U_{e1}|^2 \pm  m^3_1/m^3_2 |U_{e2}|^2$, or 
to $0.70 \pm 0.13$, i.e. by at most 18\%, 
where we have used the best fit values of $\sin^2\theta_{12}$ and 
$\sin^2\theta_{13}$. Thus,  in the interval of values 
of $m_1$ of interest, where the contribution of $| m^S_{\beta\beta,R}|$ 
is important, there can not be significant compensation between 
the three terms in the expression for $| m^S_{\beta\beta,R}|$.

From the numerical estimates of 
$| m^\nu_{\beta\beta,L}|$, $| m^N_{\beta\beta,R}|$ and $| m^S_{\beta\beta,R}|$ 
in the preceding and current sub-sections, it follows that 
in the interval of interest $m_{1} = (10^{-4} - 10^{-2})$ eV, 
we have $| m^S_{\beta\beta,R}| > (\gg) | m^\nu_{\beta\beta,L}|,| m^N_{\beta\beta,R}|$.
This is particularly important in the interval 
  $ 10^{-3}~{\rm eV} < m_{1} < 10^{-2})~{\rm eV}$, where 
$| m^\nu_{\beta\beta,L}| <  3\times 10^{-4}$ eV, 
while $| m^S_{\beta\beta,R}|\gtap 3\times 10^{-4}$ eV and, depending on 
the NME, at  $m_{1} = 10^{-3}$ eV 
can be as large as  $| m^S_{\beta\beta,R}|\cong 9.7\times 10^{-2}~{\rm eV} 
\gg | m^\nu_{\beta\beta,L}|,| m^N_{\beta\beta,R}|$.

The situation is very different in the IO case.
In the interval  $m_{3} = (10^{-4} - 10^{-2})$ eV, where 
the factor  $C^{\rm IO}_{\rm S}$ has a relatively large value,  
we have $(m_3/m_{2(1)})^3 \lesssim 7.5\times 10^{-3}$, 
which implies that actually 
$| m^S_{\beta\beta,R}| \cong C^{\rm IO}_{\rm S}\,|U_{e3}|^2\cong  
 2.2\times 10^{-2}\, C^{\rm IO}_{\rm S}$. 
As a consequence of the suppression due to 
$|U_{e3}|^2$ we have in the interval of values of 
 $m_{3}$ of interest $| m^S_{\beta\beta,R}| \ll |m^N_{\beta\beta,R}|$.

%
\subsection{The contribution of the 
interference term}
\label{sec:Nexch}
%
%
The contribution of the interference term  
$2 \text{Re} (m^{N}_{\beta \beta, R} \cdot m^{S^{*}}_{\beta \beta, R})$
in Eq. (\ref{eq:int}) in the  $0\nu\beta\beta$ decay rate
may be non-negligible only in the  interval of values of 
$m_{1(3)} = (10^{-4} - 10^{-2})$ eV,
where the new non-standard contributions are significant.
Using the analytical expressions for 
$| m^N_{\beta\beta,R}|$ and $| m^S_{\beta\beta,R}|$ 
in Eqs. (\ref{eq:M1}) - (\ref{eq:M4}) and the results 
reported in the preceding subsections, it is not difficult to 
estimate the relative magnitude of the contribution of  
this term. Our results show that 
it varies significantly with the type of neutrino mass spectrum,   
the values of the lightest neutrino mass $m_{1(3)}$
and of the Majorana phases $\alpha$ and $\beta$.

 The relative contribution of the interference term of interest 
is determined by the ratio: 
\begin{equation}
R \equiv \frac{2\text{Re} (m^{N}_{\beta \beta, R} \cdot m^{S^{*}}_{\beta \beta, R})}
{| m^\nu_{\beta\beta,L}|^2 +
|m^{N}_{\beta \beta, R}|^2 + |m^{S}_{\beta \beta, R}|^2}\,.
\label{eq:intover}
\end{equation}
%
Using the ratio $R$, the generalised effective Majorana mass 
defined in Eq. (\ref{eq:int}) can be written as:
 \begin{equation}
 m^{\nu+|N+S|}_{ee} = m^{\nu+N+S}_{ee}\,\sqrt{1 + R}\,.
\label{eq:intR}
\end{equation}
%

In the case of NO spectrum, the sign 
of the interference term of interest depends on the
Majorana phases $\alpha$ and $\beta$. 
For $\alpha = \beta = 0$, the ratio $R < 0$  
and thus, the interference terms give a negative contribution to the 
$0\nu\beta\beta$ decay rate. The magnitude of this contribution 
increases quickly when $m_1$ increases from $10^{-4}$ eV to $10^{-3}$ eV
with $R$ changing from (-0.044) to (-0.48). The effect of the 
interference term peaks at $m_1 \cong 2\times 10^{-3}$ eV  
where $R \cong - 0.85$. Thus, at this value of $m_1$ 
we have the maximal suppression of 
$m^{\nu+N+S}_{ee}$ by the factor $\sqrt{1 + R}$:
$m^{\nu+|N+S|}_{ee} \cong  0.39\,m^{\nu+N+S}_{ee}$.
The ratio $R$  decreases rapidly 
when $m_1$ increases beyond $5\times 10^{-3}$ eV 
at which $R\cong -0.48$. The quoted values of $R$ 
at  $m_1 = 10^{-4}$ eV and $m_1 = 10^{-3}$ eV  
are essentially independent of the value of the ratio 
$M^{0\nu}_N/M^{0\nu}_\nu$ lying in the reference interval 
(22.2 - 76.3). The value of $R$ quoted at  $m_1 = 5\times 10^{-3}$ eV  
corresponds to $M^{0\nu}_N/M^{0\nu}_\nu = 76.3$; 
for  $M^{0\nu}_N/M^{0\nu}_\nu = 22.2$ it is significantly  smaller 
in magnitude: $R\cong -0.089$.

The effect of the interference term is quite different 
for $\alpha =\pi$, $\beta = 0$. In this case, 
the interference terms give a positive contribution to the 
$0\nu\beta\beta$ decay rate for $m_1 < 2.26\times 10^{-3}$ eV 
where $R > 0$. At $m_1 \cong 2.26\times 10^{-3}$ eV 
it goes through zero ($R = 0$) since at this value 
$m^\nu_{\beta\beta,L} \cong 0$ and thus  $m^N_{\beta\beta,R} \cong 0$.
Correspondingly, at $m_1 \cong 2.26\times 10^{-3}$ eV 
the generalised effective Majorana mass (Eq. (\ref{eq:int}))
$m^{\nu+|N+S|}_{ee} \cong m^S_{\beta\beta,R} \cong C^{\rm NO}_{\rm S} U_{e1}^2$. 
Taking into account that 
$U_{e1}^2 \cong \cos^2\theta_{12} \cong 0.7$ 
and using Eq. (\ref{eq:CNS2}), for the reference values of 
$m_{N_{1}} = 300$ GeV, $m_{S_{3}} = 3$ TeV,  $M_{W_R} = 5.5$ TeV 
and $M^{0\nu}_N/M^{0\nu}_\nu = 22.2~(76.3)$ 
we find $m^{\nu+|N+S|}_{ee} \cong 5.3\,(18.8) \times 10^{-3}$ eV.
At  $m_1 > 2.26\times 10^{-3}$ eV the interference term is 
negative ($R < 0$). It increases in magnitude as $m_1$ increases 
in the interval $m_1 = 3.5\times 10^{-3} - 10^{-2}$ eV and, e.g.,
at $m_1 = 10^{-2}$ eV we have 
$R \cong -\,0.02~(-\,0.16)$ for 
$M^{0\nu}_N/M^{0\nu}_\nu = 22.2~(76.3)$. At  $m_1 > 10^{-2}$ eV 
we have $|R| \ll 1$ and the interference term has a sub-leading 
(practically negligible) contribution in the  $0\nu\beta\beta$ 
decay rate.

 The results for the ratio $R$ of interest are very different in the 
IO case. It is maximal in magnitude at $m_3 = 10^{-4}$ eV, 
where $R\cong -\,0.54$. However, for $m_3 \sim 10^{-4}$ eV,
the predicted values of the generalised effective Majorana mass 
$|m^{\nu+|N+S|}_{ee}|$ (see Eq. (\ref{eq:int}), 
as we will show in the next Section,  
are strongly disfavored (practically ruled out) by the existing upper limits 
from the KamLAND-Zen and GERDA experiments (see Table  \ref{tab:half-life}
and Fig. 2). In the region of values of  $m_3 \gtap 10^{-3}$ eV, 
where the predictions for the generalised effective Majorana mass 
are compatible with the current experimental upper limits 
one has  $|R| < 0.06$, with the value of 
$|R|$ decreasing rapidly with the increasing of $m_3$.
Thus, in the IO case, the interference term under discussion 
has at most, a sub-leading (practically negligible) effect  
on the  $0\nu\beta\beta$ half-life in the 
interval of values of $m_3$ where the predictions of the model considered 
are compatible with the existing lower limits on the half-life.

%
\subsection{Numerical Results}
\noindent
%
%

It follows from the analyses performed in the preceding 
four subsections in particular, that in the NO case 
the contribution due to the  $S_{1,2,3}$
exchange, $|m^S_{\beta\beta,R}|$,  
dominates over the light neutrino $\nu_i$ and 
$N_{1,2,3}$ exchange contributions 
for $10^{-4}~{\rm eV} \leq m_1 \lesssim 1.5\times 10^{-3}$ eV. 
As a consequence, in the indicated interval of values of 
$m_1$ the generalised effective Majorana mass 
$|m^{\nu+|N+S|}_{ee}|$ exhibits weak dependence
on the Majorana phases $\alpha$ and $\beta$
since $|m^S_{\beta\beta,R}|$ practically does not depend 
on these phases. At $m_1 \gtap 2\times 10^{-3}$ eV, 
for $\alpha = \beta = 0$,
the $S_{1,2,3}$ contribution is subleading and 
\begin{equation}
 m^{\nu+|N+S|}_{ee} \cong
\sqrt{| m^\nu_{\beta\beta,L}|^2 + |m^N_{\beta\beta,R}|^2}  
\cong | m^\nu_{\beta\beta,L}|\,
\left (1 + \frac{C_N}{m_{N_{1}}m_{1}}\right )^{\frac{1}{2}}\,,
~~{\rm NO}\,,~\alpha = \beta = 0\,,
\label{eq:GEMMIO}
\end{equation}
%
where we have used Eq. (\ref{eqn:mnNOIO}).  
For $\alpha = \pi$, $\beta = 0$, however, 
$|m^\nu_{\beta\beta,L}|$ is strongly suppressed 
in the interval $m_1 \cong (1.5\times 10^{-3}  - 9\times 10^{-3})$ eV 
and goes through zero at  $m_1 \cong 2.26\times 10^{-3}$ eV,
where the value of $m_1$ is obtained using 
the best fit values of the neutrino oscillations parameters. 
Therefore $|m^S_{\beta\beta,R}|$ gives significant contribution to 
$m^{\nu+|N+S|}_{ee}$ in the indicated interval and determines 
the minimal value of $m^{\nu+|N+S|}_{ee}$. 
At $m_1 \cong 2.26\times 10^{-3}$ eV, e.g., we have 
$|m^{\nu+|N+S|}_{ee}| \cong |m^S_{\beta\beta,R}| \cong C^{\rm NO}_{\rm S}|U_{e1}|^2$.

In contrast, in the IO case 
the contribution due to the  $S_{1,2,3}$ exchange   
$|m^S_{\beta\beta,R}|$ and of the interference term 
$ 2\text{Re} (m^{N}_{\beta \beta, R} \cdot m^{S^{*}}_{\beta \beta, R})$
in the interval of values of $m_3$ of interest 
are practically negligible. Thus, for 
the generalised effective Majorana mass is given by:
\begin{equation}
 m^{\nu+|N+S|}_{ee} \cong
\sqrt{| m^\nu_{\beta\beta,L}|^2 + |m^N_{\beta\beta,R}|^2}  
\cong | m^\nu_{\beta\beta,L}|\,
\left (1 + \frac{C_N}{m_{N_{3}}m_{3}}\right )^{\frac{1}{2}}\,,~~{\rm IO}\,.
\label{eq:GEMMIO}
\end{equation}
%
In this case $|m^{\nu+|N+S|}_{ee}|$ depends 
significantly on the Majorana phases.

The conclusions regarding the new non-standard contributions 
due to the exchange of virtual heavy Majorana fermions 
$N_{1,2,3}$ and $S_{1,2,3}$ to the $0\nu\beta\beta$ decay  
generalised effective Majorana mass and 
half-life reached in the phenomenological 
analysis are confirmed by our numerical results. 
\begin{figure*}[!ht]
\centering
\includegraphics[width=0.31\textwidth]{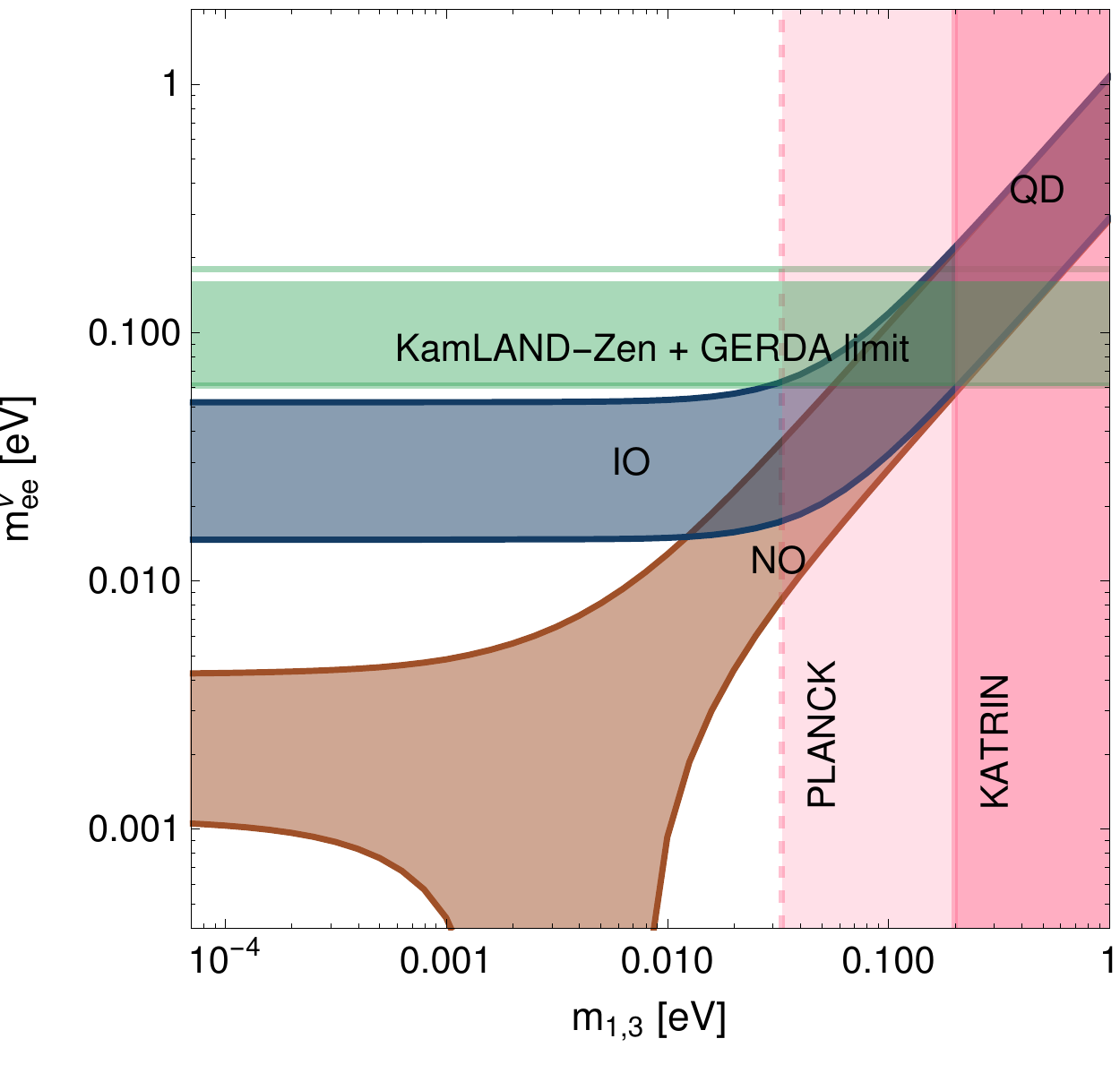} 
\hspace*{0.1cm}
\includegraphics[width=0.31\textwidth]{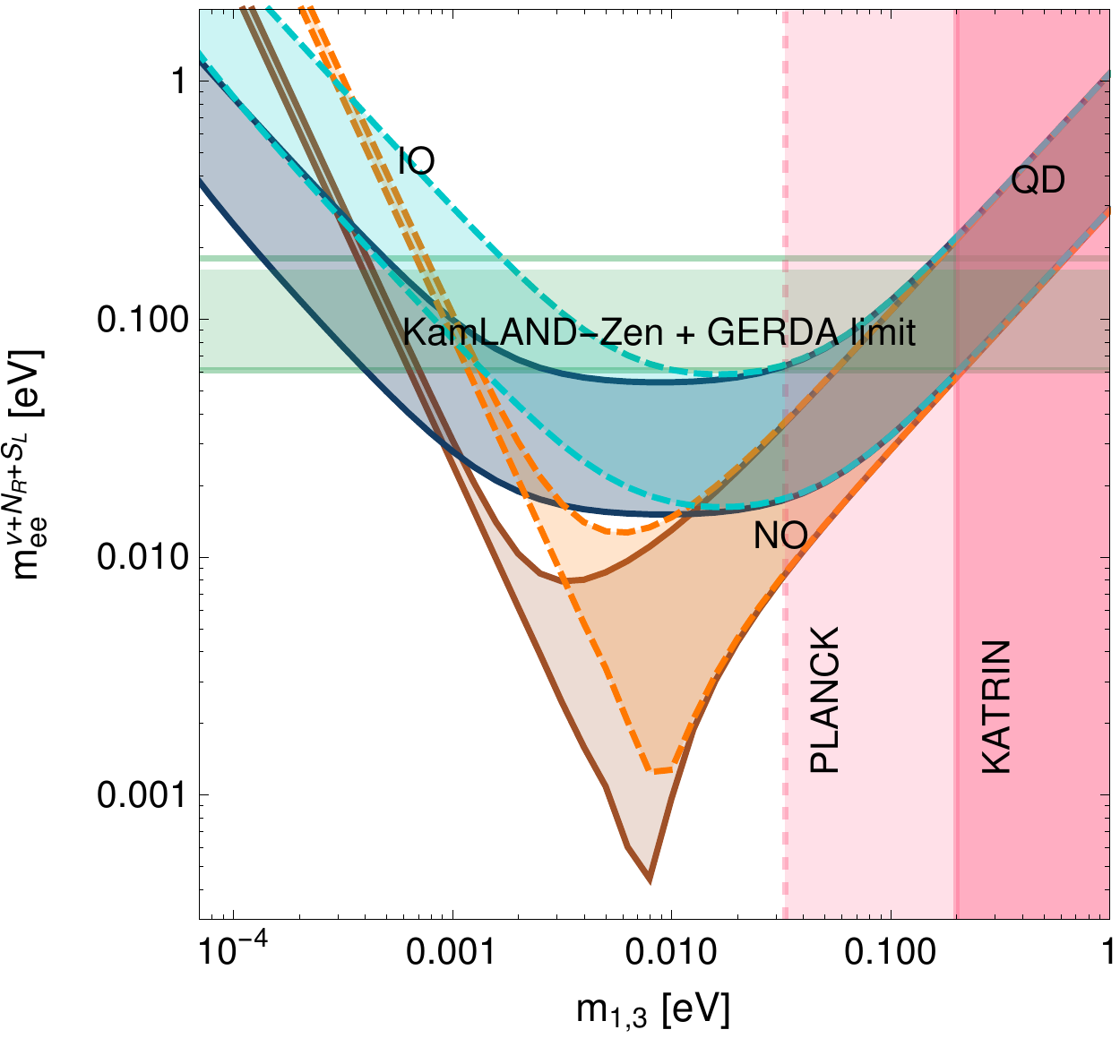} 
\hspace*{0.1cm}
 \includegraphics[width=0.31\textwidth]{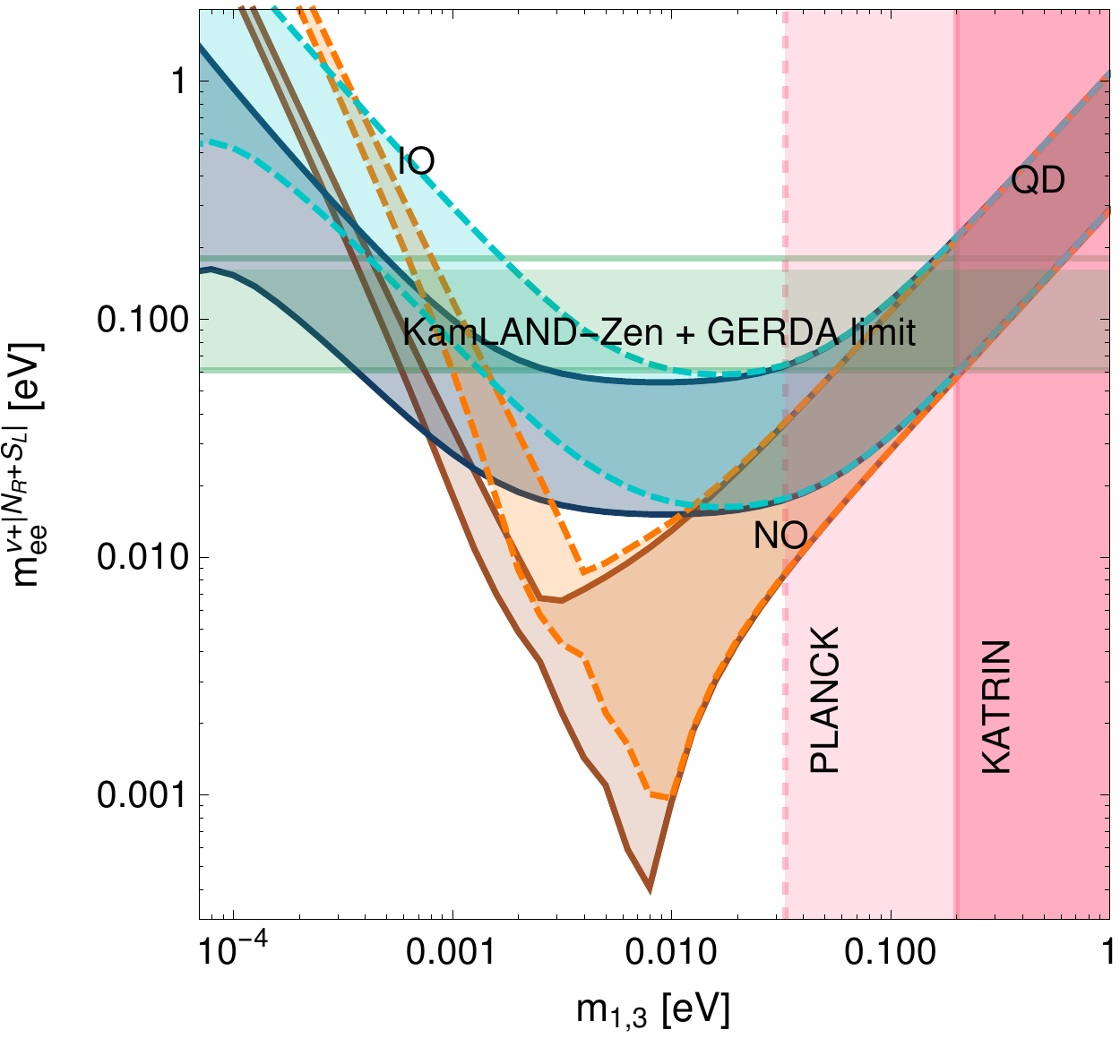} 
\hspace*{0.1cm}
\includegraphics[width=0.31\textwidth]{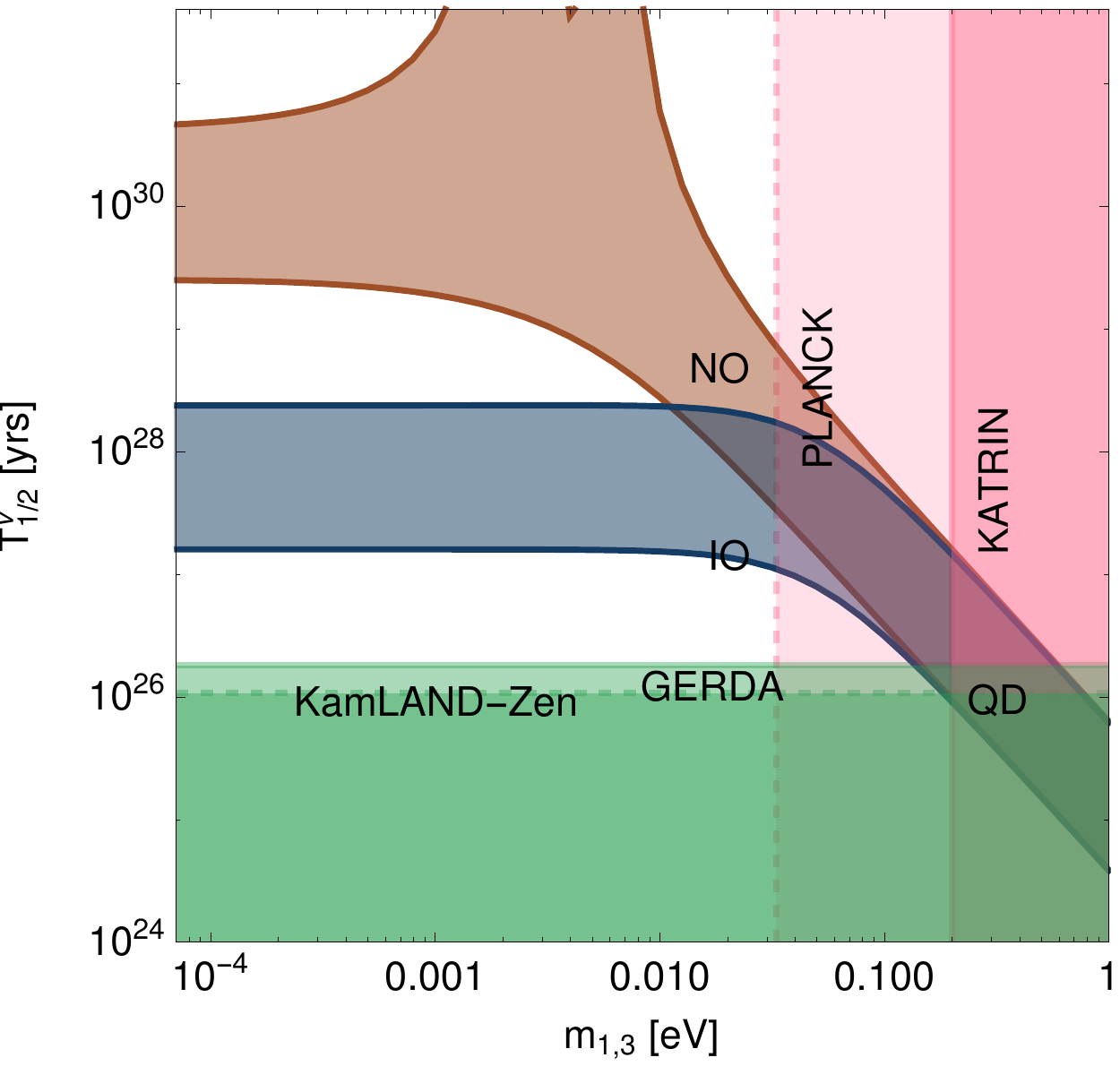}
\hspace*{0.1cm}
\includegraphics[width=0.31\textwidth]{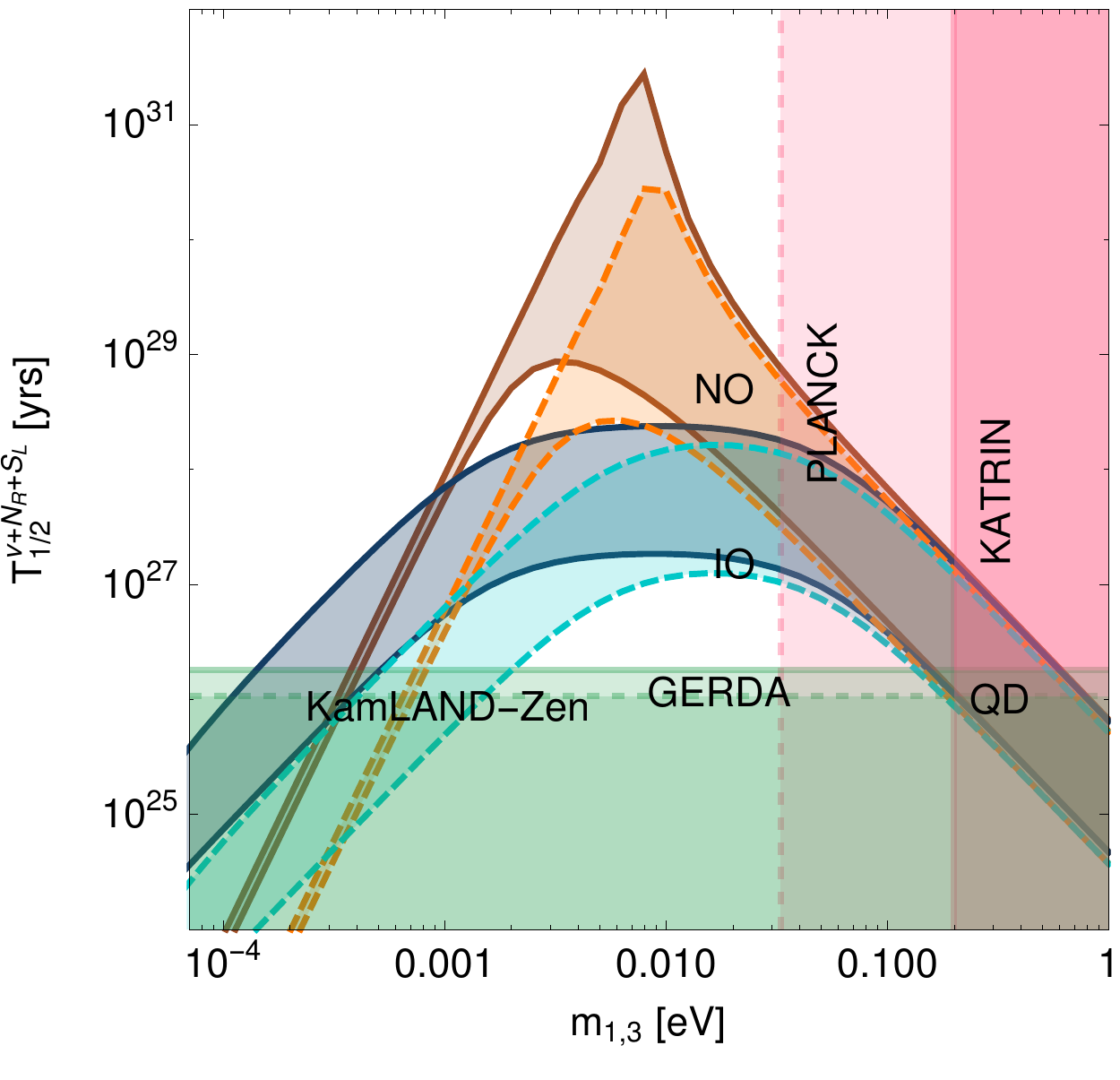} 
\hspace*{0.1cm}
\includegraphics[width=0.31\textwidth]{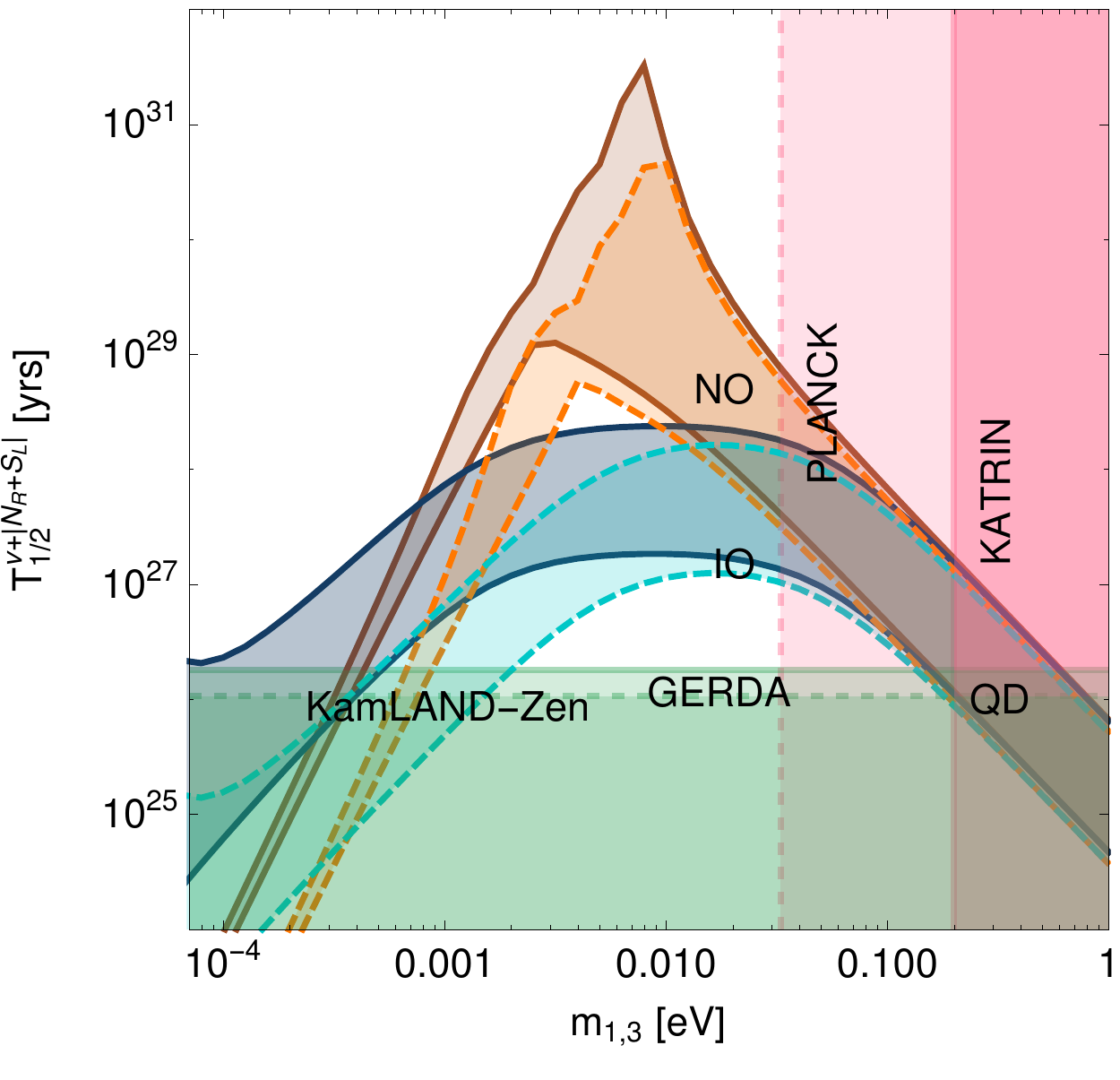}

\caption{Plots showing effective Majorana mass parameter (upper panel) 
and half-life (lower panel) of $0\nu\beta\beta$ decay as functions of 
the lightest neutrino mass $m_{1(3)}$ in the case of NO (IO) 
light neutrino mass spectrum.
The left upper panel shows the dependence on $m_{1(3)}$ of the 
standard mechanism effective Majorana mass, while in the 
middle and right upper panels, the dependence of 
the generalised effective Majorana mass in which the contributions 
due to the exchange of the heavy Majorana fermions $N_{1,2,3}$ and 
$S_{1,2,3}$ are included without accounting for (middle upper panel) 
and accounting for (upper right panel) their interference.
The brown and blue bands correspond respectively to the NO (NH) and IO (IH) 
types of light neutrino mass spectrum. The overlap of the two
bands indicates the region of the QD spectrum.
The lower panels show the dependence on $m_{1(3)}$ of the
$0\nu\beta\beta$ decay half-lives corresponding to the 
respective upper panels. The green horizontal band represents 
the current bound on effective Majorana 
mass from the experiments KamLAND-Zen and GERDA as given in 
Table~\ref{tab:half-life}, whereas the vertical pink bands represent the
bound corresponding to the upper limit 
on the sum of light neutrino masses of 0.12 eV 
reported by the Planck \cite{Planck:2018vyg} and the prospective 
bound of 0.20 eV that can be set by the  KATRIN~\cite{Aker:2019uuj}
experiment. See text for further details.
}
\label{plot:onubb2}
\end{figure*}
%
\noindent
These are illustrated in Fig. \ref{plot:onubb2}.
In the three upper panels of Fig.  \ref{plot:onubb2} 
we show i) $m^\nu_{ee} \equiv |m^\nu_{\beta\beta,L}|$ (left panel),
ii) $ m^{\nu+N+S}_{ee} \equiv \sqrt{|m^\nu_{\beta\beta,L}|^2 + 
| m^N_{\beta\beta,R}|^2 + | m^S_{\beta\beta,R}|^2}$ (middle panel), and 
iii)  $ m^{\nu+|N+S|}_{ee} \equiv
\sqrt{| m^\nu_{\beta\beta,L}|^2 + |m^N_{\beta\beta,R} + m^S_{\beta\beta,R}|^2}$  
(right panel), as functions of the lightest neutrino mass $m_{1(3)}$ 
in the case of NO (IO) light neutrino mass spectrum.
Thus, the upper left  panel shows the dependence on $m_{1(3)}$ of the 
standard mechanism effective Majorana mass, while in the upper 
middle and right panels, the dependence of 
the generalised effective Majorana mass (GEMM) in which the contributions 
due to the exchange of the heavy Majorana fermions $N_{1,2,3}$ and 
$S_{1,2,3}$ are included without accounting for (middle panel) 
and accounting for (right panel) their interference.
The brown and blue bands correspond respectively to the NO (NH) and IO (IH) 
types of light neutrino mass spectrum. The overlap of the two
bands indicates the region of the QD spectrum.
Following the discussion of NME in Section 4, 
the ratio of nuclear matrix elements 
${\cal M}^{0\nu}_N/{\cal M}^{0\nu}_\nu$ is varied in the interval 22.2 - 76.3,
as given in Eq. (\ref{eq:MnuMN}).
The minimal (maximal) value in this interval, 
${\cal M}^{0\nu}_N/{\cal M}^{0\nu}_\nu = 22.2~(76.3)$, 
corresponds to ${\cal M}^{0\nu}_\nu  = 4.68~(5.26)$. 
The results for ${\cal M}^{0\nu}_N/{\cal M}^{0\nu}_\nu = 22.2~(76.3)$ 
are indicated with solid (dashed) lines.
For the parameters  $M_{W_R}$, $m_{N_{1(3)}}$ and $m_{S_{3(2)}}$ 
the reference values of 5.5 TeV, 300 GeV, and 3TeV, respectively,
are used. All plots are obtained by varying the neutrino oscillation 
parameters in their respective 3$\sigma$ allowed ranges.  
The Majorana phases $\alpha$ and $\beta$ are varied in the interval 
$[0,\pi]$, while the Dirac phase $\delta$ is set to zero.
For both ${\cal M}^{0\nu}_N/{\cal M}^{0\nu}_\nu = 22.2$ and 76.3,
the curves showing the maximal (minimal) values of 
GEMM as functions of $m_{1(3)}$ 
correspond to $\alpha = \beta = 0$ ($\alpha =\pi$, $\beta = 0$).
The lower panels show the dependence on $m_{1(3)}$ of the
$0\nu\beta\beta$ decay half-lives corresponding to the 
respective upper panels. 
The green horizontal band represents the current bound on effective Majorana 
mass from the experiments KamLAND-Zen and GERDA as given in 
Table~\ref{tab:half-life}, whereas the vertical pink bands represent the
bound corresponding to the upper limit 
on  the sum of light neutrino masses of 0.12 eV 
reported by the Planck experiment \cite{Planck:2018vyg} and the prospective 
bound of 0.20 eV that can be set by the  KATRIN~\cite{Aker:2019uuj}
experiment. 

 A comparison between the upper left and right panels in 
Fig. \ref{plot:onubb2} shows that the presence of the new non-standard 
contributions change drastically the dependence  
of the effective Majorana mass of the standard mechanism 
$|m^\nu_{ee}|$ on the lightest neutrino mass $m_{1(3)}$ 
at $m_{1(3)} <  10^{-2}$ eV, where the new 
contributions dominate over the standard contribution.
At  $m_{1(3)} \gtap  10^{-2}$ eV the new contributions 
are strongly suppressed and practically negligible and
we have   $|m^{\nu+|N+S|}_{ee}|\cong |m^\nu_{ee}|$, 
as also is clearly seen in Fig. \ref{plot:onubb2}. 
 
 The non-standard contributions are so large 
at relatively small values of the  $m_{1(3)}$ that 
for $m_{1(3)} \lesssim 2\times 10^{-4}$ eV 
in the NO case they are ruled out even for the minimal value 
of $M^N_{0\nu}/M^\nu_{0\nu} = 22.2$ 
by the existing upper limits from the KamLAND-Zen and GERDA 
experiments. In the IO case the new contributions are 
also ruled our  for $M^N_{0\nu}/M^\nu_{0\nu} = 76.3$; 
for  $M^N_{0\nu}/M^\nu_{0\nu} = 22.2$ they are ruled out for 
$\alpha = \beta = 0$, while for $\alpha = \pi$, $\beta = 0$, 
they are compatible with the  KamLAND-Zen and GERDA 
upper limits. 

 For NO spectrum and $\alpha = \beta = 0$, 
the inequality  $|m^{\nu+|N+S|}_{ee}|> (>>) |m^\nu_{ee}|$
always holds in the interval of values of
 $m_{1} \cong  (3\times 10^{-4} - 8\times 10^{-3})$ eV 
where the non-standard contributions are significant.
In  this interval and for  $M^N_{0\nu}/M^\nu_{0\nu} = 76.3~(22.2)$, 
$|m^{\nu+|N+S|}_{ee}|\gtap 0.009~(0.007)$ eV, 
and with the exception of a very narrow interval 
around $m_{1} \cong 4.0~(2.5)\times 10^{-3}$ eV at which the 
quoted minimum of $|m^{\nu+|N+S|}_{ee}|$ takes place,  
we have $|m^{\nu+|N+S|}_{ee}|\gtap 0.010$ eV.  
In most of the considered intervals of values of 
$m_1$ the half-life $T^{\nu +|N_R + S_L|}_{1/2} \lesssim 10^{28}$ yrs. 

 In the case of NO spectrum, $\alpha =\pi$, $\beta = 0$ and 
$M^N_{0\nu}/M^\nu_{0\nu} = 76.3~(22.2)$,  
the value of $|m^{\nu+|N+S|}_{ee}|\gtap 0.010$ eV, 
and  $T^{\nu +|N_R + S_L|}_{1/2} \lesssim 10^{28}$ yrs,
at $m_{1} \lesssim 2.0~(1.5)\times 10^{-3}$ eV.
For the minimal value of  $|m^{\nu+|N+S|}_{ee}|$ 
we find ${\rm min}(|m^{\nu+|N+S|}_{ee}|) \cong 9~(3)\times 10^{-4}$ eV.
It takes place at $m_{1} \cong 9.0~(8.0)\times 10^{-3}$ eV.
We recall that  $|m^\nu_{ee}|$ goes through zero at 
$m_{1} \cong 2.26\times 10^{-3}$ eV, while 
$|m^{\nu+|N+S|}_{ee}|\cong 18.3~(5.3)\times 10^{-3}~$ eV 
at this value of $m_1$.
 
For IO spectrum, we have 
$|m^{\nu+|N+S|}_{ee}|\gtap 0.015$ eV 
for $m_3 < 9\times 10^{-3}$ eV, where 
$|m^{\nu+|N+S|}_{ee}| >  |m^\nu_{ee}|$ 
for any of the considered values of $M^N_{0\nu}/M^\nu_{0\nu}$
and of $\alpha$ and $\beta$. The approximate 
equality  $|m^{\nu+|N+S|}_{ee}| \cong  |m^\nu_{ee}|$ 
holds at $m_3\gtap 10^{-2}$ eV.
For all considered values of $m_3$,  $M^N_{0\nu}/M^\nu_{0\nu}$ and 
the Majorana phases the predicted half-life  
$T^{\nu +|N_R + S_L|}_{1/2} \lesssim 10^{28}$ yrs, while  
in the case of $\alpha = \beta = 0$, 
we have  $T^{\nu +|N_R + S_L|}_{1/2} \lesssim 2\times 10^{27}$ yrs. 

It follows from our numerical analysis  that most of the parameter 
space of the considered model, the predictions for the 
$0\nu\beta\beta$ decay generalised effective Majorana 
mass and half-life are within the sensitivity range of the 
planned next generation of neutrinoless double beta decay 
LEGEND-200 (LEGEND-1000), nEXO, KamlAND-Zen-II, CUPID, NEXT-HD 
(see \cite{Giuliani:2019uno,Agostini:2022zub} 
and references quoted therein).

\section{Comments on LFV and LHC signatures}
\label{sec:comments}

The considered model has rich lepton flavour violating (LFV) 
and collider phenomenology. A detailed investigation of the 
model's phenomenology is beyond the scope of the present study.
We limit ourselves here with a few brief comments.

  The LFV processes as like 
$\mu \rightarrow e + \gamma$, $\mu \rightarrow 3e$ decays and 
$\mu - e$ conversion in nuclei  can be mediated by heavy RH and 
sterile neutrinos $N_{1,2,3}$ and $S_{1,2,3}$. 
Although we expect the contributions due to  $N_{1,2,3}$ and especially 
due to $S_{1,2,3}$ to be rather suppressed, there might be 
a relatively large region of the 
model's parameter space where they might still be in the range of 
sensitivity of the next generation of experiments 
MEG II, Mu3e, Mu2e, COMET and PRISM/PRIME 
(see, e.g., \cite{Calibbi:2017uvl} and the references therein).

At LHC, the main channel for the production of the heavy RH neutrinos $N_{1,2,3}$ is
via on-shell $Z_R$ production and $W_R$ fusion and can be expressed as 
$p+p\rightarrow W_R^{\pm}\rightarrow l^{\pm} + N_j$, $l=e,\mu,\tau$.
This $N_j$ further decays as 
$N_j\rightarrow W_R^*\rightarrow l^{\prime \pm} + 2j$, $l^{\prime}=e,\mu,\tau$,
which is considered as the ``smoking gun'' signature of lepton number 
and lepton flavour violation at LHC. 
This rapid decay of $N_j$ happens in the case its mass is 
sufficiently large. Our model satisfies this requirement
as we have taken  ${\rm max}(M_{N_j})\sim 100$ GeV.
We recall that the mass of $W_R$ is constrained by experiments CMS,ATLAS and 
low energy precision measurements as $M_{W_R}\gtrsim 5$ TeV ~\cite{ATLAS:2018dcj,ATLAS:2019isd,CMS:2018agk,Li:2020wxi,Dekens:2021bro} and considering 
the relation $M_{Z_R}\simeq 1.2 M_{W_R}$, the mass of $Z_R$ can be constrained 
as $M_{Z_R}\gtrsim 6$ TeV. 
If the mass of $N_j$ lies in the range $5-20$ GeV, 
then it takes some time to decay and travels some distance resulting in a 
displaced vertex of leptons \cite{Helo:2013esa,Izaguirre:2015pga}. 
So, the observable in this case would be a prompt charged lepton and 
a displaced leptonic vertex. The current status of displaced vertex searches 
at LHC can be found in ref. \cite{ATLAS:2012av,CMS:2013czn,ATLAS:2012cdk}.  
Another distinguishing feature in the signatures of small mass ($<100$ GeV) 
and large mass ($\sim 800$ GeV) RH neutrinos $N_j$ is the angle between 
the produced charged leptons. In the former case, parallel tracks of charged 
leptons are expected, whereas in the later case, back-to-back emissions 
are expected \cite{Almeida:2000pz}.


\section{Summary}
\label{sec:concl}

In the present article, we have derived predictions for 
the neutrinoless double beta ($0\nu\beta\beta$)
decay generalised effective Majorana mass
and half-live in a left-right (L-R) symmetric model with 
the double seesaw mechanism at the TeV scale.
The gauge group of the model is the standard L-R symmetric 
extension of the Standard Model (SM) 
gauge group: $\mathcal{G}_{LR}\equiv SU(2)_L \times SU(2)_R \times U(1)_{B-L}$.
The fermion sector has the usual 
for the L-R symmetric models, three families of 
left-handed (LH) and right-handed (RH) 
quark and lepton fields, including right-handed neutrino fields $N_{\beta R}$,
$\beta = e,\mu,\tau$, assigned respectively to  
$SU(2)_L$  and $SU(2)_R$ doublets. It included also   
three $SU(2)_{L,R}$ singlet LH fermion fields $S_{\gamma L}$.
The Higgs sector is composed of two $SU(2)_L$  and $SU(2)_R$
Higgs doublets $H_L$ and $H_R$, and of a bi-doublet  $\Phi$.
The vacuum expectation value (VEV) of  the  $SU(2)_R$ Higgs doublet 
$H_R$ breaks the  $\mathcal{G}_{LR}$ gauge symmetry  
to the SM gauge symmetry  $SU(2)_L \times U(1)_{Y}$, 
while the VEVs of the two neutral components of the 
bi-doublet $\Phi$ break  $SU(2)_L \times U(1)_{Y}$ to $U(1)_{\rm em}$.
The Yukawa couplings of the LH and RH fermion doublets 
to the bi-doublet $\Phi$ generate (via the VEVs of 
the neutral components of $\Phi$) Dirac mass terms for the 
quarks and charged leptons, as well as 
a $\nu_{\alpha L} - N_{\beta R}$ Dirac mass term 
$M^\nu_{\rm D}$ involving the 
LH active flavor neutrino fields 
$\nu_{\alpha L}$ and the RH fields $N_{\beta R}$, 
$\alpha,\beta=e,\mu,\tau$. 
The singlet LH fermion fields $S_{\gamma L}$ are assumed to have a 
Majorana mass term 
$M_{\rm S}$ and Yukawa coupling 
with the RH doublets containing $N_{\beta R}$ which involves $H_R$. 
This Yukawa coupling produces  a $S_{\gamma L} - N_{\beta R}$ Dirac mass term 
$M_{\rm RS}$ when  $H_R$ develops a non-zero VEV.
Under the condition $|M_{\rm RS}| \ll |M_{\rm S}|$,
the RH neutrinos  $N_{\beta R}$ get a Majorana mass term 
$M_{\rm R} \cong -\,M_{\rm RS}M^{-1}_{\rm S} M^T_{\rm RS}$ 
via a seesaw-like mechanism.
This in turn generates a Majorana mass term 
for the LH flavour neutrinos 
$m_\nu \cong -\, M^\nu_{\rm D}M^{-1}_{\rm R}(M^\nu_{\rm D})^T$
via a second seesaw mechanism 
\footnote{Hence the term ``double or cascade seesaw mechanism 
of neutrino mass generation''.}. 
In such a way, the model contains in addition to the three 
light Majorana neutrinos $\nu_i$ having masses $m_i$, 
two sets of heavy Majorana 
particles $N_j$ and $S_k$, $j,k=1,2,3$, with masses 
$m_{N_j} \ll m_{S_k}$. The double seesaw scenario allows 
the RH neutrinos $N_j$ to have masses naturally  
at the GeV-TeV scale.

In our analysis of the  $0\nu\beta\beta$
decay predictions of the model, we have 
considered the case of $m_{N_j} \sim (1 - 1000)$ GeV 
and ${\rm max}(m_{S_k}) \sim (1 - 10)$ TeV, $m_{N_j} \ll m_{S_k}$. 
Working with a specific version of the model 
which can be obtained by employing symmetry arguments and 
in which the 
Dirac mass terms  $M^\nu_{\rm D}$ and $M_{\rm RS}$
are diagonal, $M^\nu_{\rm D} = k_d\,{\bf I}$, $M_{\rm RS} = k_{rs}\,{\bf I}$,
$k_d$ and $k_{RS}$ being constant mass parameters and 
{\bf I} the $3\times 3$ unit matrix, 
we have studied in detail the new ``non-standard'' contributions to the 
$0\nu\beta\beta$ decay amplitude and half-life 
arising from diagrams with an exchange of virtual $N_j$ and $S_k$.
The self-consistency of the considered set-up 
requires that the lightest neutrino mass for the neutrino mass 
spectrum with normal ordering (NO), $m_1$, 
or with inverted ordering (IO), $m_3$, 
has to be not smaller that approximately $10^{-4}$ eV.
Moreover, the RH neutrino ($N_{\beta R}$) and sterile fermion  
($S_{\gamma L}$) mixings are determined 
by the light neutrino PMNS mixing matrix.   
In the analysis of the new non-standard contributions to the 
 $0\nu\beta\beta$ decay amplitude we took into account 
the values of the nuclear matrix elements (NMEs) 
 ${\cal M}^{0\nu}_N$ and ${\cal M}^{0\nu}_\nu$
 associated with, respectively, the light and heavy Majorana neutrino 
exchange contributions, calculated for the four isotopes 
$^{76}$Ge,  $^{82}$Se,  $^{130}$Te, $^{136}$Xe
by six different groups of authors using different 
methods of NME calculation (Table \ref{tab:nucl-matrix}). 
We made use of the fact that the ratio  
${\cal M}^{0\nu}_N/{\cal M}^{0\nu}_\nu$ 
reported by each of the six cited groups 
is essentially the same for the considered four isotopes -- 
it varies with the isotope by not more than 
$\sim 15\%$. For a given isotope 
the ratio of interest obtained by the six different 
methods of NME calculation 
varies by a factor of up to $\sim 3.5$. 
In view of this, we took into account the 
uncertainties in the NME calculations  by using 
the following reference range of the ratio 
${\cal M}^{0\nu}_N/{\cal M}^{0\nu}_\nu = 22.2 - 76.3$, 
which corresponds to  $^{76}$Ge.

 We analyzed in detail the properties 
of the new non-standard contributions to the 
$0\nu\beta\beta$ decay amplitude arising due to the exchange 
of virtual heavy Majorana fermions $N_j$ and $S_k$, 
parametrized as effective Majorana masses 
$m^N_{\beta\beta,R}$ (Eq. \ref{eqn:mnNOIO}) and $m^S_{\beta\beta,R}$ 
(Eq. \ref{eq:mSNO}), respectively. 
These analyses showed that 
both $|m^N_{\beta\beta,R}|$ and $|m^S_{\beta\beta,R}|$ 
are strongly enhanced at relatively small values of 
the lightest neutrino mass $m_{1(3)}\sim (10^{-4} - 8\times 10^{-3})$ eV.
The effect of this enhancement is particularly 
important in the case of NO neutrino mass spectrum. 
The non-standard contributions are so large 
at the indicated small values of  $m_{1(3)}$ that 
for $m_{1} \lesssim 2\times 10^{-4}$ eV 
in the NO case, they are strongly disfavored (if not ruled out) 
even for the minimal value 
of $M^N_{0\nu}/M^\nu_{0\nu} = 22.2$ 
by the existing upper limits from the KamLAND-Zen and GERDA 
experiments. In the IO case, the new contributions are 
also strongly disfavored  for $M^N_{0\nu}/M^\nu_{0\nu} = 76.3$; 
for  $M^N_{0\nu}/M^\nu_{0\nu} = 22.2$ they are 
disfavored for $\alpha = \beta = 0$, while for $\alpha = \pi$, $\beta = 0$, 
they are still compatible with the  KamLAND-Zen and GERDA 
conservative upper limits. 
We find, in general,  that in both NO and IO cases
the new  non-standard contributions due to $N_j$ and $S_k$ exchange 
are dominant over the standard light neutrino exchange contribution  at 
values of the lightest neutrino mass $m_{1(3)} \sim (10^{-4} - 10^{-2})$ eV:
$|m^{\nu+|N+S|}_{ee}| > (>>) |m^\nu_{ee}|$, 
where  $m^{\nu+|N+S|}_{ee}$ is the generalised effective Majorana mass (GEMM) 
which accounts for all contributions to the $0\nu\beta\beta$ decay amplitude 
(Eqs. (\ref{halflife:no-int}),  (\ref{eqn:mee_SR})
and   (\ref{eq:int})), and $m^\nu_{ee}$ is the effective Majorana 
mass associated with the standard light neutrino exchange contribution 
(Eq. (\ref{effmassnu})). The effective Majorana mass $|m^S_{\beta\beta,R}|$ 
associated with $S_k$ exchange contribution was shown to be practically 
independent of the Majorana phases $\alpha$ and $\beta$,  
while that due to exchange of  $N_j$, $|m^N_{\beta\beta,R}|$,
exhibits strong dependence on $\alpha$ and $\beta$ 
similar to $|m^\nu_{ee}|$. 

 For NO spectrum and $\alpha = \beta = 0$, 
the inequality  $|m^{\nu+|N+S|}_{ee}|> (>>) |m^\nu_{ee}|$
always holds in the interval of values of
 $10^{-4}~{\rm eV} \lesssim m_{1} \lesssim 8\times 10^{-3}$ eV 
where the non-standard contributions are significant.
In  this interval and for  $M^N_{0\nu}/M^\nu_{0\nu} = 76.3~(22.2)$, 
$|m^{\nu+|N+S|}_{ee}|\gtap 0.009~(0.007)$ eV.
With the exception of a very narrow interval 
around $m_{1} \cong 4.0~(2.5)\times 10^{-3}$ eV at which the 
quoted minimum of $|m^{\nu+|N+S|}_{ee}|$ takes place,  
we have $|m^{\nu+|N+S|}_{ee}|\gtap 0.010$ eV.  
In most of the considered intervals of values of 
$m_1$ the $0\nu\beta\beta$ decay  half-life 
$T^{\nu +|N_R + S_L|}_{1/2} \lesssim 10^{28}$ yrs. 

In the case of NO spectrum, $\alpha =\pi$, $\beta = 0$ and 
$M^N_{0\nu}/M^\nu_{0\nu} = 76.3~(22.2)$,  
we find that $|m^{\nu+|N+S|}_{ee}|\gtap 0.010$ eV, 
and  $T^{\nu +|N_R + S_L|}_{1/2} \lesssim 10^{28}$ yrs,
at $m_{1} \lesssim 2.0~(1.5)\times 10^{-3}$ eV.
For the minimal value of  $|m^{\nu+|N+S|}_{ee}|$ 
we get ${\rm min}(|m^{\nu+|N+S|}_{ee}|) \cong 9~(3)\times 10^{-4}$ eV.
It takes place at $m_{1} \cong 9.0~(8.0)\times 10^{-3}$ eV.
We note that  $|m^\nu_{ee}|$ goes through zero at 
$m_{1} \cong 2.26\times 10^{-3}$ eV, while 
$|m^{\nu+|N+S|}_{ee}|\cong 18.8~(5.3)\times 10^{-3}~$ eV 
at this value of $m_1$. Thus, the strong suppression of the 
$0\nu\beta\beta$ decay rate at $m_{1} \sim 2.26\times 10^{-3}$ eV 
and in the interval $m_1 \cong (1.5 - 8.0)\times 10^{-3}$ eV 
in the case of only standard contribution due to the exchange of 
light Majorana neutrinos $\nu_i$ 
(Fig. \ref{plot:onubb2}, upper left panel)
is avoided due to the new non-standard contributions. 
 
For IO spectrum, we find that 
$|m^{\nu+|N+S|}_{ee}|\gtap 0.015$ eV 
for $m_3 < 9\times 10^{-3}$ eV, where 
$|m^{\nu+|N+S|}_{ee}| >  |m^\nu_{ee}|$ 
for any of the considered values of $M^N_{0\nu}/M^\nu_{0\nu}$
and of $\alpha$ and $\beta$. The approximate 
equality  $|m^{\nu+|N+S|}_{ee}| \cong  |m^\nu_{ee}|$ 
holds at $m_3\gtap 10^{-2}$ eV.
For all considered values of the parameters 
the predicted half-life  
$T^{\nu +|N_R + S_L|}_{1/2} \lesssim 10^{28}$ yrs, while  
in the case of $\alpha = \beta = 0$, 
we have  $T^{\nu +|N_R + S_L|}_{1/2} \lesssim 2\times 10^{27}$ yrs. 

It follows from our results that in most of the parameter space of 
the considered model, the predictions for the 
$0\nu\beta\beta$ decay generalised effective Majorana 
mass and half-life are within the sensitivity range of the 
planned next generation of neutrinoless double beta decay 
experiments LEGEND-200 (LEGEND-1000), nEXO, KamlAND-Zen-II, CUPID, NEXT-HD 
(see \cite{Giuliani:2019uno,Agostini:2022zub} 
and references quoted therein).

\section*{Acknowledgements}
Purushottam Sahu would like to acknowledge the Ministry of Education, 
Government of India for financial support. 
PS also acknowledges the support from the Abdus Salam International Centre 
for Theoretical Physics (ICTP) under the ``ICTP
Sandwich Training Educational Programme (STEP)'' SMR.3676 and SMR.3799.
The work of S. T. P. was supported in part by the European
Union's Horizon 2020 research and innovation program under the 
Marie Sk\l{}odowska-Curie grant agreement No.~860881-HIDDeN, by the Italian 
INFN program on Theoretical Astroparticle Physics and by the World Premier 
International Research Center Initiative (WPI Initiative, MEXT), Japan.
STP would like to thank Kavli IPMU, University of Tokyo, 
where part of this study was performed for the kind hospitality.

\section{Appendix :  \\
Derivation of neutrino masses and mixings in left-right double 
seesaw model (LRDSM)}
\label{app:lrdssm}
\subsection{LRDSM mass matrix} 
We discuss here the implementation and derivation of double seesaw mechanism in the considered left-right symmetric model. The neutral fermions needed for LRDSM are active left-handed neutrinos, $\nu_L$, active right-handed neutrinos, $N_R$ and sterile neutrinos, $S_L$. The relevant mass terms are given by 
\begin{eqnarray}
\mathcal{L}_{\rm LRDSM} &=& \mathcal{L}_{M_D}+\mathcal{L}_{M_{RS}} 
+ \mathcal{L}_{M_S} \nonumber \\
\mathcal{L}_{M_D} &=& - \sum_{\alpha, \beta} \overline{\nu_{\alpha L}} [M_D]_{\alpha \beta} N_{\beta R} \mbox{+ h.c.}\nonumber \\
\mathcal{L}_{M_{RS}} &=&  \sum_{\alpha, \beta} \overline{S_{\alpha L}} [M_{RS}]_{\alpha \beta} N_{\beta R} \mbox{+ h.c.}
\nonumber \\
\mathcal{L}_{M_{S}} &=&  \frac{1}{2} \sum_{\alpha, \beta} \overline{S^c_{\alpha L}} [M_{S}]_{\alpha \beta} S_{\beta L} \mbox{+ h.c.}
\end{eqnarray}
The flavour states for active left-handed neutrinos $\nu_{\alpha L}$, right-handed neutrinos $N_{\beta R}$ and sterile neutrinos $S_{\gamma L}$ are defined as follows
\begin{eqnarray}
\nu_{\alpha L} = \begin{pmatrix}
\nu_{eL} \\ \nu_{\mu L} \\ \nu_{\tau L}
\end{pmatrix}\, , ~ 
N_{\beta R} = \begin{pmatrix}
N_{1 R} \\ N_{2 R} \\ N_{3 R}
\end{pmatrix}\, , ~
S_{\gamma L} = \begin{pmatrix}
S_{1 L} \\ S_{2 L} \\ S_{3 L}
\end{pmatrix}\,  
\end{eqnarray}
Similarly, their mass states can be written as,
\begin{eqnarray}
\nu_{i L} = \begin{pmatrix}
\nu_{1L} \\ \nu_{2 L} \\ \nu_{3L}
\end{pmatrix}\, , ~ 
N^c_{j R} = \begin{pmatrix}
N^c_{1 R} \\ N^c_{2 R} \\ N^c_{3 R}
\end{pmatrix}\, , ~ 
S_{k L} = \begin{pmatrix}
S_{1 L} \\ S_{2 L} \\ S_{3 L}
\end{pmatrix}\,
\end{eqnarray}

The $9 \times 9$ neutral lepton mass matrix in 
the basis $\left(\nu_L, N^c_R, S_L\right)$ is given by
\begin{equation}
 \mathcal{M}_{LRDSM}=
 \left[ 
\begin{array}{c | c} 
  \begin{array}{c c} 
     \blue{\bf 0} & M_D \\ 
     M^T_D & \blue{\bf 0}
  \end{array} & 
  \begin{array}{c} 
     \red{0} \\ M_{RS}
  \end{array} \\ 
\hline 
  \begin{array}{c c} 
     \,~\red{0}  &\quad  M^T_{RS}
  \end{array} & 
  \begin{array}{c} 
     \blue{M_S}
  \end{array} \\
 \end{array} 
\right]
\label{eqn:dss-a}       
\end{equation}
where each elements of the matrix is a $3 \times 3$ matrix. Here $M_D$ is the Dirac neutrino mass matrix connecting $\nu_L - N_R$, $M_{RS}$ is the mixing matrix in the $N_R-S_L$ sector, $M_S$ is the Majorana mass matrix for sterile neutrino $S_L$. 

In order to diagonalize the above mass matrix we have used the following mass hierarchy,
\begin{equation}
 M_D < M_{RS} < M_{S} .
\end{equation}

The diagonalisation of $\mathcal{M_{\rm LRDSM}}$ after changing it from flavour basis to mass basis is done by a generalized unitary transformation as, 
\begin{eqnarray}
&&\mid \Psi \rangle_{\rm flavor} = V\mid \Psi \rangle_{\rm mass}\\
&\mbox{or,}& \begin{pmatrix}
\nu_{\alpha L}\\ N^c_{\beta R}\\ S_{\gamma L}
\end{pmatrix}
= 
\begin{pmatrix}
V_{\alpha i}^{\nu \nu} & V_{\alpha j}^{\nu N} & V_{\alpha k}^{\nu S}\\V_{\beta i}^{ N \nu} & V_{\beta j}^{NN} & V_{\beta k}^{N S}\\V_{\gamma i}^{S \nu} & V_{\gamma j}^{S N} & V_{\gamma k}^{SS}
\end{pmatrix}               
\begin{pmatrix}
\nu_{i}\\ N^c_{j}\\ S_{k}
\end{pmatrix} \, \\
&&   V^{\dagger} \mathcal{M_{\rm LRDSM}} V^* 
= \mathcal{\widehat{M}}_{\rm LRDSM} \nonumber \\
&&\hspace*{2.2cm} = \mbox{diag} \left(m_{i},m_{N_j},m_{S_k} \right) \nonumber \\
&&\hspace*{2.2cm} = \mbox{diag} \left(m_{1},m_{2},m_{3},m_{N_1},m_{N_2},m_{N_3},m_{S_1},m_{S_2},m_{S_3} \right)
\end{eqnarray}
Here the indices $\alpha, \beta, \gamma$ run over three generations of light left-handed neutrinos, heavy right-handed neutrinos and sterile neutrinos respectively, whereas the indices $i,j,k$ run over corresponding mass states. 

\subsection{Block Diagonalization of double seesaw mass matrix in LRDSM}
 
\label{app:lrdsm}
Let's write the matrix in eq.\ref{eqn:dss-a} as,

\begin{eqnarray}
\mathcal{M}_{LRDSM}=\mathcal{M_{\nu}}=\begin{pmatrix}
              \mathcal{M}_L & \mathcal{M}_D  \\
              \mathcal{M}^T_D & \mathcal{M}_S
             \end{pmatrix} \,, ~ \mbox{where,} \nonumber \\
   \mathcal{M}_L = \begin{pmatrix}
                                  0 & M_D \\
                                  M^T_D  & 0
                                 \end{pmatrix}
                           , 
                           \mathcal{M}_D=  \begin{pmatrix}
                                  0 \\ M_{RS}
                                 \end{pmatrix},\mathcal{M}_S = M_S\,
\end{eqnarray}
The complete block diagonalization is achieved in two steps by recursively integrating out the heavier modes as
\begin{eqnarray}
\mathcal{W}_{1}^{\dagger} \mathcal{M_{\nu}} \mathcal{W}_{1}^* 
      =\mathcal{\hat M_{\nu}^\prime} &\mbox{and}& \mathcal{W}_{2}^{\dagger} \mathcal{\hat M_{\nu}^\prime} \mathcal{W}_{2}^* 
      =\mathcal{\hat{M}_{\nu}}
\end{eqnarray}
where $\mathcal{\hat M_{\nu}^\prime}$ is block diagonalised $9\times9$ matrix after integrating out the heaviest mode and $\mathcal{\hat{M}_{\nu}}$
is the block diagonalised $9\times9$ matrix after integrating out the next heaviest mode. The transformation matrix $\mathcal{W}_{1}$ can be written as a general unitary matrix in the form
\begin{eqnarray}
\mathcal{W}_{1}^* &=& \begin{pmatrix}
              \sqrt{1-\mathcal{B}\mathcal{B}^{\dagger}}& \mathcal{B}  \\
              -\mathcal{B}^{\dagger}& \sqrt{1-\mathcal{B}^{\dagger}\mathcal{B}}
             \end{pmatrix} 
\end{eqnarray} where $\mathcal{B}$ is a $ 6 \times 3 $ dimensional matrix.
\begin{eqnarray}
\sqrt{1-\mathcal{B}\mathcal{B}^{\dagger}} &=& 1 - \frac{1}{2}\mathcal{B}\mathcal{B}^{\dagger}-\frac{1}{8}\left(\mathcal{B}\mathcal{B}^{\dagger}\right)^2
+ \cdots \nonumber \\
\mathcal{B} = \sum \mathcal{B}_i 
\end{eqnarray}
At leading order it looks like 
\begin{eqnarray}
\sqrt{1-\mathcal{B}\mathcal{B}^{\dagger}} &\simeq& 1 - \frac{1}{2}\mathcal{B}\mathcal{B}^{\dagger}-\frac{1}{8}\left(\mathcal{B}_1\mathcal{B}_2^{\dagger}+\mathcal{B}_2\mathcal{B}_1^{\dagger}\right)
\end{eqnarray}
The form of mixing matrix $\mathcal{B}_1^{\dagger}$ and $\mathcal{B}_1$ is given by
\begin{eqnarray}
\mathcal{B}_1^{\dagger}=\mathcal{M}_S^{-1} \cdot \mathcal{M^T_D} &=& M_S^{-1}\cdot\begin{pmatrix}
                                  0 & M_{RS}^T
                                 \end{pmatrix} 
                                 = \begin{pmatrix}
                                  0 &M^{-1}_S \cdot M_{RS}^T
                                 \end{pmatrix}  \nonumber\\
\mathcal{B}_1 &=& \begin{pmatrix}
                                  0 \\
                                M_{RS}M^{-1}_S
                                 \end{pmatrix} \nonumber\\
\sqrt{1-\mathcal{B}\mathcal{B}^{\dagger}} &\simeq& \begin{pmatrix}
                                  1& 0\\
                                  0  & 1-\frac{1}{2}M_{RS}M^{-1}_S  \cdot M^{-1}_S M^T_{RS}
                                 \end{pmatrix}\nonumber\\
\sqrt{1-\mathcal{B}^{\dagger}\mathcal{B}} &\simeq& 1-\frac{1}{2} M^{-1}_S M^T_{RS} \cdot  M_{RS}M^{-1}_S                                 
\end{eqnarray}
Thus, the first block diagonalised mixing matrix $\mathcal{W}_1$ becomes,
\begin{eqnarray}
\mathcal{W}_1 = \begin{pmatrix}
                                  1 & 0 & 0\\
                                  0 &1-\frac{1}{2}M_{RS}M^{-1}_S \cdot M^{-1}_S  M_{RS}^T& M_{RS}M^{-1}_S \\
                                  0& -M^{-1}_SM_{RS}^T & 1-\frac{1}{2} M^{-1}_SM_{RS}^T \cdot M_{RS}M^{-1}_S
                                 \end{pmatrix}
 \end{eqnarray}

After this diagonalization, $\mathcal{\hat M_{\nu}^\prime}$ has the following form.
\begin{eqnarray}
\mathcal{\hat M_{\nu}^\prime} &=& \begin{pmatrix}
                                  \mathcal{M}_{eff} & 0\\
                                  0  & \mathcal{M}_S
                                 \end{pmatrix}
 \end{eqnarray}
where,
\begin{eqnarray}
\mathcal{M}_{eff} &=& \mathcal{M}_L - \mathcal{M}_D \mathcal{M}^{-1}_S \mathcal{M}^T_D \nonumber \\
&=&\begin{pmatrix}
                     0 & M_D \\
                     M^T_D   & 0 
                    \end{pmatrix} 
    -  \begin{pmatrix}
         0 \\
         M_{RS}
        \end{pmatrix} M^{-1}_S
                  \begin{pmatrix}
                   0 & M^T_{RS} 
                  \end{pmatrix} \nonumber \\
&=&   \begin{pmatrix}
       0    & M_D  \\
       M^T_D          &  -M_{RS} M^{-1}_S M^T_{RS}
       \end{pmatrix} \, 
\end{eqnarray}
$\mathcal{M}_{eff}$ can be further diagonalised by $\mathcal{W}_2$ as,
\begin{eqnarray}
\mathcal{S}^{\dagger} \mathcal{M}_{eff} \mathcal{S}^* &=&\begin{pmatrix}
       m_{\nu}   & 0 \\
       0       &  M_R
       \end{pmatrix},
\end{eqnarray}
where, 
\begin{eqnarray}
 m_\nu &=& - M_D \left( -M_{RS} M^{-1}_S M^T_{RS} \right)^{-1} M^T_D , \nonumber \\
 && M_R = -M_{RS} M^{-1}_S M^T_{RS}.
\end{eqnarray}
The transformation matrix $\mathcal{S}$ is
\begin{eqnarray}
\mathcal{S}^* &=& \begin{pmatrix}
              \sqrt{1-\mathcal{A}\mathcal{A}^{\dagger}}& \mathcal{A}  \\
              -\mathcal{A}^{\dagger}& \sqrt{1-\mathcal{A}^{\dagger}\mathcal{A}}
             \end{pmatrix}              
\end{eqnarray}
such that, 
\begin{eqnarray}
\mathcal{A}^{\dagger}&=&\left( -M_{RS} M^{-1}_S M^T_{RS} \right)^{-1} M_D \\
&& = -M^{-1^T}_{RS} M_S M^{-1}_{RS}M_{D} = X^{\dagger}
\end{eqnarray}
\begin{eqnarray}
\text{Thus,} ~\mathcal{W}_{2} &=& \begin{pmatrix}
  \mathcal{S}  & 0\\
   0  &  1
\end{pmatrix}\\ &=& \begin{pmatrix}
1-\frac{1}{2}X X^{\dagger}  & X &0\\
   -X^{\dagger}  & 1-\frac{1}{2}X^{\dagger}X & 0 \\
   0 &0 & 1

\end{pmatrix} \nonumber \\
\end{eqnarray}

\subsection{Complete diagonalization and physical neutrino masses}
After block diagonalization, the mass matrix for the three types of neutrinos are further diagonalized by respective unitary mixing matrices $U_\nu$, $U_N$, $U_S$ resulting in physical masses for the neutrinos as follows.

\begin{eqnarray}
U_{9 \times 9} = \begin{pmatrix}
{ U_\nu}_{3 \times 3}  &  { 0}_{3 \times 3}  & { 0}_{3 \times 3}  \\
{ 0}_{3 \times 3}  &  {U_N}_{3 \times 3}    & { 0}_{3 \times 3}  \\
{ 0}_{3 \times 3}  &  { 0}_{3 \times 3}  & {U_S}_{3 \times 3}
\end{pmatrix} 
\end{eqnarray}

\begin{eqnarray}
&&U^\dagger_\nu m_\nu U^*_\nu = \hat{m}_\nu = \mbox{diag}\left(m_{\nu_1}, m_{\nu_2}, m_{\nu_3} \right) \nonumber \\
&&U^\dagger_N M_N U^*_N = \hat{M}_N = \mbox{diag}\left(M_{N_1}, M_{N_2}, M_{N_3} \right) \nonumber \\
&&U^\dagger_S M_S U^*_S = \hat{M}_S = \mbox{diag}\left(M_{S_1}, M_{S_2}, M_{S_3} \right)
\end{eqnarray}

The complete mixing matrix  now becomes,
\begin{eqnarray}
V &=& \mathcal{W}_1\cdot \mathcal{W}_2 \cdot \mathcal{U}\nonumber\\
&=& \begin{pmatrix}
1  & 0 &0\\
   0  & 1-\frac{1}{2} Y Y^{\dagger} & Y \\
   0 &-Y^{\dagger} & 1-\frac{1}{2}  Y^{\dagger} Y
 
\end{pmatrix}
\cdot
\begin{pmatrix}
1-\frac{1}{2}X X^{\dagger}  & X &0\\
   -X^{\dagger}  & 1-\frac{1}{2}X^{\dagger}X & 0 \\
   0 &0 & 1
   \end{pmatrix} 
   \cdot \begin{pmatrix}
U_{\nu}  & 0&0\\
   0 & U_N& 0 \\
   0 &0 & U_S
   \end{pmatrix}  \nonumber\\
&=& \begin{pmatrix}
U_{\nu}\left(1-\frac{1}{2}X X^{\dagger} \right)  & U_{N}X &0\\
   -U{\nu}X^{\dagger}\left(1-\frac{1}{2}YY^{\dagger} \right)  & U_N \left(1-\frac{1}{2}X^{\dagger}X \right) \left(1-\frac{1}{2}YY^{\dagger} \right) & U_{S} Y \\
   U{\nu}X^{\dagger}Y^{\dagger} &-U_NY^{\dagger} & U_S\left(1-\frac{1}{2}Y^{\dagger}Y \right)
\end{pmatrix}   
\label{mixingmatrix1}
\end{eqnarray}

where $ X^{\dagger} = -M^{-1}_{RS} M_S M^{-1}_{RS}M_{D}$, $Y = M_{RS} M^{-1}_{S}$ and fixing the typical magnitudes for $ M_D \simeq $~0.1\,MeV, $ M_{RS} \simeq $~1\,TeV, $M_{S} \simeq$ 10\,TeV we get $X \simeq 10^{-6} $, $Y\simeq 0.1$.  Since $U_{\nu}$, $U_N$ and $U_S$ are of $\mathcal{O}(1)$, the matrix elements of $\mathcal{V}$ are approximated to be
\begin{eqnarray}
\begin{pmatrix}
\mathcal{V}_{\alpha i}^{\nu \nu} & \mathcal{V}_{\alpha j}^{\nu N} & \mathcal{V}_{\alpha k}^{\nu S}\\\mathcal{V}_{\beta i}^{ N \nu} & \mathcal{V}_{\beta j}^{NN} & \mathcal{V}_{\beta k}^{NS}\\\mathcal{V}_{\gamma i}^{S \nu} & \mathcal{V}_{\gamma j}^{SN} & \mathcal{V}_{\gamma k}^{SS}
\end{pmatrix}  \simeq \begin{pmatrix}
{\cal O}(1.0)       & {\cal O}(10^{-6})     & 0  \\
{\cal O}(10^{-6})      & {\cal O}(1.0)    & {\cal O}(0.1)    \\
{\cal O}(10^{-7})   & {\cal O}(0.1)    & {\cal O}(1.0) 
\end{pmatrix}
\end{eqnarray} 
which generates sizable contribution to neutrinoless double beta decay.



\clearpage
\bibliographystyle{utcaps_mod}
\bibliography{DSS_onubb_LR}
\end{document}